   \documentclass[12pt]{article}
\if@twoside\oddsidemargin25mm
 \else\oddsidemargin25mm\evensidemargin25mm\marginparwidth25mm\fi%
\begin{document}
\renewcommand{\theequation}{\thesection.\arabic{equation}}
\csname @addtoreset\endcsname{equation}{section}
\csname @addtoreset\endcsname{theorem}{section}
\csname @addtoreset\endcsname{defn}{section}
\csname @addtoreset\endcsname{lemma}{section}
\newcommand{\eqn}[1]{(\ref{#1})}
\newcommand{\ft}[2]{{\textstyle\frac{#1}{#2}}}
\newcommand{\QED}{{\hspace*{\fill}\rule{2mm}{2mm}\linebreak}}
\newcommand{\dr}{\raise.3ex\hbox{$\stackrel{\leftarrow}{\delta}$}{}}
\newcommand{\dl}{\raise.3ex\hbox{$\stackrel{\rightarrow}{\delta}$}{}}
\newtheorem{lemma}{Lemma}
\renewcommand{\thelemma}{\thesection.\arabic{lemma}}
\newtheorem{theorem}{Theorem}
\renewcommand{\thetheorem}{\thesection.\arabic{theorem}}
\newtheorem{defn}{Definition}
\renewcommand{\thedefn}{\thesection.\arabic{defn}}
\newcommand{\dkt}{\delta _{KT}}
\newcommand{\dtx}[1]{\frac{\tilde\partial}{\partial\xi^{#1}}}
\newcommand{\mm}{m} \newcommand{\nn}{n} \newcommand{\kk}{k}
\newcommand{\sub}{{\cal C}}
\newcommand{\subtwo}[2]{{\cal C}_{(#1,#2)}}
\newcommand{\subtot}{{\cal C}}
\newcommand{\subnoa}{{\cal C}^\Phi}
\newcommand{\cc}[2]{c_{#1}^{#2}} \newcommand{\LL}[2]{L^{#1}_{#2}}
\newcommand{\AAA}{a} \newcommand{\BB}{b} \newcommand{\CC}{c}
\newcommand{\lie}[1]{{\cal L}_{#1}}
\newcommand{\THE}[1]{\Theta^{#1}}
\let\ssk=\smallskip \let\msk=\medskip \let\bsk=\bigskip
\let\qd=\quad \let\qqd=\qquad \def\qqqd{\qquad\qquad}
\let\a=\alpha \let\be=\beta \let\g=\gamma \let\de=\delta
\let\ep=\varepsilon \let\z=\zeta \let\h=\eta \let\Th=\theta
\let\dh=\vartheta \let\k=\kappa \let\la=\lambda \let\m=\mu
\let\n=\nu \let\x=\xi \let\r=\rho \let\si=\sigma
\let\om=\omega \let\ps=\psi
\let\ph=\varphi \let\Ph=\phi \let\PH=\Phi \let\Ps=\Psi
\let\Om=\Omega \let\Si=\Sigma \let\TH=\Theta
\let\La=\Lambda \let\Ga=\Gamma \let\De=\Delta
\def\da{{\dot \a}} \def\dbe{{\dot \be}}
\def\L{\de}
\def\0#1#2{\frac{#1}{#2}} \def\s0#1#2{\mbox{\small{$\frac{#1}{#2}$}}}
\def\2{{\times}} \def\3{\vec }
\def\5{\bar }  \def\6{\partial } \def\7{\hat } \def\4{\tilde }
\def\ds{\displaystyle } \def\lb{\left(} \def\rb{\right)}
\def\lra{ \leftrightarrow } \let\LRA=\Leftrightarrow
\let\then=\Rightarrow
\def\bea{\begin{eqnarray}} \def\eea{\end{eqnarray}}
\def\beann{\begin{eqnarray*}} \def\eeann{\end{eqnarray*}}
\def\beq{\begin{equation}} \def\eeq{\end{equation}}
\def\ba{\begin{array}} \def\ea{\end{array}}
\def\ben{\begin{enumerate}} \def\een{\end{enumerate}}
\def\cO{{\cal O}} \def\cB{{\cal B}} \def\cG{{\cal G}}
\def\cA{{\cal A}}  \def\cL{{\cal L}}
\def\cR{{\cal R}} \def\cN{{\cal N}} \def\cM{{\cal M}}
\def\cF{{\cal F}} \def\cT{{\cal T}} \def\cH{{\cal H}}
\def\cK{{\cal K}}
\def\cW{{\cal W}} \def\cP{{\cal P}} \def\cS{{\cal S}}
\def\Gl#1{(\ref{#1})}
\begin{titlepage}
\begin{flushright} KUL--TF--95/17 \\
   hep-th/9509035  \\
                   September 1995\\
\end{flushright}

\vfill
\begin{center}
{\LARGE
The BRST--antibracket cohomology of
$2d$ gravity\\
\vskip 1.5mm conformally coupled to scalar matter}\\
\vfill
{\Large {F}riedemann {Brandt} $^1$,
{W}alter {Troost} $^{2,3}$\\
\vskip 1.5mm and {A}ntoine {Van Proeyen}
$^{2,4}$} \\
\vfill
Instituut voor Theoretische Fysica
        \\Katholieke Universiteit Leuven
        \\Celestijnenlaan 200D
        \\B--3001 Leuven, Belgium\\
\end{center}
\vfill
\begin{center}
{\bf Abstract}
\end{center}
\begin{quote}
\small
We compute completely the BRST--antibracket cohomology on local functionals
in two-dimensional Weyl invariant gravity for given classical field content
(two dimensional metric and scalar matter fields) and gauge symmetries
(two dimensional diffeomorphisms and local Weyl transformations). This
covers the determination of all classical actions,
of all their rigid symmetries,
of all background charges and of all
candidate gauge anomalies. In particular we show that the antifield
dependence can be entirely removed from the anomalies and that,
if the target space has isometries, the condition
for the absence of matter field dependent Weyl anomalies is more general than
the familiar `dilaton equations'.
\vspace{2mm} \vfill \hrule width 3.cm
{\footnotesize
\noindent $^1$ Junior fellow of the research council (DOC) of the
K.U. Leuven;\\
\phantom{$^1$} E--mail : Friedemann.Brandt@fys.kuleuven.ac.be \\
\noindent $^2$ Onderzoeksleider, NFWO, Belgium\\
\noindent $^3$
E--mail : Walter.Troost@fys.kuleuven.ac.be\\
\noindent $^4$
E--mail : Antoine.VanProeyen@fys.kuleuven.ac.be}
\normalsize
\end{quote}
\end{titlepage}
\section{Introduction}

Wess and Zumino \cite{WZcc} have shown that
anomalies satisfy consistency conditions.
In turn, these consistency conditions can be used as a tool to classify
possible anomalies. The solution of these conditions is one instance of
a cohomology calculation: the cohomology of the BRST operator
on local functionals. In ghost number one this cohomology indeed provides
all solutions of the Wess--Zumino consistency conditions, i.e.
determines the general form of possible anomalies.
Other instances of cohomological analysis
are equally important physically. For example, at ghost number zero,
it yields the most general action that is compatible with a given
symmetry, and in ghost number $(-1)$ it provides all rigid
symmetries of the action \cite{BBH}.

The ingredients needed to perform the cohomological
analysis are: the field content,
the gauge transformation laws of the fields,
and the classical equations of motion (e.o.m.).
The e.o.m. intervene in two places: on the one hand,
the BRST operator may be nilpotent only on shell, and on the other hand,
classical observables
are physically equivalent if their difference
is proportional to e.o.m.. The BRST cohomology modulo this equivalence
is called the {\em weak} BRST cohomology.

For a class of theories which contain diffeomorphisms in the gauge
group, a general method for the analysis of the cohomology
was set up in \cite{FrBrStructure}. This class of theories
contains for instance Einstein gravity as well as supergravity
theories, but it
does not cover {\it all} diffeomorphism invariant theories.
In particular it does not include
Weyl invariant
gravity theories in two dimensions, such as the
standard bosonic string theory described at the classical level by
the Polyakov action. The reason is the absence
of an independent Weyl gauge field\footnote{One could introduce
the Weyl gauge field,
but not without further ado. See section
\ref{ss:tensors}.} in these theories whose
presence would be a crucial prerequisite for applying the methods of
\cite{FrBrStructure}. As we shall see,
this is responsible for considerable differences
in the cohomological analysis and its results for these
two dimensional models when compared with more ``standard''
gauge theories such as Yang--Mills theory or Einstein gravity.

In fact, in spite of its central importance to
string theory, the BRST cohomology on local functionals
has, to our knowledge, never been analysed
exhaustively in the literature for the case of Weyl invariant
$d=2$ gravity theories. (For recent contributions, see
\cite{BanLaz,OSS,Tataru,buchb}).
The filling of this gap is the purpose of this
paper, for the case that all matter fields are scalar fields.
The results have been announced already partly in \cite{cam,turkproc}.
Some of them are of course common knowledge. In particular this
holds for results on the strong cohomology, i.e. for the
BRST cohomology which does not take the e.o.m. into account.
We shall see however that many important aspects of the theory
show up only in the weak cohomology,
such as the rigid symmetries and the so-called background charges or
the dilaton terms which can
cancel Weyl anomalies and are well-known in string theory \cite{strinbf}.
Moreover we show in a companion paper \cite{paper2}
that the results on the weak cohomology allow one in particular cases to
construct interesting generalizations of the theory
(so-called consistent deformations \cite{BH}) which are
reminiscent of non-critical string theories
and possibly provide new models for the latter.

The necessity of re--analysing the cohomology
appears clearly if one would blindly extend
the results of \cite{FrBrStructure} to the present situation:
one would conclude, for example, that candidate anomalies can
be assumed to depend only on the undifferentiated
Weyl ghost $c$, the undifferentiated zweibeins
and on tensor fields (the two dimensional
Riemann curvature $R$, the matter fields
and their covariant derivatives) but not, e.g., on the
diffeomorphism ghosts. However, it is well known
that this is incorrect: the important Weyl anomaly
\begin{equation}
{\cal A}_0 = \int d^2x\ c\sqrt{g}R\ ,
\label{A0}
\end{equation}
can be split, after the addition
of BRST--exact terms,
in a left and right handed part, which separately
solve the consistency equations \cite{Becchi,BBBeltr}.
These two parts are
cohomologically inequivalent, involve the diffeomorphism
ghosts and cannot be written entirely
in terms of the Weyl ghost and tensor fields up to BRST--exact terms.
We will show, among other things, how the absence of the
Weyl gauge field modifies the
conclusions of \cite{FrBrStructure} in a way that implies this result.

The starting point of our analysis will be the field content, and
the symmetry transformations. These will include the diffeomorphisms,
of course, and the Weyl transformations. We will realise them on
the scalar
matter fields, and on the two-dimensional metric%
\footnote{Alternatively, one could introduce zweibeins and include
Lorentz invariance: this amounts to a technical difference only. Since for
scalars one does not need the zweibeins, we refrain from introducing them.}.
The symmetries will entail the corresponding ghosts, in our case
diffeomorphism ghosts and the Weyl ghost.

Although it may be customary, after Faddeev and Popov,
to introduce also antighosts, this is in
fact quite superfluous to investigate the classical cohomology. This is
especially obvious in the Batalin--Vilkovisky (BV) framework
\cite{BV,anombv,turkproc,GomisParis,BVboek}
(also called the field--antifield formalism)%
\footnote{We will (very summarily) introduce the
necessary ingredients, and the relation with BRST cohomology, in
section \ref{ss:action}.}.
The reason is that antighosts (as well as their antifields), and
Lagrange multiplier fields that come with gauge fixing,
are introduced as so--called {\em trivial systems}, implying
that they leave the cohomology groups unchanged.
Therefore, the antighosts will be absent from our analysis.
A related feature is that no gauge fixing
is needed: the formulation of the calculation, and  its
results, are made entirely without reference to any gauge fixing,
and are therefore at every stage manifestly gauge independent.

The other side of the coin is that {\em antifields} are present. The BV
cohomology will then have to be analysed in a space of functionals of fields
and antifields. This has, for our purposes, the additional advantage
that it automatically takes into account the {\em weak} nature of the
relevant cohomology calculation since the antifields implement
the e.o.m. in the cohomological analysis. To exploit this last feature, we
have to know the classical action. This classical action itself need
not be fixed on beforehand however: it will be determined, in an
intermediate step, from the strong BRST cohomology in the space of
integrated local functionals with ghost number zero depending
on {\em fields only}, not on antifields.

The computation of the cohomology
is carried out in three main steps. First we map the
cohomological problem on integrated local functionals to
the analogous problem on local functions of the fields and antifields.
This map is quite standard and provided by the so-called descent
equations. In the second step we isolate and eliminate successively
trivial systems. This reduces the problem to a
set of equations for ``superfields'' in the
undifferentiated matter fields and first order derivatives
of the diffeomorphism ghosts. The third step consists in
solving these equations. Here we need the
explicit form of the action which is computed in an intermediate step
by solving the ``strong'' superfield equations first.

Our analysis is local in two senses: on the one hand
we work in the space of local functionals which are, by definition,
polynomial in derivatives of all the fields and antifields,
and on the other hand we ignore global aspects of the
base and target manifold completely.
\vspace{5mm}

Let us now give an outline of the paper.
In section \ref{ss:action} we will (very briefly) introduce the
necessary elements from the BV framework, and write down the elements of
the extended action that follow directly from our assumed symmetry
transformations. We will also describe more accurately the cohomology
calculation to be performed. In section~\ref{ss:simplectr} we
make a first change of variables, showing how the determinant of the
metric and the Weyl ghost occur as a trivial system when one introduces
Beltrami variables to describe the metric. The resulting chiral splitting
\cite{Becchi,BBBeltr} runs through the rest of the paper,
and also, technically, it simplifies the calculations significantly.
In section \ref{ss:descent} we perform, following \cite{grav,FrBrStructure},
the above mentioned first step of the computation
that takes us from local functionals to local functions
via descent  equations, and also give a short discussion of the type of
global considerations that we will {\em not} take into account in the rest of
the paper. In section \ref{ss:tensors} we prepare the second step
by introducing {\em chiral tensor fields} and
{\em covariant ghost variables}, the former being a generalisation of the
usual tensor fields that we will explain and the latter
forming a subset of the derivatives of ghosts. In section
\ref{ss:strategy} we then conduct the second step which reduces
the cohomological analysis to local functions
generated by only a few chiral tensor fields and covariant ghost
variables. There the above mentioned superfields show up.
We then compute in section \ref{ss:brstcoho} the strong BRST cohomology by a
first analysis of the equations these superfields have to satisfy.
This provides in particular the most general
classical action which we discuss in detail in section \ref{ss:genclact}.
We are then in the position to finish the calculation by
solving the (weak) superfield equations
completely. This is done in section \ref{ss:fullcoho} where
we also enumerate all the resulting solutions on the level of
local functions.
In section \ref{ss:interpr} we spell out the corresponding
local functionals for the most interesting cases (ghost numbers) and
discuss their physical significance.
Although cohomologically there is a complete chiral split, for example
for the anomalies, but also for background charges and counterterms, in
many cases of practical interest only the left--right symmetric
combinations are relevant.
In section \ref{ss:Weylanom} we therefore specialize our
results to that case. This makes the connection with the case of
primary importance for string theory,
where an  anomaly is tolerated in the Weyl symmetry only, and a dilaton
field is introduced.
We conclude with a discussion, including pointers to previously published
partial results. Finally, in the appendices, we collected various
formulas and
technical results, but also a side result on the relation of target
space reparametrisations with the cohomology.
\section{Assumptions and definition of the problem}\label{ss:action}

In the BV formalism,
the fundamental object is the antibracket defined for two arbitrary
functionals $F$ and $G$ of fields $\Phi^A$ and antifields $\Phi^*_A$
by
\begin{equation}
(F,G) =  F\frac{\dr }{\delta \Phi ^A}\cdot
\frac{\dl}{\delta \Phi ^*_A} G -
F\frac{\dr}{\delta \Phi ^*_A}\cdot\frac{\dl }{\delta \Phi ^A} G\ .
\label{abracket}
\end{equation}
The consistency equation for the anomaly ${\cal A}$ is then
\begin{equation}
{\cal S}{\cal A}\equiv(S,{\cal A})=0\ ,  \label{consista}
\end{equation}
where $S$ is the extended action (which itself satisfies the BV
'master' equation $(S, S)=0$).
Two solutions ${\cal A}$ and ${\cal A}'$  of \eqn{consista} are
equivalent anomalies (related by field redefinitions, change of
regularization, or local counterterms) iff
\begin{equation}
{\cal A}'-{\cal A}={\cal S}M \label{equiva}
\end{equation}
where $M$ is the integral of a local function (a `local functional').
Now, anomalies normally have ghost number 1.
Hence, what we have to solve for their
classification is the cohomology of ${\cal S}$ with ghost number~1
on local functionals. As we mentioned already we will not
restrict ourselves to this case but perform the
analysis for all other ghost numbers as well.
For ghost number 0 this is relevant e.g. for the
renormalization problem.

If the gauge algebra is `closed', antifields enter
only linearly in $S$. That will be the case here. The
`Slavnov' operator ${\cal S}$ can then be split in a `Koszul--Tate' operator
and the remaining part which we call the `BRST' operator $s$:
\begin{equation}
{\cal S}=\dkt + s\ .\label{defdkts}
\end{equation}
This splitting is related to the antifield number. The latter is
defined to be zero for fields (which have non--negative ghost
numbers), and minus the ghost number for antifields. $\dkt$ is the part
of ${\cal S}$ which lowers the antifield number (by 1), while $s$ is the
part which does not change the antifield number. For general gauge
theories with an `open' gauge algebra
there are also terms which raise the antifield number,
but not in the cases treated in this paper.
Note that on the fields we thus have
${\cal S}=s$ whereas on the antifields both $\dkt$ and $s$ are nonvanishing.
The expansion of ${\cal S}^2=0$ in antifield number implies
that $\dkt$ and $s$ are separately nilpotent and anticommute:
\begin{equation}
\dkt^2=s\dkt+\dkt s=s^2=0\ .
\label{nilpot}\end{equation}
The equation $s^2=0$ holds only due to the lack of further terms in
\eqn{defdkts} and is not true for gauge theories with
on open algebra where $s^2$ vanishes
only weakly, i.e. `up to field equations'.
\vspace{5mm}

We consider scalar fields $X^\mu$, $\mu=1,\ldots,D$ in interaction
with the $d=2$ metric fields $g_{\alpha\beta}=g_{\beta\alpha}$ with
$\alpha,\beta\in\{+,-\}$.
The coupling of the $X^\mu$ and $g_{\alpha\beta}$ is assumed to
be generally covariant and Weyl invariant at the classical
level. More precisely we require
the classical action, denoted by $S_{cl}(X^\mu, g_{\alpha\beta})$,
to be invariant under two dimensional diffeomorphisms and local dilatations
so that the extended action $S$ reads
\begin{eqnarray}
S=S_{cl}(X^\mu, g_{\alpha\beta}) - \int d^2x
\left( {\cal S}\Phi^A\right) \Phi^*_A\ , \label{proper}
\end{eqnarray}
where ${\cal S}\Phi^A$ denotes the BRST--transformations of
the fields corresponding to their transformations under
two dimensional diffeomorphisms and local dilatations,
\footnote{Differentiations will always act on everything to their right,
unless the scope is limited by the "$\cdot$" punctuation mark. }
\bea {\cal S}g_{\a\be}&=&\xi^\gamma \6_\gamma g_{\a\be}+
\6_\a \xi^\gamma\cdot g_{\gamma\be}+
\6_\be \xi^\gamma\cdot g_{\a\gamma}+c\, g_{\a\be}\ ;
\nonumber\\
{\cal S}X^\m &=&\xi^\a \6_\a X^\m\ ;\qd
{\cal S}\xi^\be =\xi^\a \6_\a \xi^\be\ ;\qd
{\cal S}c =\xi^\a \6_\a c\ .\label{gaugetrafo}\eea
Here $\xi^\alpha$ are the ghosts for general
coordinate transformations and
$c$ is the ghost for local dilatations.
The sets of fields and
antifields are accordingly given by
\[\{\Phi^A\}=\{X^\mu, g_{\alpha\beta}, \xi^\alpha, c\}\ ;
\qquad \{\Phi^*_A\}=\{X^*_\mu, g^{*\alpha\beta}, \xi^*_\alpha, c^*\}\ .\]
The ghosts and the antifields
$X^*_\mu$ and $g^{*\alpha\beta}$ are odd--graded whereas
$X^\mu$, $g_{\alpha\beta}$, $\xi^*_\mu$ and $c^*$ are
even graded.
With no loss of generality, $g^{*\alpha\beta}$ is  taken to be symmetric
since $g_{\alpha\beta}$ is symmetric
too%
\footnote{When treating symmetric tensors and their antifields
one may sum over $\alpha\geq\beta$, or one can work
symmetrically---which
we will do. Then we have to  take $\left(
g_{\alpha\beta}, g^{*\gamma\delta}\right) =\ft12\left(
\delta_\alpha^\gamma \delta_\beta^\delta
+\delta_\alpha^\delta\delta_\beta^\gamma\right) $. }.
The ghost number is zero for $X^\mu$ and $g_{\alpha\beta}$, $(-1)$ for
their antifields, one for the ghosts $\xi^\alpha$ and $c$, and $(-2)$
for the antifields of the ghosts.

We do not impose any restriction on
the classical action $S_{cl}$, except that it is a
local and regular\footnote{Regular dependence on the
fields $X^\mu$ and $g_{\alpha\beta}$ requires the action
to be well--defined within the allowed range of values these fields
may take. In the case of the metric this range is restricted by
$\det (g_{\alpha\beta})< 0$; in the case of the matter fields
we do not specify the range
since anyhow we neglect topological aspects, i.e. actually the
regularity requirement will not matter in the subsequent
analysis (see section~\ref{ss:descent}). Regularity also includes the
requirement that there exist solutions to the field equations,
differentiability of the action, and that in the set of local functions
that we consider all functions that vanish when the field equations
are satisfied are actually a linear combination of the field
equations.}
functional of the fields $X^\mu$ and $g_{\alpha\beta}$
and that
\eqn{proper} extends it to a proper (minimal) solution of the
BV master equation in the sense of \cite{BV}. This requires that
\begin{enumerate}
\item[(i)] the integrand of $S_{cl}$ is regular and depends polynomially on
the partial derivatives
of $X^\mu$ and $g_{\alpha\beta}$;
\item[(ii)] $S_{cl}$ is invariant under \eqn{gaugetrafo},
 i.e. it should just satisfy ${\cal S}S_{cl}=0$;
\item[(iii)] $S_{cl}$ has no nontrivial local symmetries apart from those
imposed by (ii).
\end{enumerate}
An extension of the requirement imposed by (i) on the integrand of
$S_{cl}$ serves as definition of local functions throughout the
paper and fixes thereby the space of functions and functionals
on which we will perform the cohomological analysis. Namely
a local function depends by definition polynomially on
the derivatives of the $X^\mu$ and $g_{\alpha\beta}$ and
on the (undifferentiated) ghosts, antifields and their
partial derivatives, whereas we allow for nonpolynomial
dependence on the undifferentiated $X^\mu$ and $g_{\alpha\beta}$.
Furthermore we allow a local function to depend explicitly
on the two dimensional coordinates $x^\a$
(see section \ref{ss:descent} for remarks on this point).
A local functional is by definition
an integrated local function of the fields and antifields.

The condition (ii) just requires $S_{cl}$ to be invariant
under diffeomorphisms and local Weyl transformations.

(iii) guarantees the
properness of ${\cal S}$, i.e. the
completeness of our approach in the sense that
the BRST operator encodes all (nontrivial)
local symmetries of the classical theory (additional local symmetries
of $S_{cl}$ would make the introduction of further ghost fields
necessary).

Of course the requirements
(i)--(iii) characterize the models to which our analysis apply
only indirectly through the symmetries and field content
of $S_{cl}$ (and through the locality requirement). The
derivation of its most general explicit form will in fact
be part of our results, see section \ref{ss:genclact}.
A simple example for a functional satisfying (i)--(iii) is of course
\begin{equation}
S_{cl} = -\int d^2x \ \ft{1}{2}\sqrt{g}\,
g^{\alpha\beta}\partial_\alpha X^\mu
\cdot \partial_\beta X^\nu\eta_{\mu\nu} \label{S0simple}
\end{equation}
where $\eta_{\mu\nu}$ is a constant symmetric
non-degenerate matrix.

Our aim is to compute the cohomology of the operator ${\cal S}$ in
the space of local functionals.
\begin{equation}
{\cal S}W^g=0. \label{cohoS}
\end{equation}
Two solutions represent the same class and are called equivalent,
$ \ W^g\approx W'^g\ $, if they differ by an ${\cal S}$--exact
functional, i.e. if $W^g- W'^g={\cal S}M^{g-1} $ holds for
for some local functional $M^{g-1}$.  The ${\cal S}$-operation increases
the ghost number by one and is nilpotent.
\section{A simplifying canonical transformation}\label{ss:simplectr}

There is a field redefinition
that simplifies our problem
considerably. It will eliminate some fields from the cohomology,
and cause a chiral
split of the transformation laws \cite{Becchi,BBBeltr}.
Expressed in BV language, we take as the generating fermionic functional
\begin{eqnarray}
F&=&\int d^2x\left[ e^*\sqrt{g}
 +h^{++\,*}
\frac{g_{++}}{g_{+-}+\sqrt{g}}
 +h^{--\,*}
 \frac{g_{--}}{g_{+-}+\sqrt{g}}
+\tilde c^*\left( c\sqrt{g}
 +\partial_\alpha \xi^\alpha \sqrt{g}
 \right)\right.\nonumber\\
&&\left.+c_+^* \left( \xi^+ +
 \frac{g_{--}}{g_{+-}+\sqrt{g}}\xi^-
\right)
+c_-^* \left( \xi^- +
 \frac{g_{++}}{g_{+-}+\sqrt{g}}\xi^+
\right) +\tilde X^*_\mu X^\mu \right]\ . \label{Fchiral}
\end{eqnarray}
where $g=|det\ g_{\alpha\beta}|= g_{+-}^2-g_{++}g_{--}$.
This generates a canonical transformation from fields and antifields
$\{\Phi,\Phi^*\}$ to $\{\tilde\Phi, \tilde\Phi^*\}$ through
\begin{equation} \tilde\Phi^A=\frac{\delta F(\Phi , \tilde\Phi^*)
}{\delta \tilde\Phi^*_A}\hspace{2cm}
\Phi^*_A=\frac{\delta F(\Phi , \tilde\Phi^*)}{\delta \Phi ^A}\ .
\label{Fcan}\end{equation}
In our case,
$\{\tilde\Phi\}= \{X^\mu, h_{++}, h_{--}, e, c^\alpha, \tilde
c\}$.
We have changed from  the three fields $g_{++},\, g_{--}$ and $g_{+-}$
to $e,\, h_{++}$ and $h_{--}$,  the last two being the `Beltrami
variables'. This transformation
becomes singular for $g_{++}g_{--}=0$ and $g_{+-}<0$ (simultaneously).
However, as we shall discuss in section \ref{ss:descent}, the
singularity can become important at most for global
considerations and is thus negligible for our purposes.
At the same time we have introduced more convenient
combinations of the ghost fields, but their explicit relation to
the original diffeomorphism ghosts $\xi$ will remain important
in the sequel. Also, it should be noted that the transformation
between the old set of fields and the new set does not involve the
antifields, so that the ${\cal S}$--cohomology
in the antifield independent sector does not change.

After this canonical transformation
the extended action takes the form
\begin{eqnarray}
\lefteqn{S=S_{cl}(X^\mu,h_{++},h_{--})+\int d^2x(
\ft{1}{1-y}X^*_\mu
 c^\alpha \nabla_\alpha X^\mu }  \nonumber\\
&&+ h^{++\,*}\nabla _+ c^- + h^{--\,*}\nabla _- c^+
+e^*\tilde c - c^*_+ c^+\partial _+ c^+
- c^*_- c^-\partial _- c^-)\ ,
\label{Sgtilde}\end{eqnarray}
where the covariant derivative $\nabla $ is defined in
appendix~\ref{app:useform}, and $y$ is the abbreviation
\begin{equation}
y\equiv h_{++}h_{--}\ .
\end{equation}
The inverse of the above field transformation
is given in appendix~\ref{app:useform}.

Note that we have claimed in \eqn{Sgtilde} that $S_{cl}$ does not
depend on $e$ when written in terms of the new fields.
This can be checked in particular for \eqn{S0simple} which would lead
to
\begin{equation}
S_{cl}(X^\mu,h_{++},h_{--})=-\int d^2x\,
\frac{1}{1-y}\nabla _+ X^\mu \cdot
\nabla _- X^\nu \eta _{\mu \nu }\ .
\end{equation}
Indeed
the master equation requires in particular
$\delta S_{cl}/\delta e=0$ which means that the integrand of
$S_{cl}$ is independent of $e$ (up to a total derivative which we neglect).
This implies that $e$ and $\tilde c$
become a so--called trivial system since they have the simple
transformation property ${\cal S}e=\tilde c$ and do not occur
in the ${\cal S}$--transformation of the other new fields and antifields.
Hence, $e$ and $\tilde c$ can be omitted for any
cohomology considerations and with no loss of generality we
can assume, whenever we work with the new fields and antifields,
that the complete set of fields and antifields is
\begin{equation} \{ \Phi^A, \Phi^*_A\}\ ;\qquad
\{ \Phi^A\}= \{ X^\mu, h_{++}, h_{--}, c^+, c^- \}\ .
\label{allFields}\end{equation}

We shall see that the use of
the new variables has additional advantages. In particular, apart
from eliminating $e$ and $c$,
we have obtained that ${\cal S}h_{++}$ and ${\cal S}c^-$
involve only $c^-$ but not $c^+$,
which is the chiral splitting announced before.

We note that
a similar simplifying canonical transformation can be done in
the zweibein formulation in order to eliminate the
Weyl and local Lorentz ghosts present in the zweibein formulation.
Then $e^+_+$ and $e^-_-$ become trivial together with
(appropriate redefinitions of)
these ghosts, and one is left with
the matter fields, two
functions of the vielbein components given by
$h_{++}$ and $h_{--}$, the diffeomorphism ghosts and with the corresponding
antifields.
\section{Descent equations and their integration}\label{ss:descent}

The first step towards a solution of \eqn{cohoS} consists in an
analysis of the descent equations arising from it. This traces
our problem back to the
${\cal S}$-cohomology on local functions rather than on local
functionals (integrals of local functions). The analysis
of the descent equations is
independent of the form of $S_{cl}$ and has been first performed
in this form in \cite{grav} (see also
\cite{FrBrStructure})\footnote{The antifields
which are not considered in \cite{grav,FrBrStructure} can
be treated on an equal footing with the fields as far as the analysis
of the descent equations is concerned because \eqn{reprd} holds
on fields and antifields.}. We can adopt it since we are not
interested in global aspects of the target manifold and the
two dimensional base manifold. What this
means is spelled out in the following, together with
a discussion of the singularity in the transformation to the
Beltrami variables defined by \eqn{Fchiral} and \eqn{Fcan}.

A crucial tool within the analysis of the descent equations
performed in \cite{grav,FrBrStructure} is the `algebraic Poincar\'e
lemma' describing the cohomology of the
exterior derivative $d$ in the space of
{\it local differential forms}. The latter are by definition
forms $\om_p=dx^{\alpha_1}\wedge\ldots\wedge dx^{\alpha_p}
\omega_{\alpha_1\ldots\alpha_p}$ where $\omega_{\alpha_1\ldots\alpha_p}$
are local functions (see section \ref{ss:action}).
The lemma has been derived by various
authors independently (cf. e.g. \cite{APL} and references
in \cite{BBH}).
It states that the closed forms which are not locally exact
are exhausted by the constant 0--forms and by volume forms which
have non-vanishing Euler--Lagrange derivative with respect to
at least one field or antifield. Here a form $\omega_p$ is
called {\it locally exact} if it can be written as
$d\eta_{p-1}$ for some local form $\eta_{p-1}$ locally, i.e. in any
(sufficiently small)
local neighbourhood in ${\cal M}\times{\cal T}$
where ${\cal M}$ and ${\cal T}$ denote the base and
target space manifold respectively. The latter is
the space in which (all) the fields and antifields take their values.
Of course, a locally exact form can fail to be globally exact
in ${\cal M}\times{\cal T}$.

The general version of the algebraic Poincar\'e
lemma, taking global properties of
${\cal M}\times{\cal T}$ into account, has been derived in \cite{Takens}.
Famous examples for locally but not globally exact local forms
are the integrands of characteristic classes
(of nontrivial bundles).
In two dimensional gravity this is in particular
the integrand $d^2x\sqrt{g}R$ of the two dimensional Einstein action.
Other examples for closed but globally non--exact forms
present in $d=2k$ dimensional gravitational theories are
$(2k-1)$--forms in the metric components and their
first derivatives discussed in \cite{Torre}. The latter stem from the
nontrivial De Rham cohomology
of the target space of the metric components which itself originates
in the requirement that the metric has Minkowskian signature. In
our case there exists therefore a closed 1--form which
generically fails to be globally exact if $g_{\a\be}$ has
signature $(-,+)$. Further closed local forms which fail
to be globally exact can of course arise from
nontrivial De Rham cohomology
of the target space of the matter fields $X^\m$. A refinement
of the analysis of the descent equations which takes into account the
global properties of ${\cal T}$ has been given recently in \cite{bbhgrav}.

In this paper we will completely neglect global aspects of
the two dimensional base manifold and of the
target manifold. This means that whenever we call a functional,
form or function ${\cal S}$-- or $d$--exact (`trivial'),
we have in mind that it is locally exact in
${\cal M}\times {\cal T}$
which does not necessarily imply that
it is globally exact as well.

For our purposes the singularity in the canonical transformation
performed in section \ref{ss:simplectr} is therefore
harmless since it occurs only on
the 2-dimensional subspace ${\cal T}_s=
\{(g_{++},g_{--},g_{+-}):\ g_{++}g_{--}=0,\  g_{+-}<0\}$ of the
3-dimensional target space of the metric components
given by ${\cal T}_g=\{(g_{++},g_{--},g_{+-}):\
g_{++}g_{--}-(g_{+-})^2<0\}$ where we
assumed $g_{\a\be}$ to have signature $(-,+)$.
When using Beltrami variables, one
thus actually works in a target space of metric components
given by ${\cal T}_g- {\cal T}_s$ rather than by ${\cal T}_g$.
We note that
${\cal T}_g-{\cal T}_s$ and ${\cal T}_g$ indeed have different
de Rham cohomology.
Hence, if one wants to consider seriously global
aspects of ${\cal T}\times{\cal M}$ using Beltrami variables,
the singularity in the transformation
to these variables has to be taken into account.

Let us now turn to the discussion of the descent equations.
The analysis takes advantage of
the fact that a necessary condition for a local functional $W^g$
to be a solution of \eqn{cohoS} is that
the ${\cal S}$--transformation of its
integrand is a total derivative. If one views the integrand
as a local 2--form with ghost number $g$,
\begin{equation}
W^g=\int \omega_2^g\label{D2}\ ,
\end{equation}
this requires ${\cal S}\omega_2^g+d\omega_1^{g+1}=0$ for some
local form $\omega_1^{g+1}$, where $d=dx^\alpha\partial_\alpha$ is the
exterior derivative operator. Using
now\footnote{The differentials $dx^\alpha$ are
treated as odd graded variables which implies
${\cal S}dx^\alpha=-dx^\alpha{\cal S}$ and $dx^\alpha dx^\beta=
-dx^\beta dx^\alpha $. The latter allows to omit the wedge product
symbol.}
\begin{equation}
{\cal S}^2=d^2={\cal S}d+d{\cal S}=0
\label{sdalg}\end{equation}
one derives by means of the algebraic Poincar\'e lemma the descent equations
\begin{equation}
{\cal S}\omega_2^g+d\omega_1^{g+1}=0\ ; \qquad {\cal
S}\omega_1^{g+1}+d\omega_0^{g+2}=0\
;\qquad {\cal S}\omega_0^{g+2}=0\ .\label{D1}
\end{equation}
The analysis of \eqn{D1} performed in \cite{grav,FrBrStructure}
shows that the local function (zero--form)
$\omega_0^{g+2}$ occurring here is nontrivial
($\omega_0^{g+2}\neq{\cal S}\omega_0^{g+1}$) and
does not involve explicitly the coordinates $x^\alpha$ whenever
$\omega_2^g$ is nontrivial ($\omega_2^g\neq {\cal S}\omega_2^{g-1}
+d\omega_1^{g}$). Conversely, any nontrivial
$x$--independent solution of the last equation \eqn{D1} apart from
the constant gives rise to a
nontrivial solution of \eqn{cohoS} whose integrand
can be obtained from it according to
\begin{equation}
\omega_2^g =\ft12 dx^\alpha
dx^\beta
\frac{\partial}{\partial \xi^\beta} \frac{\partial}
{\partial \xi^\alpha}
\omega_0^{g+2}
\label{D3}
\end{equation}
where ${\partial}/{\partial \xi^\alpha}$ indicates an ordinary
derivative with respect to {\em undifferentiated}
ghosts $\xi^\alpha$ ({\it not} the functional or Euler--Lagrange
derivative).
It is important here to use the ghosts $\xi^\alpha$ and {\it not}
the ghosts
$c^\alpha$ arising from \eqn{Fchiral}. Namely, \eqn{D3} originates in the
property of ${\cal S}$ that one can represent the exterior
derivative on the fields and antifields (and their
derivatives) by
\begin{equation}
d= b{\cal S}-{\cal S}b\ \ \ ;\ \ \
\qquad b\equiv dx^\alpha\frac{\partial}{\partial \xi^\alpha}\ .
\label{reprd}\end{equation}
\eqn{reprd} simply reflects that
diffeomorphisms are encoded in ${\cal S}$ and does not hold
on the coordinates $x^\alpha$ themselves.
It is therefore important that
$\omega_0^{g+2}$ depends only on the
fields and antifields and their derivatives
but not explicitly on the coordinates, as shown in
\cite{grav,FrBrStructure}. Using \eqn{reprd} (and its consequence
$bd-db=0$), as well as \eqn{sdalg},
it is then straightforward to show
that one can `integrate' the descent equations \eqn{D1} in the
form $\omega_1^{g+1}=b\omega_0^{g+2}$ and $\omega_2^{g}=
\ft12 b^2\omega_0^{g+2}$, the latter being just \eqn{D3}.

We conclude that, neglecting global properties of the base and
target space,
the cohomology of ${\cal S}$ on local functionals with
ghost number $g$ is isomorphic to its cohomology
on those local functions with ghost number $(g+2)$
which do not depend explicitly on the $x^\alpha$.
On the representatives $\int\om_2^g$ resp. $\om_0^{g+2}$
of the corresponding cohomology classes
this isomorphism is explicitly established through the substitution
$\xi^\alpha\rightarrow \xi^\alpha+dx^\alpha$ which converts
$\omega_0^{g+2}$ to $\omega_0^{g+2}+\omega_1^{g+1}+\omega_2^g$,
cf. \eqn{D3}. Since we will compute the cohomology of
${\cal S}$ using the variables introduced
in section \ref{ss:simplectr}, we note that this substitution rule
translates into
\begin{equation}
c^\pm\ \rightarrow\ c^\pm+dx^\pm+h_{\mp\mp}dx^\mp\ ;\qquad
\partial_\alpha c^\pm\rightarrow  \partial_\alpha c^\pm  +
\partial_\alpha h_{\mp\mp}\cdot dx^\mp\qquad
\mbox{etc.,}               \label{substccd}
\end{equation}
where we used $ \partial_\alpha dx^\beta=0$.
Note that
these results imply already that the integrands of the solutions
of \eqn{cohoS} do not depend explicitly on the $x^\alpha$, up to trivial
contributions of course. We stress however that for the validity
of the final result it is nevertheless important to allow for
the presence of local functionals whose integrands depend
explicitly on the $x^\alpha$
since otherwise there would be {\it more} nontrivial solutions of
\eqn{cohoS}. Indeed, one would find additional solutions
whose integrands are $x$--independent and trivial
in the space of local $x$--dependent forms but nontrivial
in the space of local $x$--independent forms.
A typical example for such integrands is
\beq \xi^\alpha{\cL}=
{\cal S}(-x^\alpha{\cL})+\partial_\beta(x^\alpha \xi^\beta{\cL})\ ,
\label{counterexample}\eeq
where $\cL$ denotes a Weyl--invariant density such as, e.g.,
${\cL}=\sqrt{g}g^{\alpha\beta}
\partial_\alpha X^\mu\cdot \partial_\beta X^\nu\eta_{\mu\nu}$.
The occurrence of these additional solutions originates in a seeming
harmless change of the
algebraic Poincar\'e lemma when one formulates it in the space of
local $x$--independent forms: in that space the
differentials $dx^\a$ are not exact and therefore
the descent equations do not always terminate with
a zero-form! For instance, the descent equations
arising from \eqn{counterexample} terminate with the
one-form $dx^\a \xi^+\xi^-\cL$ which is trivial in the
space of $x$-dependent forms but not necessarily in the space
of $x$-independent forms.

We finally  mention that there are in principle two
modifications of the results
if the investigation is restricted to the space of
forms which are globally defined
on ${\cal M}\times{\cal T}$ rather than only locally:
(a) those solutions which are only locally but not globally defined,
disappear from the list of solutions we will find;
(b) globally defined solutions $\omega_2^g$ which can locally be
written as ${\cal S}\omega_2^{g-1}+
d\omega_1^{g}$ have to be added to that list if they fail to be
globally of this form.
\section{Chiral tensor fields}
\label{ss:tensors}

We have seen in the previous section that the
${\cal S}$--cohomology on local functionals with ghost number $g$ can
be obtained from the ${\cal S}$--cohomology
on local $x$--independent functions with ghost number $(g+2)$.
In the next section we show that the latter cohomology
can be reduced to
the ${\cal S}$--cohomology in a particular subspace of
local functions generated by quantities
which we will call {\it covariant ghost variables}
and {\it chiral tensor fields}.
This section is devoted to prepare this result
by introducing these quantities.

Usually, tensor fields are defined by their transformation laws under
the  symmetries of interest.
This can be expressed just as well with the help of their BRST
transformation, which gives a more convenient formulation for
the analysis of the BRST cohomology.
In many cases one finds
that (components of) the gauge fields occur in trivial pairs
together with all the
derivatives of the ghost fields. They can therefore be eliminated from
the BRST cohomology on local functions.
The gauge fields and their derivatives
then only remain in restricted combinations
which are `tensor fields'. Their BRST transformation involves
only the undifferentiated ghosts.
As a result, the representatives of the cohomology classes (of
the BRST cohomology on local functions) can
be expressed entirely in terms of
tensor fields and the undifferentiated ghosts \cite{FrBrStructure}.
Well--known examples for such theories
are Yang--Mills theories \cite{grav}, ordinary (non-Weyl invariant) gravity
in the vielbein formulation
\cite{grav,bbhgrav} and supergravity theories \cite{sugra}.

Let us clarify this feature with the simplest example, Maxwell theory \cite{let}.
Consider local functions of the gauge
field $A_\mu$, the ghost
$C$ and their derivatives. The BRST transformations read just
$s A_\mu = \partial_\mu C$, $sC=0$.
A first set of trivial pairs is thus
$(A_\mu,\partial_\mu C)$. With one derivative more, there are the trivial
pairs $(\partial_{(\mu}A_{\nu)}, \partial_\mu\partial_\nu C)$.
This leaves the
combinations $F_{\mu\nu}=2\partial_{[\mu}A_{\nu]}$ unpaired.
Obviously one can continue this separation to higher order
derivatives. One then
changes variables from $\{C;A_\mu,\partial_\mu C;
\partial_\mu A_\nu, \partial_\mu\partial_\nu C; \dots\}$ to
$\{C;A_\mu,\partial_\mu C;
\partial_{(\mu}A_{\nu)}, F_{\mu\nu}, \partial_\mu\partial_\nu C; \dots\}$,
subdivided in the trivial pairs $\{(A_\mu,\partial_\mu C),
(\partial_{(\mu}A_{\nu)},$ $\partial_\mu\partial_\nu C), \dots\}$ and the
unpaired variables $\{C, F_{\mu\nu},\dots \}$.
The choice of the remaining
combinations like $F_{\mu\nu}$ is dictated by the requirement that only
unpaired (undifferentiated) ghost variables may appear in their BRST
transformation. These remaining combinations are the tensor fields,
($F_{\m\n}$,
$\6_{(\m}F_{\n)\rho}$, etc.) and the undifferentiated ghost $C$.

Of course one should not expect that one can eliminate {\it all}
derivatives of the ghosts from the cohomology in any gauge theory.
That can be done if all ghosts are independent (which
is also true in our case), and if there is a gauge field for each
symmetry (which is not).

A well--known counterexample is provided already by ordinary
gravity in the metric formulation where one can eliminate
all derivatives of the diffeomorphism ghosts
of second and higher order but not all of their first order derivatives:
e.g. in two dimensions,
it is not possible to pair off the three components of the metric
 with the four components of the gradients of the diffeomorphism
ghosts
(the remaining first order derivatives then play a role analogous to
the undifferentiated Lorentz ghosts in the vielbein formalism).
For the case treated in this paper, the situation is even more
subtle since, apart from using the metric formulation of gravity,
we do not introduce a Weyl gauge field.
As a consequence,
there are {\it infinitely many} derivatives of the ghosts
which do not occur in trivial pairs and thus cannot be eliminated
through the procedure sketched above\footnote{Nevertheless
it will turn out in the end
that again all derivatives of order $>2$ disappear, but the
argument is more sophisticated than that of eliminating trivial
pairs.}. This is easily checked by the following simple
counting argument. Analogously to the above example of Maxwell theory
we consider the BRST transformations of the
derivatives of $g_{\a\be}$ of fixed order (``level'') $n$. They
contain as leading terms
derivatives of order $(n+1)$ of the diffeomorphism ghosts
$\xi^\a$ and $n$th order derivatives of the Weyl ghost $c$, cf.
\eqn{gaugetrafo}. This suggests to assign level
$(-1)$ to the two undifferentiated ghosts $\xi^\a$ which are clearly
unpaired. At level 0 there are
the three components of the
undifferentiated metric $g_{\a\be}$ but
four components $\6_\a \xi^\be$ and
the undifferentiated Weyl ghost, i.e. two ghost variables remain unpaired.
Similarly, at level 1, the 6 algebraically independent first order
derivatives
$\6_\gamma g_{\a\be}$ cannot be paired with the 6
second order derivatives of the $\xi^\a$ together with the
2 first order derivatives of $c$. Analogously
one easily verifies that  at all higher levels precisely two
derivatives of the ghosts remain unpaired.

One may also check that the same feature occurs in the zweibein formulation.
In this formulation one introduces zweibeins, but also the
Lorentz ghost apart from the diffeomorphism and dilatation ghosts.
The zweibeins $e_\alpha^a$ transform into the gradients $\partial_\alpha
c^a$ of the diffeomorphism ghosts, leaving at
level~0 the undifferentiated ghosts $c$ and $c'$ of local
dilatations and local Lorentz transformations.
At level 1 one has 8 components
$\partial_\alpha e_\beta^a$ versus  the 6 components
$\partial_\alpha\partial_\beta c^a$ plus the 4 components
$\partial_\alpha c$ and $\partial_\alpha c'$ and so on.
Therefore
we cannot directly adopt the methods and results developed in
\cite{grav,FrBrStructure,bbhgrav} for non--Weyl invariant gravity,
or Weyl invariant gravity with Weyl gauge field.

There seems to be a way around this mismatch of
derivatives of the ghosts and the
gauge fields: one could introduce an extra gauge field
$b_\a$ for Weyl transformations.
Indeed, in presence of $b_\a$
the mismatch disappears since all derivatives of the
Weyl ghost can be paired with the $b_\a$ and their derivatives
as in the above example of Maxwell theory.
As a consequence the cohomology
problem could be treated as in \cite{FrBrStructure}.
Then however, one  would be computing a different
cohomology, including $b_\a$ dependence in the
functionals. One could eliminate this dependence by
requiring invariance under arbitrary shifts $\delta
b_\a=\Lambda_\a$, which expresses the absence of $b_\a$.
The mismatch then remains.
Alternatively, this new invariance brings in another gauge
field, and so on. Continuing in this way, one
would get an infinite set of gauge invariances and gauge fields.
In fact, this would amount to  gauging two copies of the subalgebra
$\{L_n|n\geq -1\}$ of the Virasoro algebra, as  in \cite{infSCd2}.
We will denote these two copies
henceforth by $\{L^+_n\}$ and $\{L^-_n\}$ respectively.
Wishing to  avoid the approach with an infinite tower of symmetries
and gauge  fields, we will not introduce a gauge field for the Weyl
transformations. Of course we will then have to
adapt the methods of  \cite{grav,FrBrStructure,bbhgrav}.
\bsk

In our approach
we only introduce $g_{\alpha\beta}$ as
gauge fields. As a consequence one {\it cannot} reduce the
cohomology to a problem involving only undifferentiated
ghost fields, or derivatives of ghosts up to some finite order,
by the standard argument sketched above.
However, we can still use this argument to get rid of all derivatives
of the ghosts except for two at every level, as the above counting
suggests. In particular, all `mixed' derivatives, namely
$\partial_\pm c^\mp$ and their derivatives, can be
eliminated by the standard argument.
The remaining derivatives can be chosen to be
$(\partial_+)^{m+1} c^+$ and $(\partial_-)^{m+1} c^-$,
where $m$ is the level used in the above counting
and runs from $(-1)$ to infinity. These derivatives of $c^+$ and $c^-$
are called the {\em covariant ghost variables}.

The more difficult task is to construct the
quantities which take over the role usually played by tensor fields.
We call them {\em chiral tensor fields}. Their
characteristic property is that their
BRST transformation may contain the covariant ghost
variables, but no other derivatives of the ghosts.
The fact that in our case the set of covariant ghost variables
is infinite
corresponds to the infinite set of (undifferentiated) ghosts in
the approach using an infinite tower of gauge symmetries.

In the remainder of this section we will
explicitly construct an appropriate
basis for the chiral tensor fields, denoted by
$\{\cB^i\}$. This construction is slightly involved but a
crucial and necessary step within the computation of the
${\cal S}$--cohomology.
It is also interesting in itself since
it shows how the above mentioned subalgebras $\{L^+_\mm\}$ and
$\{L^-_\mm\}$ ($\mm\geq -1$) of the
Virasoro algebra come into play and are represented
on the $\cB^i$.
In particular it turns out that the $\cB^i$ can be chosen as eigenfunctions
of $L^+_0$ and $L^-_0$. This will be very useful
in the next section, since it will eventually allow to reduce
the cohomological analysis to a problem where only
a small finite subset of $\{\cB^i\}$ and those six
covariant ghost variables enter which correspond to the
$sl(2)$ subalgebras $\{L^\pm_{-1},L^\pm_0,L^\pm_1\}$.

It is understood in the following
that all functions that occur are functions of the fields introduced
in section \ref{ss:simplectr} and of their derivatives.
Treating the ghosts separately, we
will use for the remaining variables the collective notation
\[ [Z]\equiv \left( (\partial_+)^m(\partial_-)^nZ^i:\
m,n=0,1,2,\ldots\right)\ ,\quad
\{Z^i\}=\{h_{\pm\pm},\, X^\mu,\,
h^{*\pm\pm},\, X^*_\mu,\, c^*_\pm\}\ .\]

We introduce the following notation for the
above mentioned covariant ghost variables:
\begin{equation}
\cc \pm\mm= \frac{1}{(\mm+1)!} \left( \partial_\pm\right) ^{\mm+1}c^\pm\
;\quad
\mm=-1,0,1,\ldots\ .
\label{S4} \end{equation}
Using \eqn{s00}
one easily verifies that their BRST transformations read
\begin{equation}
{\cal S}\cc \pm\mm=\ft12 f^\mm{}_{\nn\kk}\cc \pm\kk \cc \pm\nn\ ,
\label{S11}\end{equation}
where $f^\mm{}_{\nn\kk}$ are the structure constants of the Virasoro algebra:
\begin{equation}
f^\mm{}_{\nn\kk}=(\nn-\kk)\delta^\mm_{\nn+\kk}\ .
\end{equation}
Note that the sum in \eqn{S11} is finite due to
$\mm,\nn,\kk\geq -1$  and that the covariant ghost variables
transform among themselves, i.e. that no other derivatives of the ghosts
occur in  ${\cal S}\cc \pm\mm$.

Having defined the covariant ghost variables, we are now in a position to
give a precise definition of chiral tensor fields.
The differential ${\cal S}$ decomposes
into a `Koszul--Tate part' $\dkt$ \cite{Henncoho} and a
`BRST'--part $s$ (see \eqn{defdkts}). On the fields one has
${\cal S}=s$, as $\dkt$ has nonvanishing action only
on the antifields. On the antifields $\dkt$ collects that part of the
${\cal S}$--transformation which does not involve the ghosts:
\begin{eqnarray}
& &\dkt c^*_-=-\nabla_+ h^{*++}+\7X^*_\mu\nabla_- X^\mu\ ;
\label{delta1a}\\
& &\dkt h^{*++}=S_{cl}\frac{\dr }{\delta h_{++}}\ ;\quad
\dkt \7X^*_\mu=S_{cl}\frac{\dr }{\delta X^\mu}\, \frac 1{1-y}
\label{delta2a}
\end{eqnarray}
where, for reasons which will become clear soon, we have introduced
\begin{equation}\7X^*_\mu=\frac 1{1-y}X^*_\mu\ .
\label{defhat}\end{equation}

 Note that the change from
$X^*_\m$ to $\7X^*_\m$ becomes singular for $y=1$.
This singularity is actually the same that occurred
already in section \ref{ss:simplectr} in the field
redefinitions leading
to Beltrami variables since $y=1$ is equivalent to
$\sqrt g(g_{+-}+\sqrt g)=0$. Hence, we do not introduce
further singularities here.

Explicit expressions for $s$ are given in appendix~\ref{ss:brsttrafos}.
Note that only \eqn{delta2a} involves explicitly the classical
action.
The precise form of that action however does not matter in the following
since chiral tensor fields are identified by their
$s$--transformation. Now we give the definition: a chiral tensor field
is a local function $T([Z])$ such that
\begin{equation}
sT([Z])=\cc \alpha\mm \, T^\alpha_\mm ([Z])\ .
\label{deftensor} \end{equation}
The sum on the r.h.s.
of \eqn{deftensor} contains actually only finitely
many nonvanishing $T^\alpha_\mm$ since $s$ is a local
operator and $T$ is
by definition a local function.

The nilpotency of $s$  guarantees that
the $T^\alpha_\mm $ in \eqn{deftensor} are chiral tensor
fields as well.
To prove this, one applies $s$ to \eqn{deftensor} and concludes from
$s^2=0$ that
$sT^\alpha_\mm $ cannot involve mixed derivatives of the ghosts.
Also the full operation ${\cal S}$ on $T$ leads to chiral tensor
fields, as we will now prove. This will be true if
$T':=\dkt T$ is automatically a chiral tensor field when this
holds already for $T$. To prove this, we have to show that
$sT'$ does not involve mixed
derivatives of the ghosts. Now, $s\dkt+\dkt s=0$ and
$\dkt \cc \alpha\mm=0$ imply $sT'=-\dkt sT=
\cc \alpha\mm  \dkt T^\alpha_\mm$, which evidently does not involve mixed
derivatives of the ghosts.
Therefore, ${\cal S}T$ depends only on tensor fields
and ghosts $c_\pm^m$. This finishes the proof.

Note that the undifferentiated fields
$X^\mu,$ $ \7X^*_\mu,$ $ h^{*\pm\pm}$ and $c^*_\pm$
are  chiral tensor fields according to \eqn{s2}--\eqn{s5}
($X^*_\mu$ itself is not a chiral tensor field). The partial
derivatives of a chiral tensor field however are in general not
chiral tensor fields:  we
have to complete them to covariant ones.
To that end we define `Virasoro' operators $\LL \alpha\mm$ ($\mm\geq -1$,
$\a=+,-$)
on chiral tensor fields (and, for later purpose,
on the covariant ghost variables as well)
through the anticommutators
\begin{equation}
\LL \alpha\mm:= \frac{\partial}{\partial \cc \alpha\mm}\, {\cal S}
+{\cal S}\, \frac{\partial}{\partial \cc \alpha\mm}
= \frac{\partial}{\partial \cc \alpha\mm}\,  s
+s\, \frac{\partial}{\partial \cc \alpha\mm}\ ,
\quad \mm=-1,0,1,\ldots  \ ,
\label{virasoro}\end{equation}
where one can use both $s$ and ${\cal S}$ since $\dkt$ vanishes
on $\cc \alpha\mm$.
Using the notation of \eqn{deftensor}, we obtain
\[ \LL \alpha\mm T=T^\alpha_\mm \ ;\quad
\LL \pm\mm \cc \pm\nn=f^\nn{}_{\kk\mm}
\cc \pm\kk\ ;\quad \LL \pm\mm \cc \mp\nn=0\ .\]
Note that the $\LL \alpha\mm$ are derivations, i.e. they satisfy the
product rule, since they are defined as anticommutators of two
antiderivations. Furthermore
their algebra closes on (functions of) chiral tensor fields and the
covariant ghost variables and is isomorphic to the algebra of vector
fields $z^{\mm+1}\frac{d}{dz}$ that are regular at $z=0$:
\begin{equation}
\left[\LL \pm\mm ,\LL \pm\nn\right]=(\mm-\nn)\LL \pm{\mm+\nn}\
;\qquad
\left[\LL +\mm ,\LL -\nn\right]=0\ .\label{S7}
\end{equation}
\eqn{S7} is easily verified on the ghosts $\cc \alpha\mm$ using
the Jacobi identity for the structure constants
$f^\kk{}_{\mm\nn}$. One verifies it on any
chiral tensor field
$T$ by evaluating $s^2T=0$ using \eqn{S11}. Indeed, since $T^\alpha_\mm$
is a chiral tensor field (see above), we have
\[ sT^\alpha_\mm=s (\LL \alpha\mm T)
=\cc \beta\nn\LL \beta\nn\LL \alpha\mm T\]
which requires \eqn{S7} to hold on $T$ in order to be consistent
with $s^2T=0$.

We can now describe and construct the generators replacing
in the new basis the $X^\mu,$ $X^*_\mu,$ $ h^{*\pm\pm},$ $c^*_\pm$
and their partial derivatives.
We denote the set of these new generators by
$\{\cB^i\}$ and require
\ben  \renewcommand{\theenumi}{(\Roman{enumi})}
\item $\{\cB^i\}$ consists of
`covariant derivatives' of the fields
$X^\mu,$ $ \7X^*_\mu,$ $ h^{*\pm\pm},$ $ c^*_\pm$
which complete (all) their partial derivatives to chiral
tensor fields;
\item each $\cB^i$ is an eigenfunction
of $L^+_0$ and $L^-_0$.
\een
(II) is not really needed for the construction of a basis for the
chiral tensor fields
but can be imposed and will be useful later, as mentioned already above.
It is indeed fulfilled for the undifferentiated fields
$X^\mu,$ $ \7X^*_\mu,$ $ h^{*\pm\pm},$ $ c^*_\pm$. This is
evident from \eqn{s2}--\eqn{s5} which also yields
the $\LL \alpha{0}$--eigenvalues of these fields
(`weights'), denoted by $w_\alpha$:
\begin{equation}
\begin{array}{c|cccccc}
Z^i & X^\mu & \7X^*_\mu & h^{*++} & h^{*--} & c^*_-  & c^*_+\\
\hline
(w_+,w_-) & (0,0) & (1,1) & (0,2) & (2,0) & (0,2) & (2,0) \\
\end{array}\ .
\label{weights}\end{equation}
We now observe that the operators $\LL \alpha{-1}$
already serve as covariant derivatives of the matter fields
which we denote by $X_{m,n}^\mu$:
\begin{equation}
X_{m,n}^\mu=\left( \LL +{-1}\right) ^m  \left( \LL -{-1}\right) ^n X^\mu
=\left( \frac{\partial}{\partial c^+}\, {\cal S}\right) ^m
\left( \frac{\partial}{\partial c^-}\, {\cal S}\right) ^n X^\mu\ ,
 \quad m,n=0,1,2,\ldots
\label{S8}
\end{equation}
where the second equality holds due to ${\cal S}^2=0$.
In order to see that the $X_{m,n}^\mu$ indeed complete the
partial derivatives to covariant derivatives of the $X^\mu$
one verifies that
\beq X_{m,n}^\mu = \frac{1}{\left( 1-y\right) ^{m+n}}
\left( \nabla_+\right) ^m \left( \nabla_-\right) ^n X^\mu +{\cal
O}(m+n-1)\ ,\label{order}\eeq
where ${\cal O}(m+n-1)$ denotes a complicated function of
$h_{++}$, $h_{--}$, $X^\m$
involving only their derivatives of $(m+n-1)$th and lower order.
The first few $X_{m,n}^\mu$ are given in appendix~\ref{app:useform}
(in fact only $X_{0,0}^\mu$, $X_{1,0}^\mu$, $X_{0,1}^\mu$
and $X_{1,1}^\mu$
will ultimately be needed for the cohomology).
The action of $L^\pm_\mm$ and $s$ on $X_{m, n}^\mu$
can be obtained using
the algebra \eqn{S7} and the fact that $X^\mu_{0,0}=X^\mu$ has
`highest weight'
\begin{equation}
L^\pm_\mm X^\mu  =0\ \qquad \forall \mm\geq 0\ .
\label{S77}
\end{equation}
In particular, one easily verifies that $X_{m,n}^\mu$ is an
eigenfunction of the $\LL \alpha{0}$ with weights $(m,n)$, using
\eqn{S7} and \eqn{weights}.
In the next section we will show in detail that the change of
generators from the $\partial_+^m \partial_-^n X^\mu$ to the
$X^\mu_{m, n}$ is in fact local and invertible except where
the transformation to the Beltrami variables itself
becomes singular (cf. proof of lemma \ref{lem:basiscT}).

Analogously one checks that a basis $\{\7X^*_{\mu\, m,n}\}$
for the covariant derivatives of the antifields $\7X^*_\mu$ is given by
\begin{equation}
\7X^*_{\mu\, m,n}=\left( \LL +{-1}\right) ^m  \left( \LL -{-1}\right) ^n
\7X^*_\mu\ , \quad m,n=0,1,2,\ldots \ ,
\label{S8a}
\end{equation}
and that $\7X^*_{\mu\, m,n}$ has weights $(m+1,n+1)$.
Again, the change from $X^*_{\mu}$ and its partial derivatives to
the $\7X^*_{\mu\, m,n}$ is local and (locally) invertible, see next section.
The action of $L^\pm_\mm$ and $s$ on $\7X^*_{\mu\, m,n}$ can be obtained
using the algebra \eqn{S7} and
\begin{equation}
L^\pm_\mm \7X^*_\mu  =0\ \qquad \forall \mm\geq 1\ ,
\label{S78}
\end{equation}
which follows from \eqn{s3}.

The construction of a complete basis for the covariant derivatives of
the remaining antifields $h^{*\pm\pm}$ and $c^*_{\pm}$ is more
subtle since \eqn{s4} and \eqn{s5} show that $\LL +{-1}$ does not serve
as an appropriate covariant derivative operator on
$h^{*++}$ or $c^*_-$ due to
$\LL +{-1}h^{*++}=\7X^*_\mu X^\mu_{0,1}$ and $\LL +{-1}c^*_-=0$.
Analogous statements hold for $\LL -{-1}$ on $h^{*--}$ and
$c^*_+$ of course. We therefore have to look
for an alternative construction of covariant derivatives. It can be
easily found. Namely the operators
\begin{equation}
D_\pm=\6_\pm-\sum_{\mm\geq -1}\frac{1}{(\mm+1)!}
(\partial_\mp)^{\mm+1}h_{\pm\pm}\cdot \LL \mp\mm\label{covder}
\end{equation}
provide covariant derivatives $D_\alpha T$ of
an arbitrary chiral tensor field $T$ since they are constructed
such that $sD_\alpha T$ does not contain $\partial_- c^+$,
$\partial_+ c^-$ or any of their derivatives
(again, the sum appearing in $D_\alpha T$ contains
only finitely many nonvanishing terms since $T$ is local by assumption).
In fact we could have used the $D_\alpha$
to construct a basis for the covariant derivatives
of all fields $Z\in\{X^\mu,$ $ \7X^*_\mu,$ $ h^{*\pm\pm},$ $ c^*_\pm\}$
through $(D_+)^m(D_-)^nZ$. However that basis would not
satisfy requirement (II)
since the operators $D_\pm$ have the
following commutation relations with the $\LL \alpha{0}$:
\begin{equation}
[\LL \pm{0},D_\pm ]=\LL \pm{-1}\ ;\quad [\LL \pm{0},D_\mp]=0\ .
\label{comm1}
\end{equation}
On the other hand \eqn{comm1} and \eqn{S7} imply
\begin{equation}
[\LL \alpha{0},\tilde D_\beta]=0
\label{comm2}
\end{equation}
where
\begin{equation}
\tilde D_\alpha=D_\alpha-\LL \alpha{-1}\ .
\label{comm3}
\end{equation}
We note that the $\tilde D_\alpha$ vanish on $X^\mu$ and $\7X^*_\mu$,
i.e. on these fields one actually has $D_\alpha=\LL \alpha{-1}$.
In contrast, $\tilde D_+h^{*++}$ and
$\tilde D_+c^*_-$ do not vanish and complete $\6_+h^{*++}$ and
$\6_+c^*_-$ to chiral tensor fields. In particular we have
$\tilde D_+ c^*_-=D_+ c^*_-$ due to $\LL +{-1}c^*_-=0$ and, more
generally,
\begin{equation}
(\tilde D_\pm)^m c^*_\mp=(D_\pm)^m c^*_\mp\ .
\label{remark}
\end{equation}
It is now straightforward to construct a basis for the
covariant derivatives of $h^{*++}$ and $c^*_-$ with definite weights
through
\begin{equation}
c^*_{-\, m,n}= (\LL -{-1})^n (D_+)^m c^*_-\ ;\quad
h^{*++}_{m,n}= (\LL -{-1})^n (\tilde D_+)^m h^{*++}\ ;\quad
                m,n=0,1,2,\dots \ ,   \label{basis2}
\end{equation}
and analogously one constructs covariant derivatives
$c^*_{+\, m,n}$ and $h^{*--}_{m,n}$ of $h^{*++}$ and $c^*_-$.
We note that one has
\beq L^\a_\mm c^*_{\pm}  =L^\a_\mm h^{*\pm\pm}=0\
\qquad \forall \mm\geq 1\ .\eeq
One now checks again that all the derivatives of the
$c^*_{\pm}$ and $h^{*\pm\pm}$
appear as leading terms (highest derivatives) in the new variables.
Furthermore
the change from the $c^*_\pm$, $h^{*\pm\pm}$ and their
partial derivatives to the $c^*_{\pm\, m,n}$ and $h^{*\pm\pm}_{m,n}$
is local and invertible, see next section.
E.g. we have, using \eqn{s4} and \eqn{s5},
\begin{eqnarray}
h^{*++}_{0,1} &=& L^-_{-1}h^{*++} =\partial_-h^{*++}
-h_{--}\hat X^*_\mu X^\mu_{0,1}  \nonumber\\
h^{*++}_{1,0} &=& \tilde D_+ h^{*++} =\nabla_+ h^{*++}
-(1-y)\hat X^*_\mu X^\mu_{0,1}  \nonumber\\
c^*_{-\,0,1} &=& L^-_{-1}c^*_-=\partial_- c^*_- \nonumber\\
c^*_{-\,1,0} &=&  D_+ c^*_-=
(\6_+ -h_{++}\6_- -2\6_-h_{++}\cdot)c^*_-\ .
\end{eqnarray}
\vspace{5mm}

This completes the construction of $\{\cB^i\}$.
The complete list of the $\cB^i$ and their weights is given by
\begin{equation}
\begin{array}{c|cccccc}
\cB^i & X^\mu_{m,n} & \7X^*_{\mu\, m,n} &
h^{*++}_{m,n} & h^{*--}_{m,n} & c^*_{-\, m,n}  & c^*_{+\, m,n}\\
\hline
(w_+,w_-) & (m,n) & (m+1,n+1) & (0,n+2) & (m+2,0) & (0,n+2) & (m+2,0) \\
\multicolumn{7}{r}
{(m,n=0,1,2,\ldots)} \ .
\end{array}
\label{weights1}\end{equation}
We finally give the weights of the covariant
ghost variables:
\begin{equation}
\begin{array}{c|cc}
\cc \alpha\mm & \cc +\mm & \cc -\mm \\
\hline
(w_+,w_-) & (\mm,0) & (0,\mm)
\end{array}\quad (\mm=-1,0,1,\ldots)\ .
\label{weights2}\end{equation}

As a side comment we remark that one has
$h^{*\pm\pm}_{m+1,n}=-\dkt c^*_{\mp\, m,n}$
for $m,n\geq 0$ which can easily be checked explicitly for $m=n=0$
and then extended to $m,n\geq 0$ using $[\LL \alpha\mm,\dkt]=0$.
This illustrates a general property of $\dkt$ explained above,
namely that it maps chiral tensor fields to chiral tensor fields.
\section{Reduction  to $H^*({\cal S}, \sub)$}\label{ss:strategy}
In this section we prove that
one can contract the ${\cal S}$--cohomology in the full space
of local functions to a particular
subspace which we will denote by $\sub$. In the first
step we will perform a reduction to the space
of local functions $\om(\cc \a{m},\cB^i)$ of
the chiral tensor fields $\cB^i$ and the covariant ghost variables
$\cc \a{m}$ introduced in the previous section,
and in the second step a reduction to
the space of local functions
$\om(\cc \a{m},\cB^i)$ with vanishing
$L_0^+$ and $L_0^-$ weights.
The latter is the above mentioned subspace $\sub$.
The cohomology of ${\cal S}$ in
$\sub$ is denoted by $H^*({\cal S}, \sub)$. In the third step we will
give a basis for the functions in $\sub$, which will
be described in terms
of a finite number of `superfields'  in the
undifferentiated matter fields $X^\m$ and the ghosts $\cc \alpha{0}$.
The subsequent cohomological
analysis will be in terms of those superfields.
The final step, the computation of $H^*({\cal S}, \sub)$,
can then be done by a direct calculation, which will be
carried out in section~\ref{ss:fullcoho}.

In order to compute the
${\cal S}$-cohomology in the space of local $x$-independent
functions, we have to solve
\begin{equation}
{\cal S}\omega^G([\Phi,\Phi^*])=0
\label{S1}\end{equation}
where $\omega^G$ has arbitrary ghost number $G$ and its argument
$[\Phi,\Phi^*]$ indicates the local%
\footnote{A local
function is polynomial in all these
variables except for the undifferentiated $h_{\pm\pm}$ and $X^\m$.}
dependence on {\it all}
fields, antifields and their partial derivatives collectively,
\[
[\Phi,\Phi^*]\equiv\left\{
(\partial_+)^m(\partial_-)^n\Phi^A,\, (\partial_+)^m(\partial_-)^n\Phi^*_A |
\ m,n=0,1,\ldots\right\}\ .
\]
The set of fields $\Phi$ was given in \eqn{allFields}.
Two solutions of \eqn{S1} are called equivalent if they differ by an
${\cal S}$--exact solution or a constant. The latter can
occur only in the ghost number--zero section due to the absence of constant
ghosts,
\begin{equation}
\omega^G ([\Phi,\Phi^*]) \approx \omega'{}^G ([\Phi,\Phi^*])
\ \Leftrightarrow\
\omega'{}^G - \omega^G =
{\cal S}\eta^{G-1}([\Phi,\Phi^*]) +\lambda\, \delta_0^G \ ,
\end{equation}
where $\eta^{G-1}$ is a local function with ghost number $(G-1)$ and
$\lambda$ is a constant.
This definition of equivalence is motivated by the fact that
${\cal S}$--exact solutions of \eqn{S1},  and the constants,
correspond via the descent equations
to (locally) trivial functionals, see section \ref{ss:descent}.

We will now
isolate trivial pairs, which we can then remove from the
computation of the cohomology.
Trivial pairs are doublets of generators  $(U,V)$ satisfying
\ben
\item[(a)] $U$ and $V$ have the simple transformations
${\cal S}U=V$, ${\cal S}V=0$;
\item[(b)] $U$ and $V$ do not occur in the ${\cal S}$-transformation
of any other generator;
\item[(c)] $U$ and $V$ generate the algebra of functions of
$U$ and $V$ {\em freely}, i.e. there are no
extra%
\footnote{i.e. other than  the  Grassman algebra relations. This third
condition is usually satisfied automatically, and will therefore be
left out of focus. In section \ref{ss:fullcoho} we will meet an example
where it is {\em not} valid.}
relations.
\een
By a standard argument, using a contracting homotopy,
one then easily shows that such trivial pairs of
generators indeed do not contribute nontrivially to
the ${\cal S}$-cohomology (neglecting global properties of the target
manifold). This reduces
the computation of the ${\cal S}$-cohomology in the algebra
of all generators to the same problem in the algebra of
those generators which remain after the trivial pairs have been
removed. The difficult part in this step is in fact
not that of finding  $U$'s and $V$'s satisfying (a) but
the construction of a complete set of complementary
generators, since they are conditioned by (b).

We used elimination of trivial pairs already before
to remove the
fields $(e,\4c)$ and  their antifields.
Indeed, these pairs satisfy conditions (a)--(c) in the basis of
generators introduced in section~\ref{ss:simplectr}
 (this was in fact one of the reasons for introducing that basis).
Other trivial pairs of generators  are
the antighosts and corresponding Lagrange multiplier fields (and
their derivatives) which
one introduces for gauge fixing. They
also satisfy evidently (a)--(c) and therefore have been omitted
from the very beginning.
 In the cases just cited one can eliminate sets of fields
completely from the cohomology since two {\em undifferentiated} fields
(or antifields) group in trivial pairs respectively. Therefore
all their derivatives group in trivial pairs as well and these
fields disappear completely from the cohomology (both on local functions
and on local functionals). This is different in the cases
considered in the following since not all derivatives of the involved fields
(ghosts) occur in trivial pairs.
\bsk

Let us now derive the reduction to functions of  the
chiral tensor fields ${\cal B}^i$
and the covariant ghost variables $\cc \alpha\mm$
introduced in section \ref{ss:tensors}.
The transformation laws ${\cal S}h_{\pm\pm}
=\nabla_\pm c^\mp$ suggest that trivial pairs
are given by
\beq (\mu_\ell,{\cal S}\mu_\ell)\qd\mbox{where}\qd
\{\mu_\ell \} =\left\{  (\partial_+)^m (\partial_-)^n h_{++},
(\partial_+)^m (\partial_-)^n h_{--}|\
m,n=0,1,\ldots\right\}\ .
 \label{trivpairs}\label{defmul}\eeq
These pairs evidently satisfy property (a) but the fulfillment
of (b) is not straightforward. Rather, we first have to
complete $\{\mu_\ell,{\cal S}\mu_\ell\}$ to a new basis of
generators satisfying (b) and replacing  the old generators
(field, antifields and their derivatives)
in order to be able to remove the $\mu$'s and $({\cal S}\mu)$'s.
Of course we require the change of basis from the old to the
new generators to be invertible and local.

The new generators ${\cal S}\mu_\ell$ replace
the `mixed' derivatives of the ghosts $c^\a$, i.e.
$\6_+c^-$, $\6_-c^+$ and derivatives thereof, as
one has ${\cal S}h_{\pm\pm} =\6_\pm c^\mp+\ldots$. Hence,
we can replace the
ghosts $c^\a$ and all their derivatives by the
${\cal S}\mu_\ell$ and the  $\cc \pm\mm$ .

Now,
a set of generators completing $\{\mu_\ell,{\cal S}\mu_\ell\}$
to a basis with the desired property (b)
is given by $\{\cc \a{m},\cB^i\}$.
This follows from the facts that by construction both ${\cal S}\cc \a{m}$
and ${\cal S}\cB^i$ can be written entirely in terms of the
$\cc \a{m}$ and $\cB^i$ again, and that the change of basis is indeed local
and (locally) invertible due to the following lemma:
\begin{lemma} Any local
function of $X^\mu$, $h_{\pm\pm}$, $c^\pm$, $X^*_\mu$,
$h^*_{\pm\pm}$, $c^*_\pm$ and their derivatives
can be written as a local
function of  the $\cc \pm\mm$,  ${\cal B}^i$,
$\mu_\ell$ and
${\cal S}\mu_\ell$  and vice versa\footnote{A local
function of the $\cc \pm\mm$,  ${\cal B}^i$, $\mu_\ell$ and
${\cal S}\mu_\ell$ depends polynomially on all these generators
except possibly on the undifferentiated fields $X^\m$, $h_{++}$ and
$h_{--}$.}.
\label{lem:basiscT}\end{lemma}
{\bf Proof:} One easily verifies that
the lemma is implied by the facts that (i) the undifferentiated
fields $X^\m$, $h_{++}$ and $h_{--}$ are elements both of the old
and of the new basis, and (ii) all other generators of the old basis
can be written as local functions of the new generators and vice versa.
(i) is relevant since
$X^\m$, $h_{++}$ and $h_{--}$ can occur nonpolynomially in
local functions, contrary to all other generators.

Hence, all we have to prove is (ii). To that end
we assign a level to each generator given by the
highest order of derivatives
of fields or antifields occurring in them.
The proof can then be performed inductively. First
one verifies that (ii) holds at
level 0. This is obvious since
the new generators with this level just agree with old ones
(undifferentiated fields and antifields) up to the
replacements $X^*_\m\leftrightarrow \7X^*_\m$.
In the second step of the induction one shows that (ii)
holds at level $n$ if it holds at all smaller levels.

For the derivatives of the $X^\m$ that second step of the induction
can be performed as follows.
Consider the set of $n$th order derivatives of the $X^\m$,
i.e. $\{(\6_+)^{n-p}(\6_-)^pX^\m|p=0,\ldots,n\}$.
The corresponding set of new generators with the same
level is $\{X^\m_{n-p,p}|
p=0,\ldots,n\}$. Due to \eqn{order} (and \eqn{defnabla}), one has
\beq X^\m_{n-p,p}=\sum_{q=0}^n M_{p,q}^{(n)}(h)\, (\6_+)^{n-p}(\6_-)^p
X^\m+ {\cal O}(n-1)\ .
\eeq
Here ${\cal O}(n-1)$ denotes a local function of
generators with levels $k\leq n-1$,
cf. \eqn{order} and \eqn{defnabla}. Since (ii) is supposed to
hold at all levels smaller than $n$ we don't have to worry about this term.
The question then is whether $M_{p,q}^{(n)}$ is invertible. This can be seen by
considering the transformation of independent variables from $x^\pm$ to
$y^\pm=(1-y)^{-1}(x^\pm+h_{\mp\mp}x^\mp)$. With constant $h_{\mp\mp}$
it is obvious that this leads to the same matrix for the transformation
between $x^\pm$ derivatives and $y^\pm$ derivatives, since
$\partial/\partial y^\pm=\nabla_\pm$. From this one easily sees that
$\det M^{(n)}=(1-y)^{-n(n+1)/2}$.
This proves that all
$(\6_+)^m(\6_-)^n X^\m$ are indeed local functions of the new generators
and that the change from the $(\6_+)^m(\6_-)^n X^\m$ to
the $X_{m,n}^\m$ is invertible except for $y=1$.
The latter is the same
singularity that occurred already in the change to
the Beltrami variables themselves, cf. remark after \eqn{defhat}.
Note that if we would have encountered here infinitely many
further singularities (e.g. at any level a different one), then
the change to the new generators would not have been allowed.

Analogously one checks that
the change from the fields $X^*_\mu, h^{*\pm \pm },c^*_\pm $ and
their derivatives
to the corresponding new generators
is also local and invertible except for $y=1$.

The $\cc\pm\mm$ contain the ghosts and their `unmixed' derivatives.
Using
${\cal S}h_{\pm\pm}=\nabla_\pm c^\mp=\partial_\pm c^\mp+\ldots$,
cf. \eqn{s1}, we see that the mixed derivatives of the ghosts are the
highest derivative parts of ${\cal S}\mu_\ell$. Therefore all
the ghosts and their derivatives and all (derivatives of) $h_{\pm \pm }$
are replaced by  $\{\cc\pm\mm , \mu_\ell, {\cal S}\mu_\ell\}$.
\QED

Since the new basis of generators has been constructed such that
it satisfies requirements (a) and (b), we can now conclude that
the trivial pairs of generators can be removed from the
cohomology:
\begin{lemma}
(i) Any solution of \eqn{S1} can be expressed entirely in terms
of the $\cc \alpha\mm$ and $\cB^i$ modulo an ${\cal S}$--exact
contribution,
\begin{equation}
{\cal S}\omega^G([\Phi,\Phi^*])=0\ \Rightarrow\ \omega^G([\Phi,\Phi^*])
=\hat\omega^G
(\cc \alpha\mm,{\cal B}^i)+{\cal S}\eta^{G-1}([\Phi,\Phi^*])  \ .\label{S3}
\end{equation}
(ii) A function of the variables $\cc \alpha\mm$ and $\cB^i$
is ${\cal S}$--exact iff it is the ${\cal S}$--transformation
of a another function of these variables,
\begin{equation}
\hat\omega^G(\cc \alpha\mm,{\cal B}^i)={\cal S}\eta^{G-1}([\Phi,\Phi^*])\
\Leftrightarrow\
\hat\omega^G(\cc \alpha\mm,{\cal B}^i)={\cal S}\hat\eta^{G-1}
(\cc \alpha\mm,{\cal B}^i)\ .\label{S3a}
\end{equation}
\label{lem:newlemma}\end{lemma}

Now we go to step 2: the reduction to the space $\sub$, which includes
only the zero eigenspaces of $L_0^\alpha$. Since all variables
$\cc \alpha\mm$ and $\cB^i$ have by construction definite
weights (cf. \eqn{weights1} and \eqn{weights2}),
we can decompose any function
$\hat\omega^G(\cc \alpha\mm,{\cal B}^i)$ into parts
with definite weights,
\begin{equation}
\hat\omega^G(\cc \alpha\mm,{\cal B}^i)=\sum_{m,n}
\hat\omega^G_{m,n}(\cc \alpha\mm,{\cal B}^i),\quad
\LL +0\hat\omega^G_{m,n}=m\hat\omega^G_{m,n},\
\LL -0\hat\omega^G_{m,n}=n\hat\omega^G_{m,n}\ .\label{S3b}
\end{equation}
By their very definition
\eqn{virasoro}, the $\LL \alpha{0}$ can be represented as
anticommutators of ${\cal S}$ with other operators (namely
with the derivatives with respect to $\cc \alpha0$). Together
with \eqn{S3b} this is already sufficient to conclude
from ${\cal S}\hat\omega^G=0$ the
${\cal S}$--exactness of all $\hat\omega^G_{m,n}$ apart from
$\hat\omega^G_{0,0}$. Namely, \eqn{virasoro} implies that
${\cal S}$ commutes with the $\LL \alpha{0}$ and therefore
leaves their eigenspaces invariant. Hence,
${\cal S}\hat\omega^G=0$ requires
all $\hat\omega^G_{m,n}$ to be separately ${\cal S}$--invariant.
This implies in turn that $\hat\omega^G_{m,n}$ is ${\cal S}$--exact
unless both $m$ and $n$ vanish,
cf. e.g. `basic lemma' in \cite{let}. Using the
${\cal S}$--invariance of the $\LL \alpha{0}$--eigenspaces
again, we can further conclude that $\hat\omega^G_{0,0}$ is
${\cal S}$--exact if and only if it is ${\cal S}$--exact in
the space of functions with weights $(0,0)$.
This is the space $\sub$ of functions
mentioned in the beginning of this section:
\begin{equation}
\sub=\{\omega(\cc \alpha\mm,{\cal B}^i):\ \LL +{0}\omega=
\LL -{0}\omega=0\}\ .
\label{sub1}\end{equation}
We have therefore shown that the computation of all solutions
of \eqn{S1} can be reduced to the computation of $H^*({\cal S},\sub)$:
\begin{lemma}
(i) Any solution of \eqn{S1} is a function in $\sub$
up to an ${\cal S}$--exact part,
\begin{eqnarray}
{\cal S}\omega^G ([\Phi,\Phi^*])  =0 \ \Rightarrow\
 \omega^G ([\Phi,\Phi^*]) =\bar \omega^G
 +{\cal S}\eta^{G-1} ([\Phi,\Phi^*]),\quad \bar \omega^G\in\sub\ .
\end{eqnarray}
(ii) A function in $\sub$ is ${\cal S}$--exact in the
space of local functions iff it is ${\cal S}$--exact
in $\sub$,
\begin{eqnarray}
\bar\omega^G  ={\cal S}\eta^{G-1}([\Phi,\Phi^*]),\
\bar \omega^G\in\sub \ \Leftrightarrow\
 \bar \omega^G  ={\cal S}\bar \eta^{G-1},\ \bar \eta^{G-1}\in\sub\ .
\end{eqnarray}
\label{lem:L00}\end{lemma}

For later purposes, in step 3 we now characterize $\sub$ more explicitly.
Consider first
the variables with weights $(0,0)$.
They are collectively denoted by $z^A$:
\beq \{z^A\}=\{X^\m,\, \THE +,\, \THE -\}\qd\mbox{where}\qd
\THE \pm\equiv \cc \pm{0}\ .
\label{sub2a}\eeq
We  interpret $\{z^A\}$ as coordinates of a superspace.
A generic superfield $\cH(z)$ is then a function of the form
\beq \cH(z)=H(X)+H^+(X)\THE + +H^-(X)\THE -+H^{+-}(X)\THE +\THE -\ .
\label{superfield}\eeq
The functions $H(X),\ldots,H^{+-}(X)$  in
the expansion \eqn{superfield} will be called the component
fields of $\cH(z)$.

Now, recall that among all
generators $\cc \alpha\mm$ and $\cB^i$ only the undifferentiated
ghosts $\cc +{-1}=c^+$ and
$\cc -{-1}=c^-$ have negative weights, given by
$(-1,0)$ and $(0,-1)$ respectively, cf.
\eqn{weights1} and \eqn{weights2}. The nilpotency
of $c^+$ and $c^-$ implies
that functions in $\sub$ cannot involve
generators whose $\LL +0$-- or
$\LL -0$--weight exceeds $1$ (recall that we are dealing with local
functions and therefore a function in $\sub$ is polynomial in all
variables $\cc \alpha\mm$ and $\cB^i$ except for the $X^\mu_{0,0}=X^\mu$).
Furthermore, whenever a
variable with $\LL +0$--($\LL -0$--) weight 1 occurs, it must appear
necessarily together with $c^+$ ($c^-$).   It is then
easy to verify that any function
in $\sub$ can be expressed in terms of the $z^A$ which have
weights $(0,0)$ and in terms of
the following zero-modes of the $\LL \a{0}$:
\beq\ba{ll}   T^\m=c^+X^\m_{1,0}+c^-X^\m_{0,1}\ ;&
         R^\m=c^+X^\m_{1,0}-c^-X^\m_{0,1}\ ;\\
T^+=2\cc +{-1}\cc +{1}=c^+\6_+^2c^+ \ ; &
T^-=2\cc -{-1}\cc -{1}=c^-\6_-^2c^-\ ;\\
T_{+-}^\mu= c^+c^- X^\mu_{1,1}\ ; &
T^*_\mu=c^+c^-\hat X^*_\mu\ .
\ea\label{sub3}\label{neu2}\eeq
The motivation for introducing the linear combinations $T^\m$
and $R^\m$ of $c^+X^\m_{1,0}$ and $c^-X^\m_{0,1}$ is
that $T^\mu$ and $T^\pm$ group naturally in a ``super-multiplet''
corresponding to $\{z^A\}$ via the BRST operator (see below),
\beq\{ T^A\}=\{T^\m,\, T^+,\, T^-\} \ .
\label{new2}\eeq
Note that $T^A$ is a vector of fermionic type, i.e. the first components
$T^\m$ are fermionic, while the others are bosonic.  On the $z$'s
and the quantities
\eqn{neu2}, ${\cal S}$ takes the simple form
\beq\ba{ll} {\cal S}z^A=T^A\ ;& {\cal S}T^A=0\ ;\\
{\cal S}R^\m=-2T^\m_{+-}\ ;& {\cal S}T^\m_{+-}=0\ ;\\
{\cal S}T^*_\mu =\dkt T^*_\mu \ .&
\ea\label{sub4new}\eeq
The explicit form of $\dkt T^*_\mu$ depends on the classical action.
Therefore, we  have to determine this action
before we can completely compute $H^*({\cal S},\sub)$.

Due to the composite nature of the  quantities
\eqn{neu2}, involving nilpotent ghosts, their algebra
is not freely generated but subject to the
following identities:
\bea &&
T^\m T^\n=-T^\n T^\m=-R^\m R^\n\ ;\qd
R^\m T^\n =R^\n T^\m\ ;\qd T^\m T^\pm=\mp R^\m T^\pm\ ; \nonumber\\
&& T^\pm T^\pm=T^A T^B T^C=
R^\m T^A T^B=R^\m R^\n T^A=R^\m R^\n R^\rho=0\ ; \nonumber\\
&& T^A T^\m_{+-}=R^\n T^\m_{+-}=T^A T^*_\m=R^\n T^*_\m=0\ ; \nonumber\\
&& T^\m_{+-}T^\n_{+-}=T^\m_{+-}T^*_\n=T^*_\m T^*_\n=0\ .
\label{new3b}
\eea

Taking these identities into account it is now
straightforward to write down the most general function in $\sub$.
It can be parametrized by superfields multiplying the
various non-vanishing monomials in the quantities \eqn{neu2}.
The parametrization we will use in the following sections is described by
the following lemma.
\begin{lemma} Any function $\om\in\sub$ can be uniquely
written in the form
\bea \lefteqn{
\om[\cA,\cB_A,\cF_{AB},{\cal C}_\mu,\cN_{\mu\nu},\cK_\mu,\cH_\mu]}
\nonumber\\
&=&\cA(z)+T^A\cB_A(z)-\ft12 T^AT^B\cF_{BA}(z)(-)^A\nonumber\\ &&
+R^\m {\cal C}_\m(z)+
\ft12R^\m T^\n \cN_{\m\n}(z)+\left( T^\m_{+-}+\ft12 R^\mu
T^A\partial_A\right) \cK_\m(z)
+ T^*_\m \cH^\m(z)
\label{new1}\eea
where $\cA(z),\ldots,\cH^\m(z)$ are superfields of the
form \eqn{superfield} and $\cF_{AB}$ and $\cN_{\m\n}$ satisfy
\beq
\cF_{++} =\cF_{--}=0\ ;\qd
\cF_{AB}=-(-)^{AB}\cF_{BA}\ ;\qd
\cN_{\m\n}=\cN_{\n\m}\ .
\label{new4}\eeq
Here $(-)^A$ denotes the grading of $z^A$, i.e.
$(-)^\m=1$ and $(-)^\a=-1$.
\label{lem:superfields}
\end{lemma}
That $\cN_{\m\n}$ can be assumed to be symmetric follows from
$R^\m T^\n=R^\n T^\m$, cf. (\ref{new3b}).
The graded antisymmetry of
$\cF_{AB}$ follows from the commutation relations of the $T^A$
\footnote{The sign factor for the
corresponding term in $\omega$ is the natural one.}.
Note that two components, $\cF_{++}$ and $\cF_{--}$, are missing
because of the identities $T^\pm T^\pm=0$ occurring in
(\ref{new3b}).

The proof of the lemma is straightforward. We just note
that the decomposition
\eqn{new1} is indeed unique since \eqn{new4} implies that the
$\cF_{\m\n}$ are antisymmetric whereas the $\cN_{\m\n}$ are symmetric
under exchange of their indices, i.e.
$T^\m T^\n \cF_{\m\n}(z)$ and $R^\m T^\n \cN_{\m\n}(z)$ are clearly
independent functions in $\sub$.
\QED

Although it is not necessary to include
the term $R^\mu T^A\partial_A \cK_\m(z)$ in
\eqn{new1} (it can be absorbed in the
$\cF_{A\m}$--terms), we have introduced it for later
convenience.

In the final step for the determination of the cohomology,
we explicitly compute ${\cal S}$ on the function \eqn{new1},
and identify the kernel and the image of this operation.
That is done
first for the antifield--independent part in section~\ref{ss:brstcoho},
leading  in section~\ref{ss:genclact} to the classical action.
{}From that action, we know also $\dkt T^*_\mu$ in \eqn{sub4new},
and can then
make the analysis in full generality in section~\ref{ss:fullcoho}.
${\cal S}\omega=0$ and
$\omega\neq {\cal S}\eta$ will impose conditions on the superfields
$\cA(z),\ldots,\cH^\m(z)$ occurring in \eqn{new1}. In particular
these conditions will involve derivatives of the superfields
with respect to the $z^A$. Therefore it is convenient
to introduce
the following shorthand notations for these
derivatives:\footnote{%
Note the double use of the symbol $\6_\a$; we hope
it is  clear from the context whether this stands for
$\6/\6\Theta^\a$ or $\6/\6x^\a$.}
\beq  \{\partial_A\}=\left\{ \6_\m=\frac {\6}{\6X^\m},\,
 \6_\alpha =\frac {\6}{\6\Theta^\alpha }\right\}\ .
\label{sub6}\eeq
This allows to express the ${\cal S}$--transformation of an
arbitrary function of the $z$'s through
\beq {\cal S}\, F(z)=T^A\partial_A F(z) \ .\label{sub5}\eeq

We end this section with three remarks.
\ben
\item Notice that all quantities \eqn{neu2} occur in pairs
$(A,{\cal S} A)$ except for the $T^*_\mu$. However
this does not imply a trivial cohomology on functions
of the $A$'s and  $({\cal S}A)$'s since  their algebra
is not a free one due to \eqn{new3b}. In particular
$z^A$ and $T^A$ do not form a trivial pair, notwithstanding
eq.(\ref{sub4new}): they do not obey
the condition (c) (see the beginning of this section).

\item Note that functions in $\sub$ involve only the six
covariant ghost generators
 $\cc \pm{-1}$, $\cc \pm{0}$ and
$\cc \pm{1}$. They correspond to the two $sl(2)$-copies $\{
\LL \pm{-1},$ $\LL \pm{0},$ $\LL \pm{1}\}$. Furthermore,
one easily checks that all $\LL \pm\mm$ with $\mm=2,3,\ldots$
vanish on the generators  occurring in functions in $\sub$.
Hence, lemma \ref{lem:L00} can be viewed as a reduction of the
${\cal S}$-cohomology in the space of local functions to the
``weak $sl(2)$-Lie algebra cohomology" in $\sub$. However, we cannot
use the standard results on the Lie algebra cohomology
here since the $sl(2)$-representations on the generators are not finite
dimensional (recall that $\LL +{-1}$ and $\LL -{-1}$ act like derivatives
on the generators which leads to infinite multiplets).\footnote{Although
only a finite number of generators contributes
to $\sub$, the usual results on the Lie algebra cohomology do not
apply here since, by setting to zero the other generators, one
would violate \eqn{S7} (nevertheless
${\cal S}\sub\subset \sub$ holds
since the nilpotency of the ghosts prevents those generators which
do not occur in functions $\om\in\sub$ from contributing to
${\cal S}\om$).}
\item Since $\sub$ contains only functions with ghost numbers ranging
from 0 to 6, we conclude  that the cohomology of ${\cal S}$
on local functions is trivial for all other ghost numbers.
According to section \ref{ss:descent} this implies that
the cohomology of ${\cal S}$
on local functionals can be nontrivial at most for ghost numbers
ranging from $-2$ to 4 (in fact the value $-2$ does not occur
since $H^0({\cal S},\sub)$ is representated by a constant as one
can easily check already at this stage).
\een
\section{Strong BRST cohomology on antifield independent functions}
\label{ss:brstcoho}

We have shown in sections \ref{ss:tensors} and
\ref{ss:strategy} that the computation
of the ${\cal S}$-cohomology on local functions
can be reduced to the computation of $H^*({\cal S},\sub)$ which is
the ${\cal S}$-cohomology in the subspace of local functions
described by lemma \ref{lem:superfields}.
As a first step towards the computation of this cohomology we
will now compute the ${\cal S}$-cohomology in a subspace $\subnoa$ of
$\sub$ given by the antifield independent functions.
(We denote this cohomology by $H^*({\cal S},\subnoa)$.)
This can be done consistently, since the
closure of the algebra (absence of quadratic terms in antifields in
the extended action) implies that
the ${\cal S}$-transformation of any function in
$\subnoa$ is again contained in $\subnoa$.
Note that the resulting cohomology classes are not a subset
of the cohomology classes of ${\cal S}$
in the  space of local functions of fields {\em and} antifields:
the image of ${\cal S}$ acting on that space contains functions in
$\subnoa$. Therefore it can happen that an
${\cal S}$-invariant function in $\subnoa$ is
trivial in $H^*({\cal S},\sub)$ although it is
nontrivial $H^*({\cal S},\subnoa)$.
Functions with this property always contain the field equations.
Whereas $H^*({\cal S},\sub)$, to be computed in section \ref{ss:fullcoho},
is a {\em weak} cohomology, $H^*({\cal S},\subnoa)$ is
the strong BRST cohomology,
since on fields the operation ${\cal S}$ is
the BRST operator.

The main reason for computing
$H^*({\cal S},\subnoa)$ first is that it provides, for ghost number 2,
the general classical action $S_{cl}$ described in section
\ref{ss:action}. The latter has to be determined
before we can compute $H^*({\cal S},\sub)$
completely, since it fixes the
${\cal S}$-transformation of the quantities $T^*_\m$, cf.
\eqn{sub4new}.

Now, any function in
$\subnoa$ takes the form \eqn{new1} with $\cH^\m=0$.
Using \eqn{sub4new} and the identities \eqn{new3b}
we obtain for the ${\cal S}$-transformation of
a generic element of $\subnoa$:
\begin{eqnarray}
{\cal S}\omega[\cA,\cB_A,\cF_{AB},{\cal C}_\mu,\cN_{\mu\nu},\cK_\mu,0]=
\omega[0,\tilde \cB_A,\tilde \cF_{AB},0,0,\tilde \cK_\mu, 0]
\label{new6a}\end{eqnarray}
where
\begin{eqnarray}
\tilde \cB_A(z)&=& \partial_A \cA(z)\ ; \nonumber\\
\tilde \cF_{AB}(z)&=& \partial_A \cB_B(z) -(-)^{AB}\partial_B \cB_A(z)
\qd \mbox{for}\qd [AB]\neq[++]\mbox{ or } [--]\ ;
\nonumber\\
\tilde \cK_\mu(z)&=& -2{\cal C}_\mu(z) \ .
\label{new6}\end{eqnarray}
Recall that $\tilde \cF_{++}$ and $\tilde \cF_{--}$
do not occur in $\omega$, cf. \eqn{new4}.

We now analyse the implications for the cohomology.
${\cal S}\omega=0$ requires all superfields
\eqn{new6} to vanish.
The last equation in \eqn{new6} shows
(i) that ${\cal S}\omega=0$ requires
${\cal C}_\mu=0$, and (ii) that the superfield $\cK_\mu$
can be always removed from $\omega$ by subtracting an ${\cal S}$-exact
function.
Next we observe that ${\cal N}_{\mu\nu}$ is not restricted by
\eqn{new6} and does not occur in the image of $\subnoa$ under ${\cal S}$.
Hence, the terms in $\omega$ involving the superfields
${\cal N}_{\mu\nu}$ clearly represent nontrivial
cohomology classes of $H^*({\cal S},\subnoa)$.

The remaining functions are the (graded) antisymmetric
$\cF_{AB}(z)$, $\cB_A(z)$ and $\cA(z)$. They form a super
2--form, 1--form and 0--form, on which ${\cal S}$ acts as a
superderivative.
The first two equations \eqn{new6} show that
${\cal S}\omega=0$ restricts the superfields $\cA$ and $\cB_A$,
and that $\cF_{AB}(z)$ is trivial if it is of the form
$\partial_A \cB'_B -(-)^{AB}\partial_B \cB'_A$ for some superfields
$\cB'_A$. The condition on $\cA$, namely $\6_A\cA=0$,
clearly implies that $\cA$ is constant. The same equation implies
that a super--one--form $\cB_A$  which is exact is in the image of
${\cal S}$, while
${\cal S}\omega=0$ requires $\cB_A$ to be
``almost closed" in superspace.
It would be  closed if
also the conditions
$\tilde \cF_{\pm\pm}=0$ were present,  but this is not the case,
as stated already. This is the reason why the
super-one-form defined through the $\cB_A(z)$ is not
exact in superspace\footnote{If it were really closed then it would
be exact as well---this is easily proved, for all non-vanishing
super-form degrees, just like the usual
Poincar\'e lemma, using that
the superspace coordinates $z^A$ and
the corresponding superspace differentials group in trivial pairs.}.
The extent to which this super--one--form is only ``almost exact"
is described in the following
``super--Poincar\'e lemma for almost closed super--one--forms":
\begin{lemma}
The general solution of
\begin{equation}
\partial_A \cB_B(z) -(-)^{AB}\partial_B \cB_A(z)=0 \quad\mbox{for}\quad
[AB]\neq[++]\mbox{ or } [--],
\label{almostclosed}
\end{equation}
is given by $\6_A\cA'(z)+\de_A^+a_{++} \Theta^++
\de_A^-a_{--}\Theta^-$, i.e.
\beq \cB_\m(z)=\6_\m \cA'(z),\qd
\cB_\pm(z)=\6_\pm \cA'(z)+a_{\pm\pm} \THE \pm \ .
 \label{Bpmconst}\eeq
where $a_{++}$ and $a_{--}$ are arbitrary ($X$--independent) constants.
\label{lem:superPoincare}
\end{lemma}
{\bf Proof:} Explicitly, the equations \eqn{almostclosed} read
\begin{equation}
\partial_{[\mu}\cB_{\nu]}(z)=0\ ;\qd
\partial_{\mu}\cB_\a(z)-\6_\a \cB_{\mu}(z)=0\ ;\qd
\6_{+}\cB_{-}(z)+\6_{-}\cB_{+}(z)=0 \ ,
\label{new7b}
\end{equation}
and we have to prove that this implies \eqn{Bpmconst}
for some  $\cA'(z)$ and $a_{\pm\pm}$.
{}From $\partial_{[\mu}\cB_{\nu]}(z)=0$ we conclude
$\cB_\mu=\partial_\mu \cB(z)$ for some superfield $\cB(z)$, using
the usual Poincar\'e lemma.
The second set of equations \eqn{new7b}
then  yields $\6_\m(\cB_\a-\6_\a \cB)=0$. Using the usual
Poincar\'e lemma again (this time for zero-forms), we conclude
$\cB_\a=\6_\a \cB+\rho_\a+a_{\a\be} \THE \be+
d_\a \THE +\THE -$ where $\rho_\a$, $a_{\a\be}$ and $d_\a$ are
constants and summation over
$\be$ is understood. This implies
\begin{equation}
\6_{\a}\cB_{\be}+\6_{\be}\cB_{\a}=
2a_{(\alpha\beta)}+
2d_{(\beta}\epsilon_{\alpha)\gamma}\Theta^\gamma\ ,
\end{equation}
and the last equation \eqn{new7b}
leads to $a_{(+-)}=0$ and $d_\alpha=0$.
One now easily verifies that this
implies \eqn{Bpmconst}
by setting $\cA'(z)=\cB(z)+\rho_\alpha\Theta^\alpha
+a_{+-}\THE +\THE -$. \QED
The fact that some nontrivial  solutions remain
is due to the {\em absence} of the equations $\tilde\cF_{++}=0$
and $\tilde\cF_{--}=0$: adding these also would kill the solutions.

Therefore we conclude:
\begin{lemma} The BRST-cohomology in $\subnoa$ is given by
\bea  {\cal S}\om=0,\qd \om\in\subnoa
\qd\LRA \qd \om=\omega^0+a_{++} T^+\THE ++a_{--} T^-\THE - \nonumber\\
-\ft12 T^AT^B\cF_{BA}(z)(-)^A+\ft12 R^\m T^\n \cN_{\m\n}(z)
+{\cal S}\eta,\qd
\eta\in\subnoa\label{new9}\eea
where $\omega^0$, $a_{++}$ and $a_{--}$ are constants.
The functions $\cF_{AB}(z)$ that give non-vanishing contributions
are defined only up to ``super-curls", i.e. up to
\beq \cF_{AB}(z)\rightarrow \cF_{AB}(z)+
\6_A \cB'_B(z)-(-)^{AB}\6_B \cB'_A(z)\ .
\label{new10}\eeq
\label{lem:brstcoho}
\end{lemma}
\section{General classical action}\label{ss:genclact}

We are now in the position to determine
the general classical action $S_{cl}$ described
in section \ref{ss:action}. Indeed, according to
section \ref{ss:descent}, $S_{cl}$
can be obtained
from the most general ${\cal S}$-invariant antifield
independent function with ghost number 2. Hence,
it is provided by $H^2({\cal S},\subnoa)$, i.e. by
lemma \ref{lem:brstcoho} for ghost number 2.

Now, up to trivial solutions,
the only parts with ghost number 2 contained in
\eqn{new9} are given by $\ft12 T^\m T^\n B_{\m\n}(X)$ and
$\ft12 R^\m T^\n G_{\m\n}(X)$ where
$B_{\m\n}(X)$ and $G_{\m\n}(X)$ are the
lowest component fields of the superfields
$\cF_{\m\n}(z)$ and $\cN_{\m\n}(z)$ respectively.
Note that they are antisymmetric and symmetric respectively
due to \eqn{new4}. Using \eqn{neu2} we obtain
\bea \omega_0^2&=&\ft12 T^\m T^\n B_{\m\n}(X)+\ft 12 R^\m T^\n G_{\m\n}(X)
\nonumber\\
&=& c^+ c^-X_{1,0}^\mu X_{0,1}^\nu
[G_{\mu\nu}(X)+B_{\m\n}(X)]\label{cla5}\eea
where we have specified the form degree and ghost number
of $\om$ again in order to make contact with the notation used in
section \ref{ss:descent}.
The remaining arbitrariness given by \eqn{new10}
affects only $B_{\m\n}(X)$ and reads
\beq B_{\mu\nu}(X)\ \rightarrow\ B_{\mu\nu}(X)+
2\6_{[\mu}B'_{\nu]}(X)
\label{cla7}\eeq
where $B'_{\m}(X)$ is the lowest component field of the
superfield $\cB'_{\m}(z)$ occurring in \eqn{new10}.

It is now straightforward to evaluate $S_{cl}$
from \eqn{cla5} using the prescription given in section \ref{ss:descent}
which converts invariant functions to invariant functionals.
Applying \eqn{D3} resp. the substitution rule \eqn{substccd} to
\eqn{cla5} results in the 2--form
\begin{equation}
\omega_2^0= dx^+dx^-(1-y)X_{1,0}^\mu X_{0,1}^\nu
[G_{\mu\nu}(X)+B_{\m\n}(X)]\ .
\label{twoform}
\end{equation}
This is the integrand of the most general classical action.
Using \eqn{actsas} it can be cast in a more familiar form:
\begin{equation}
S_{cl}=\int d^2x\, \left( \ft12\sqrt{g}\, g^{\alpha\beta}G_{\mu\nu}(X)
\partial_\alpha X^\mu
\cdot \partial_\beta X^\nu
+B_{\mu\nu}(X)\partial_+X^{\mu}\cdot \partial_- X^{\nu } \right)\ .
\label{genclact}\end{equation}
Notice that adding trivial contributions ${\cal S}\eta^1_0$ with
$\eta^1_0\in\subnoa$ to
\eqn{cla5} results in adding (locally) exact forms to \eqn{twoform},
i.e. total derivatives to the integrand of \eqn{genclact}. Indeed,
since $\eta^1_0\in\subnoa$ does not involve antifields,
the application of \eqn{D3} to
${\cal S}\eta^1_0$ cannot give rise to a 2--form
${\cal S}\eta_2^{-1}$ but only to
$d\eta_1^0$.
Since we neglect total derivatives
whether or not they are total differentials globally,
this does not change \eqn{genclact}. In particular,
a change of $B_{\mu\nu}$ as in \eqn{cla7} gives indeed rise
to a total derivative term as
\begin{equation}
2\int d^2 x\, \partial_+ X^{\mu}\cdot \partial_- X^{\nu }\cdot
\partial_{[\mu}B'_{\nu]}(X)= \int d^2 x\, \left[ \partial_+\left(
B'_\mu \partial_- X^\mu \right) -\partial_-\left( B'_\mu\partial_+
X^\mu\right) \right]\ .  \label{clactBdE}
\end{equation}

We obtain thus the well-known actions of the non--linear $\sigma$--models.
Examples are the WZNW--models where the $X^\mu$
parametrize some Lie group manifold with group elements $G=\exp (X^\mu
T_\mu)$, where $T_\mu$ is a suitable matrix representation of the Lie
algebra. Then $g^{\alpha\beta}\partial_\alpha X^\mu
\cdot \partial_\beta X^\nu\cdot G_{\mu\nu}$ equals $g^{\alpha\beta}
Tr\left( G^{-1}\partial_\alpha G\cdot G^{-1}\partial_\beta G\right) $.
Similarly the second contribution to the classical action can then be
written as a topological term in 3 dimensions.

With formulas of appendix~\ref{app:useform}, the general classical
action \eqn{genclact} can also be written as
\begin{eqnarray}
\lefteqn{S_{cl}=\int d^2x\, \left\{ \ft1{1-y}G_{\mu\nu}(X)[
(1+y)\partial_+
X^\mu \cdot \partial_- X^\nu
-h_{++} \partial_-X^\mu \cdot \partial_-X^\nu\right.}\nonumber\\
& &\left.
\phantom{\ft1{1-y}}-h_{--} \partial_+X^\mu \cdot \partial_+ X^\nu]
+B_{\mu\nu}(X)\partial_+ X^\mu\cdot
\partial_- X^\nu\right\} \ ,
\end{eqnarray}
which gives a suitable form for the equations of motion for $X^\mu$,
providing the $\dkt$--trans\-for\-ma\-tion of $X^*_\mu$ given in
\eqn{delta1}. This results in the following ${\cal S}$--transformation
of the quantity $T^*_\mu$ defined in \eqn{sub3}:
\bea {\cal S}T^*_\mu&=&
-2G_{\mu\nu}T^\nu_{+-}-2\Gamma_{\rho\si,\mu}^-
c^+ c^-X_{1,0}^\rho X_{0,1}^\sigma\nonumber\\
&=&-2G_{\mu\nu}T^\nu_{+-}-2\Gamma_{\rho\si,\mu}^+
c^+ c^-X_{1,0}^\sigma X_{0,1}^\rho
\nonumber\\
&=&-2G_{\mu\nu}T^\nu_{+-}-\Gamma_{\rho\si,\mu}R^\rho T^\si
+\s0 12 H_{\rho\si\mu}T^\rho T^\si
\label{st*}\eea
where  the notations of \eqn{gammadef}--\eqn{Gamma+-def}
have been used.
Note that \eqn{st*} would not change even if we took into
account global properties of the base or target manifold since
the equations of motion for $X^\mu$ remain the same whether or
not the total derivative terms one adds to $S_{cl}$
are globally exact.

\eqn{genclact} is the most general functional satisfying requirements
(i) and (ii) imposed on the classical action in section \ref{ss:action}.
We shall carry out the analysis in the following for this
general form of the classical action. That means that we will
not assume any particular properties of the functions
$G_{\m\n}(X)$ and $B_{\m\n}(X)$, not even invertibility of
$G_{\m\n}(X)$ (This is also the reason why we use the `Levi--Civita
connection' in the form with all indices down.).
The only non-degeneracy restriction we  impose is the implication
\beq G_{\m\n}(X) \, h^\n(X) =\Gamma_{\m\n,\rho}(X) \, h^\rho (X) =
H_{\m\n\rho}(X) \, h^\rho(X) =0
\qd\then\qd h^\m=0\label{localsymm}\eeq
since otherwise requirement (iii) imposed on
$S_{cl}$ in section \ref{ss:action} would be violated.
Indeed, the presence of a non--vanishing solution $h^\m$ of
$G_{\m\n}  h^\n =\Gamma_{\m\n,\rho}  h^\rho  =
H_{\m\n\rho}  h^\rho=0$
would give rise to an additional gauge symmetry of $S_{cl}$ generated
by $\delta_\epsilon X^\m=\epsilon\,  h^\m(X)$ and
$\delta_\epsilon\, g_{\a\be}=0$ where
$\epsilon$ is an arbitrary function (on the two dimensional
base manifold).

As already mentioned, some of the solutions \eqn{genclact} can still
be cohomologically trivial when they can be written as
\beq \int({\cal S}\eta_2^{-1}+d\eta_1^0)\label{repara1}\eeq
for some 2--forms $\eta_2^{-1}$
and $\eta_1^{0}$ involving nontrivially the antifields.
This may look strange at first since
\eqn{genclact} itself is needed to define the ${\cal S}$--transformation
of the antifields. Nevertheless some functionals \eqn{repara1}
connect two different twodimensional actions, which are
then physically equivalent. These connections have a
natural interpretation in terms of (infinitesimal) target
space reparametrizations.  They
have been called sigma model symmetries or
pseudo--symmetries \cite{pssym}, and
occur naturally in the cohomological analysis which we perform.
A generalisation of this statement, concerning field
redefinitions in general can be found in
appendix \ref{ss:repara}.
\section{Complete computation of $H^*({\cal S},\sub)$}\label{ss:fullcoho}

After this intermezzo, which was necessary to determine the full
${\cal S}$ transformation law \eqn{st*} of $T^*_\mu$,
we  come back to the computation of $H^*({\cal S},\sub)$.
We will compute the most general ${\cal S}$--invariant
function \eqn{new1} modulo trivial solutions. We work
henceforth with  a given
action, i.e. for given functions $G_{\m\n}(X)$ and $B_{\m\n}(X)$.
Fixing  these functions is needed, because ${\cal S}T^*_\mu$
depends on them.
Nevertheless we will not have to impose restrictions on
these functions, i.e.
we will compute $H^*({\cal S},\sub)$ completely for {\it any}
given choice of them. In particular
we do not assume $G_{\m\n}(X)$ to be invertible.

We present the result of the calculation of $H^*({\cal S},\sub)$ in the
form of \eqn{new1}. It is more convenient however to express
it in terms of
\begin{equation}
\hat {\cal C}_\mu= {\cal C}_\mu+{\cal H}_\mu\ ,  \label{new11}
\end{equation}
where $\cH_\m$ is obtained from the superfield $\cH^\m$
occurring in \eqn{new1} by lowering its index with $G_{\m\n}$
\beq \cH_\m(z)=G_{\m\n}(X)\cH^\n (z)\ .\label{new12}\eeq
Hence, the  general expression will contain $\hat {\cal C}_\mu$ instead of
${\cal C}_\mu$,  i.e.  the terms
containing  $\hat {\cal C}_\mu$ and
$\cH^\m$ are given by
\beq R^\mu\hat {\cal C}_\mu(z)-R^\mu \cH_\m (z)+
T^*_\m \cH^\m (z)\ .\label{new13}\eeq

With this choice of  basis  for the superfields, closed and exact
functions can be easily identified.
Using \eqn{sub4new} and \eqn{st*} one easily verifies that
the result of ${\cal S}\omega$, gets modified from \eqn{new6} to
\begin{eqnarray}
{\cal S}\omega[\cA,\cB_A,\cF_{AB},{\cal C}_\mu,
\cN_{\mu\nu},\cK_\mu,\cH^\mu]=
\omega[0,\tilde \cB_A,\tilde \cF_{AB},0,
\tilde \cN_{\mu\nu},\tilde \cK_\mu,0]
\label{image}
\end{eqnarray}
with
\begin{eqnarray}
& &\tilde \cB_A(z)= \partial_A \cA(z)\ ;\qd
\tilde{\cal N}_{\mu\nu}=\cF'_{(\mu\nu)}\ ;\qd
\tilde \cK_\mu(z)= -2\hat {\cal C}_\mu(z)\ ; \nonumber\\
& &\tilde \cF_{AB}(z)= \partial_A \cB_B(z) -(-)^{AB}\partial_B \cB_A(z)
+\ft12 (\cF'_{AB}-(-)^{AB}\cF'_{BA}) \ ,
\label{c0new}\end{eqnarray}
where $\cF'_{AB}$ are auxiliary quantities  defined by
\beq\cF'_{\mu\nu}=
D_\mu^- \cH_\nu +D_\nu^+ \cH_\mu \ ;\qd
\cF'_{\mu\pm }=-\cF'_{\pm\mu}=\mp \partial_\pm \cH_\mu \ ;\qd
\cF'_{+-}=\cF'_{-+}=0\ .
\eeq
Here we have used the covariant derivatives \eqn{covdertsp}.
Note that only the symmetric (antisymmetric) part of
$\cF'_{\mu\nu}$ enters in $\tilde \cN_{\mu\nu}$ ($\tilde \cF_{\mu\nu}$)
and that one has
\beq \cF'_{(\mu\nu)}=\6_{\mu}\cH_{\nu} +\6_{\nu}\cH_{\mu}
-2\Gamma_{\mu\nu,\rho}\cH^\rho\ ;\qquad \cF'_{[\mu\nu]}=
H_{\mu\nu\rho}\cH^\rho\ .
\label{parts}\eeq

The functions $\cK_\mu$ and $\hat {\cal C}_\mu$
disappear from the cohomology,
just as $\cK_\mu$ and ${\cal C}_\mu$ in section~\ref{ss:brstcoho}.
We now address the changes in the analysis of  that  section.
The inclusion of the antifield dependent terms,
i.e. the presence of the superfields $\cH^\m(z)$,  modifies
the result of section \ref{ss:brstcoho} in two ways:
\begin{enumerate}
\item New solutions of ${\cal S}\om=0$
involving non--vanishing $\cH^\m(z)$ may exist. As
 one has $\tilde \cH^\mu=0$ in
\eqn{image}, any ${\cal S}$-invariant
function of this type gives a new solution of the cohomology problem.
\item Some of the solutions provided by
lemma \ref{lem:brstcoho} become trivial.
\end{enumerate}
We see immediately from \eqn{c0new} that the second modification
applies only to solutions involving the
superfields $\cF_{AB}(z)$ and the $\cN_{\m\n}(z)$,
whereas the constant solutions and the two solutions
$T^+\THE +$ and $T^-\THE -$ occurring
in \eqn{new9} remain nontrivial. We  postpone
a further specification
until we analyse the cohomology at specific ghost numbers,
and now elaborate on the first modification.

We now investigate whether there are `new' solutions
involving ${\cal H}^\mu$.
The equation $\tilde \cF_{\mu\nu}+\tilde{\cal N}_{\mu\nu}=0$
takes the form
\begin{equation}
0= D_\mu^- \cH_\nu +D_\nu^+ \cH_\mu +
\partial_\mu \cB_\nu -\partial_\nu \cB_\mu\ ,\label{isom}
\end{equation}
which may be decomposed into a symmetric and an antisymmetric
part, corresponding to $\tilde{\cal N}_{\mu\nu}=0$ and
$\tilde \cF_{\mu\nu}=0$ respectively. \eqn{isom}
is the Killing equation for $\cH^\mu$.
The new solutions therefore correspond to  isometries of the
target space.%
\footnote{%
In appendix~\ref{ss:killing} some properties
of Killing vectors and Lie--derivatives are given,
always allowing  a degenerate metric.} We will see that they also encode
the rigid symmetries of the sigma model. Apart from solving the
Killing equation,
there are no more conditions for the part of $\cH^\mu$
which is independent of $\Theta^\pm $ and denoted%
\footnote{We use the notation introduced in \eqn{superfield}.}
by $H^\mu$, and thus has to satisfy
\begin{equation}
D_\mu^- H_\nu +D_\nu^+ H_\mu +
\partial_\mu B_\nu -\partial_\nu B_\mu=0 \ .\label{Hcomp0}
\end{equation}
We call
the non--vanishing $H^\m$ solving these equations {\it
Killing vectors}, and denote a basis for them by
$\{\zeta^\m_\AAA(X)\}$. The corresponding
vectors $B_\m$ in the Killing equation \eqn{Hcomp0} are denoted by
$b_{\mu\AAA}(X)$,
\begin{equation}
 \{\zeta^\m_\AAA(X),b_{\m\AAA}(X):\ \AAA=1,\ldots,N\}\ .
\label{Killingbasis}
\end{equation}

The conditions  $\tilde \cF_{\mu \pm }=0$ require
\begin{equation}
0=\partial_\mu \cB_\pm  -\partial_\pm
(\cB_\mu\pm \cH_\mu) \label{tilFmupm}\ .
\end{equation}
We can solve these equations for
$\partial_+ \cB_\mu$ and $\partial_- \cB_\mu$ and insert the
result in (\ref{isom}) after applying $\6_+$ or $\6_-$ to the latter.
Using \eqn{antisymcovder}, this results in
\begin{equation}
 D_\mu^\pm(\partial_\pm \cH_\nu)=0 \ .  \label{delpmcovconst}
\end{equation}
Hence, $\partial_\pm \cH_\nu$ should be  ``covariantly constant''
vectors.
Such vectors are analysed in section~\ref{ss:covconstKil}, where it
is shown that they are related to the
chiral symmetries, which for the example of
WZW models are the  Ka\v{c}--Moody symmetries.
In particular \eqn{delpmcovconst} requires the component fields
$H^{\mu\, \pm}(X)$ of $\cH^{\m}(z)$ to solve
\begin{equation}
D_\mu^+ H^+_\nu=0 \ ;\qquad D_\mu^- H^-_\nu=0 \ .\label{Hcomp1}
\end{equation}
We denote a basis for these special Killing vectors by
\begin{equation}
\{\zeta_{\AAA^+}^\m(X):\ \AAA^+=1,\ldots,N^+\};\qd
\{\zeta_{\AAA^-}^\m(X):\ \AAA^-=1,\ldots,N^-\}\ .
\label{basisspecialK}
\end{equation}
The numbers $N^+$ and $N^-$ of
$\zeta_{\AAA^+}$'s and $\zeta_{\AAA^-}$'s
are in general different.
As shown in appendix~\ref{ss:covconstKil}, \eqn{Hcomp1} implies
that they satisfy \eqn{isom} with $B_\mu^\pm=\mp H_\mu^\pm$.
Therefore \eqn{basisspecialK} are subsets of \eqn{Killingbasis}.
This implies $N^c\equiv N^++N^-\leq N\leq
D(D+1)/2$, since the latter  is the maximal value of
linearly independent
Killing vectors ($D$ being the range of $\mu$).

The final equation $\tilde \cF_{+-}=0$ gives restrictions on the
possible new solutions only through
the component field $H^{+-}_\mu(X)$ of $\cH_{\m}(z)$.
The conditions \eqn{delpmcovconst} imply that $H^{+-}_\mu$
should be covariantly
constant for both signs of the torsion, i.e.
$D_\mu^+ H^{+-}_\nu=D_\mu^- H^{+-}_\nu=0$. Such vectors are considered
in section~\ref{ss:nonchconstKill}, where we find that $H^{+-}_\mu
=\6_\m\Lambda$ for some ``scalar" $\Lambda(X)$, see \eqn{d2Lambda0}.
However, \eqn{tilFmupm} and $\tilde \cF_{+-}=0$ imply $\Lambda=constant$
and thus $H^{+-}_\mu=0$.
As argued in appendix~\ref{ss:nonchconstKill},
this is only possible if $H^{\mu\,+-}$
generates an extra gauge symmetry distinct from diffeomorphisms and
Weyl transformations. (We obtain the equations \eqn{localsymm}).
We exclude this possibility using assumption (iii) of
section \ref{ss:action} and conclude  $H^{\mu\,+-}=0$.
Including them we would have local symmetries which are not included
in the BRST operator. If we would  include these symmetries
in the BRST operator with new ghosts $c_{+-}$,
the vectors $H^{\mu\,+-}$ would not be cohomological solutions,
but rather
determine the extra term in the extended action at antifield
number~1: $S_{extra}= X^*_\mu H^{\mu\,+-}c_{+-}$.

It is interesting to note how the different types
of symmetries are all organised in terms of the new solutions
$\cH$: all the  rigid symmetries make use of the  $H(X)$ component,
those rigid symmetries related to
the chiral symmetries occur in $H^\pm(X)$, and the extra gauge symmetries
would show up in $H^{+-}(X)$.

We have now analysed all conditions imposed by ${\cal S}\omega=0$
and have used part of the freedom to add trivial solutions for
fixing the form of $\omega$.
We give a summary of all solutions in the form of a theorem:
\begin{theorem} \label{summary theorem}
The  cohomology of ${\cal S}$ on local
functions is given by
\bea {\cal S}\om=0\ \LRA\ \om&=& {\cal S}\eta+
\omega^0+a_{++} T^+\Theta^ ++a_{--} T^-\Theta^ - \nonumber\\
& &-\ft12 T^AT^B\cF_{BA}(z) (-)^A
+\ft12 R^\m T^\n \cN_{\m\n}(z)\nonumber\\
& &+T^\m \cB_\m(z)-R^\mu \cH_\mu (z)+T^*_\m \cH^\mu (z)
\label{new33}\eea
where $\omega^0$, $a_{++}$ and $a_{--}$ are constants and
the superfields $\cB_\m(z)$, $\cH^\m (z)$ and $\cH_\m (z)$ are given
in terms of the solutions of \eqn{Hcomp0} and \eqn{Hcomp1} according to
\bea
 \cB_\m(z)&=&\la^\AAA b_{\m \AAA}(X)-\la^{\AAA^+}\zeta_{\m\AAA^+}(X)\Theta^+
 +\la^{\AAA^-}\zeta_{\m\AAA^-}(X)\Theta^-\ ;
\nonumber\\
\cH^\m(z)&=&\la^\AAA
\zeta^\m_\AAA(X)+\la^{\AAA^+}\zeta^\m_{\AAA^+}(X)\Theta^+
 +\la^{\AAA^-}\zeta^\m_{\AAA^-}(X)\Theta^-\ ,
\label{new34}
\eea
where the $\la$'s are arbitrary constants.
There are still trivial
solutions which can be added to
\eqn{new33} without changing its form for fixed choices
\eqn{Killingbasis} and \eqn{basisspecialK}. \eqn{c0new} shows that
they are given by
\beq {\cal S}[T^A\cB'_A(z)+T^*_\m \cH'^\m(z)-R^\m \cH'_\m(z)]
\label{counterfunction}\eeq
and give rise to the following redefinitions of the superfields
in \eqn{new33}:
\begin{eqnarray}
{\cal N}_{\mu\nu}(z)+\cF_{\mu\nu}(z)&\rightarrow &
{\cal N}_{\mu\nu}(z)+\cF_{\mu\nu}(z)+  2\partial_{[\mu}\cB'_{\nu]}(z)+
D_\mu^- \cH'_\nu (z) +D_\nu^+ \cH'_\mu (z)\ ;
\nonumber\\
\cF_{\mu\pm }(z)&\rightarrow &
\cF_{\mu\pm }(z) +\partial_\mu \cB'_\pm (z)- \partial_\pm\cB'_\mu (z)
\mp \partial_\pm \cH'_\mu (z)\ ;
\nonumber\\
\cF_{+-}(z)&\rightarrow &
\cF_{+-}(z)+ \partial_+ \cB'_-(z) +\partial_- \cB'_+(z)\ ,
\label{trivsolutions}  \label{new35}
\end{eqnarray}
where $\cB'_\mu (z) $ and $\cH'{}^\m (z)$ denote arbitrary superfields and
$\cH'_\nu (z)=G_{\m\n}(X)\cH'{}^\m (z)$.

Hence, the different inequivalent types of solution are:
\begin{enumerate}
\item \label{it:om0}
The constants  $\omega=\omega^0$.
\item  \label{it:apm}
The two solutions $T^+\Theta^ +$ and $T^-\Theta^ -$
stemming from lemma \ref{lem:superPoincare}.
\item \label{it:FAB}
The terms involving the superfields
$\cF_{AB}$ and $\cN_{\mu\nu}$, in so far as they
cannot be put to zero by redefinitions \eqn{trivsolutions}.
\item  \label{it:isom}
The solutions involving the $N$ Killing vectors $\zeta^\m_{\AAA}$
and the respective $b_{\m \AAA}$.
\item  \label{it:constvect}
The terms involving the $N^c$ covariantly constant
Killing vectors $\zeta^\m_{\AAA^+}$, $\zeta^\m_{\AAA^-}$.
\end{enumerate}
The numbers $N^c\leq N$ can be zero.
\end{theorem}

We will now order the solutions according to ghost number
and reduce the remaining arbitrariness by removing trivial solutions.
Recall that a generic superfield $\cF(z)$
contains parts with ghost number ranging from $0$ to $2$.
This is due to the nilpotency of the $\THE \a=\cc \a{0}$.
Since $T ^\mu$, $R^\mu$ and $T_\mu^*$ have ghost number 1
respectively and $T^\pm $ has ghost number 2,
the various superfields and
constants occurring in \eqn{new33} contribute only
to solutions $\omega$ with specific ghost number $G$:
\beq\ba{r|c|c|c|c|c|c}
 & \omega^0 &a_{\pm\pm}  & \cF_{\m\n},\cN_{\mu\nu}
& \cF_{\pm \mu} & \cF_{+-} &\cB_\m,\cH^\m \\ \hline
G & 0&3  & 2,3,4 & 3,4,5 & 4,5,6 &1,2
\ea\ .\label{ghsol}\eeq

Note that the cohomology groups
$H^G({\cal S},\sub)$ are infinite dimensional
for $G=2,\ldots,6$ due to the presence of arbitrary
functions of the $X$'s in the results for these ghost numbers.
It is therefore more instructive to compare the number
of arbitrary functions occurring for the various values of $G$ rather
than the dimensions of the $H^G({\cal S},\sub)$ themselves. Of course
one should subtract from this number the number of
arbitrary functions contained in the remaining
trivial solutions, and add again zero modes of the
trivial solutions. In addition there are
extra solutions or zero modes. An overview is given in
table~\ref{tbl:numbersol}.
\begin{table}[ht]\caption{Overview of the cohomology at fixed ghost number.
The upper indices $\pm$ and $+-$ refer to the component of the superfield
as in \eqn{superfield}.
The numbers indicate the number of arbitrary functions
that characterise the solution. The numbers in square
brackets refer to the number of extra constants. In the counting
we assumed an invertible target space metric (otherwise e.g. $H'_\mu$
does not subtract 2D solutions).}
\tabcolsep 5pt
\label{tbl:numbersol}\begin{center}\begin{tabular}{|c|lr|lr|lr|r|}\hline
G&soln.  & number & zero  & number & zero for& number &
result\\
& & & modes&&zero&&\\ \hline
0&$\omega^0$  & [1] & & & & &[1]\\  \hline
1& $(\zeta_a^\mu,b_{\mu a})$ & $[N]$     & & & & &$[N]$\\ \hline
2& $ \zeta_{a^\pm}^\mu $ & $[N^c ]$ & & & & & $[N^c ]$   \\
 & $F_{\mu\nu}+N_{\mu\nu}$& $D^2$ &$ B'{}_\mu,H'{}_\mu$&$2D$ &$A'',
 \zeta^\mu _a$&
 $ 1+[N]$& $(D-1)^2 +[N]$\\ \hline
3& $a_{\pm\pm} $& [2]& & & & &[2]\\
& $ F_{\mu\nu}^{\pm}+N_{\mu\nu}^{\pm}$& $2D^2$ & $ B'{}_{\mu}^{\pm},
H'{}^\pm_\mu$
 & $4D$ & $A''{}^\pm $&
$2$&$2D(D-1)$\\
 & $F_{\mu\pm }$&$2D$& $B'{}_{\pm }$&2&
$ \zeta^\mu_{a^\pm}$  &$[N^c ]$&$[N^c ]$\\ \hline
4&$F_{\mu\nu}^{+-}+N_{\mu\nu}^{+-}$& $D^2$ & $ H'{}_{\mu}^{+-}$ & $D$ & &   &  \\
 &$ F_{\mu\pm}^{ \pm}      $& $4D$  & $ B'{}_{\mu}^{+-}$ & $D$ & &   &  \\
 &$F_{+-}$& 1 & $ B'{}_{\pm}^{\pm} $ & 4  & $A''{}^{+-}$ & 1 & $D^2+2(D-1)$ \\ \hline
5& $ F_{\mu\pm}^{ +-} $& $2D$& &   &&& $2D$\\
 &$F_{+-}^{\pm}$  &  2  &$ B'{}_{\pm}^{ +-} $ & 2 &&&      \\ \hline
6&$F_{+-}^{+-}$ &1&&&&&1 \\ \hline
\end{tabular}\end{center}\end{table}    \tabcolsep 6pt

We now present the explicit solution for each ghost number.

\underline{$G=0$.} In this case the only solution is
$ \om^0=constant$.
\bsk

\underline{$G=1$.}
The possible solutions are those of
type~\ref{it:isom}
in theorem \ref{summary theorem}.
We can write the result for $\omega^1$ in terms of the
Killing vectors as
\beq \om^1=\la^\AAA\om_\AAA^1\label{G=1a}\ ,\eeq
where $\la^\AAA$ are arbitrary constants and
\beq \om_\AAA^1=T^\m b_{\m \AAA}
            -R^\m G_{\m\n}\zeta^\n_\AAA+
           T^*_\m \zeta^\m_\AAA\ .\label{G=1b}\eeq
\bsk

\underline{$G=2$.} There are two types of solutions with $G=2$:
those of type~\ref{it:FAB} involving the
component fields $N_{\mu\nu}$ and $F_{\mu\nu}$ of $\cN_{\mu\nu}$ and
$\cF_{\mu\nu}$, and
secondly there are the possible solutions of type~\ref{it:constvect}.

Up to trivial solutions we therefore obtain in the case $G=2$
\beq \om^2=\om^2_{(0)}+\la^{\AAA^+}\om_{\AAA^+}^2+\la^{\AAA^-}\om_{\AAA^-}^2
\label{G=20}\eeq
where
\bea
\omega^2_{(0)} &=&
\ft 12 T^\m T^\n F_{\m\n}(X)+\ft 12 R^\m T^\n N_{\m\n}(X)
\nonumber \\ &=&
c^+ c^-X_{1,0}^\mu X_{0,1}^\nu
[F_{\mu\nu}(X)+N_{\m\n}(X)]\ ;
\label{G=2}\\
\om_{\AAA^\pm}^2&=&
[T^*_\mu-(R^\n\pm T^\n)G_{\n\m}]\zeta^\m_{a^\pm}\THE \pm
\nonumber\\
\LRA & &\left\{\ba{l}
\om_{\AAA^+}^2=(T^*_\mu-2c^+X^\n_{1,0}G_{\n\m})\zeta^\m_{a^+}\THE +
\\[.5ex]
\om_{\AAA^-}^2=(T^*_\mu+2c^-X^\n_{0,1}G_{\n\m})\zeta^\m_{a^-}\THE -\ .
\ea\right.
\label{G=2b}\eea
Specialising to $N_{\m\n}=G_{\m\n}$, $F_{\m\n}=B_{\m\n}$ and
$\la^{\AAA^\pm}=0$, this reproduces the result derived in
section~\ref{ss:genclact} for the classical action.
The remaining arbitrariness resulting from
\eqn{trivsolutions} reads:
\bea
\lefteqn{N_{\m\n}(X)+F_{\m\n}(X)\qd \rightarrow}\nonumber\\
& & N_{\m\n}(X)+F_{\m\n}(X)+2\partial_{[\mu}B'_{\nu]}(X)+
D_\mu^- H'_\nu(X) +D_\nu^+ H'_\mu(X)
\label{freedom2}\eea
where $H'{}^\m(X)$ and $B'_\m(X)$ are arbitrary
functions (they are the lowest component fields
of the superfields $\cH'{}^\m (z)$ and $\cB'_\m (z)$
occurring in \eqn{trivsolutions}).
Note that \eqn{freedom2} represents a larger arbitrariness
than its analog \eqn{cla7} found in section \ref{ss:genclact} since
there we did not take the antifields into account (see the
remark at the end of section \ref{ss:genclact} for the interpretation
of this additional freedom in the case that
$F_{\m\n}=B_{\m\n}$ and $N_{\m\n}=G_{\m\n}$).
\bsk

\underline{$G=3$.} There are two types
of solutions with $G=3$ arising from \eqn{new33}: first there are
the solutions of type~\ref{it:apm}
and secondly there are solutions
of type~\ref{it:FAB}
containing the component fields $F_{\m\n}^{\pm}$,
$N_{\m\n}^{\pm}$ and $F_{\m\pm}$ of the
superfields $\cF_{\m\n}$, $\cN_{\m\n}$ and
$\cF_{\m\pm}$. In fact the $F_{\m\pm}$
can be assumed to be zero with no
loss of generality since they can be removed by choosing
\beq B'{}^\pm_\m= -F_{\m\pm}+\6_\m B'_\pm\mp H'{}^\pm_\m
\label{additionaleq}\eeq
in \eqn{trivsolutions}. Here $B'_\pm$ is
irrelevant since contributions $\6_\m B'_\pm$ to $B'{}^\pm_\m$
drop out of \eqn{trivsolutions}, and thus this term can  actually
be omitted in \eqn{additionaleq}
(it corresponds to the zero for zero entry $A''{}^\pm$ in
table~\ref{tbl:numbersol}).
On the other hand $H'{}^\pm_\m $ will
appear again below.
Up to trivial contributions
the solution for $G=3$ thus reads
\beq \om^3=a_{++}\om^3_++a_{--}\om^3_-+\om^3_{X+}+\om^3_{X-}
\ ,\label{G=3a}\eeq
where $a_{\pm\pm}$ are the constants occurring in \eqn{new33} and
\bea \om^3_\pm&=&T^\pm\Theta^ \pm=
-  c^\pm \, \6_\pm c^\pm\cdot
\6_\pm^2 c^\pm \ ;\label{G=3b}\\
\om^3_{X\pm}&=&
\ft 12 R^\m T^\n N_{\m\n}^\pm(X)\THE \pm
+\ft 12 T^\m T^\n F_{\m\n}^\pm(X)\THE \pm
\nonumber \\
&=& c^+ c^-X_{1,0}^\mu X_{0,1}^\nu
[F_{\mu\nu}^\pm(X)+N_{\m\n}^\pm(X)]\THE \pm\ .
\label{G=3c}\eea
The solutions $\om^3_{+}$ and $\om^3_{-}$ are nontrivial, but
$\omega^3_{X\pm }$ can still have trivial contributions.
The remaining arbitrariness is given by those
transformations \eqn{trivsolutions} which
preserve the form \eqn{G=3a}, i.e. which do not reintroduce
$F_{\m\pm}$. These transformations involve only $H'{}^\pm_\m$
since $B'{}^\pm_\m$ is
completely determined in terms of $H'{}^\pm_\m$
according to \eqn{additionaleq} which yields, setting
$F_{\m\pm}=0$ and dropping $\6_\m B'_\pm$,
\beq B'{}^\pm_\m=\mp H'{}^\pm_\m\ .\label{BHrel}\eeq
One easily verifies
that therefore $\om^3_{X\pm}$ are determined only up to
\bea
N_{\m\n}^+(X)+F_{\m\n}^+(X) &\rightarrow & N_{\m\n}^+(X)+F_{\m\n}^+(X)
+2D^+_\n H'{}_{\m }^+(X)\ ;
\nonumber\\
N_{\m\n}^-(X)+F_{\m\n}^-(X) & \rightarrow & N_{\m\n}^-(X)+F_{\m\n}^-(X)
+2D^-_\m H'{}_{\n }^-(X)
\label{freedom3}\eea
where $H'{}^{\m\pm}(X)$ are arbitrary functions. They trivialize
parts of the solutions $N_{\mu\nu}^\pm +F_{\mu\nu}^\pm$, unless they
are themselves covariantly constant Killing vectors,
in which case they do not contribute to \eqn{freedom3}.
\bsk

\underline{$G=4$.}
All solutions are of type~\ref{it:FAB} and involve
the component fields $F_{\m\n}^{+-}$, $N_{\m\n}^{+-}$,
$F_{\m\pm}^\pm$, $F_{\m\pm}^\mp$ and $F_{+-}$
of the corresponding superfields
respectively. Using \eqn{trivsolutions} one verifies that
one can always achieve
\beq F_{+-}=0\ ;\qd F_{\m +}^-=F_{\m -}^+\equiv F_\m
\label{choicemade}\eeq
by choosing
$B'{}_+^-$, $B'{}_-^+$ and $B'{}_\m^{+-}$ appropriately. Note that, again,
only $D+1$ out of the $D+2$ component fields $B'{}_+^-$,
$B'{}_-^+$ and $B'{}_\m^{+-}$
are needed for the choice \eqn{choicemade},
which is related to the zero for zero mode $A''{}^{+-}$ in
table~\ref{tbl:numbersol}.
Hence, one finds up to trivial solutions
\bea \om^4&=&\ft 12  \THE +\THE -
[R^\m T^\n N_{\m\n}^{+-}(X)+T^\m T^\n F_{\m\n}^{+-}(X)]
+T^\m (T^-\THE ++T^+\THE -)F_\m(X)\nonumber\\
& &+T^\m [T^-\THE -F_{\m -}^-(X)+T^+\THE +F_{\m +}^+(X)]\ .
\label{G=4gen}\eea
The remaining arbitrariness is given by
\bea
N_{\m\n}^{+-}(X)+F_{\m\n}^{+-}(X) &\rightarrow&
N_{\m\n}^{+-}(X)+F_{\m\n}^{+-}(X)
+D^-_\m H'{}_{\n }^{+-}(X)+D^+_\n H'{}_{\m }^{+-}(X)\ ;\nonumber\\
F_\m (X)      &\rightarrow &
              F_\m(X)-H'{}_{\m }^{+-}(X)\ ;\nonumber\\
F_{\m \pm}^\pm (X)        &\rightarrow & F_{\m \pm}^\pm (X)
+\6_\m B'{}^\pm_\pm (X)
\label{freedom4gen}\eea
where $H'{}^{\m +-}(X)$ and $B'{}^\pm_\pm(X)$ are
arbitrary functions. If $G_{\m\n}$ is invertible, we can
simplify the result and simultaneously reduce the remaining
freedom. Namely in that case we can remove $F_\m$ by choosing
$H'{}^{\m +-}=G^{\m\n}F_\n$.
Since this fixes $H'{}^{\m +-}$ completely, we are then left with
\bea \det (G_{\m\n})\neq 0:\qd
\om^4&=&\ft 12  \THE +\THE -
[R^\m T^\n N_{\m\n}^{+-}(X)+T^\m T^\n F_{\m\n}^{+-}(X)]
\nonumber\\
& & +T^\m [T^-\THE -F_{\m -}^-(X)+T^+\THE +F_{\m +}^+(X)]
\label{G=4spec}\eea
with the only remaining arbitrariness
\beq
F_{\m \pm}^\pm (X)      \qd  \rightarrow \qd F_{\m \pm}^\pm (X)
+\6_\m B'{}^\pm_\pm(X)\ .
\label{freedom4spec}\eeq
\bsk

\underline{$G=5$.}
Analogously one verifies that the result is,
up to trivial solutions,
\beq \om^5=\THE +\THE - T^\m
\left[ T^-F_{\m -}^{+-}(X)+T^+F_{\m +}^{+-}(X)\right]\ .
\label{G=5}\eeq
No arbitrariness is left, i.e. \eqn{G=5} is nontrivial for any
non--vanishing choice of $F_{\m \pm}^{+-}(X)$.
\bsk

\underline{$G=6$.} Any non--vanishing
function in $\sub$ with ghost number 6 is
${\cal S}$--invariant, nontrivial and of the form
\beq \om^6= T^+T^-\THE +\THE -F^{+-}_{+-}(X)\label{G=6}\eeq
where  $F_{+-}^{+-}(X)$ is an arbitrary (non--vanishing) function.
\section{Results and their interpretation}\label{ss:interpr}

In this section we spell out the results for the antibracket cohomology
on local functionals with ghost numbers $g=-1,0,1$ implied by
the computation of the previous sections. We give their
physical interpretation too.
Of course the results of the previous sections provide also
a complete list of solutions of the cohomology problem for
functionals of all other ghost numbers but no
physical interpretation of them is known yet. We just recall here
that our results imply the absence of such functionals for all
ghost numbers $g<-1$ and $g>4$, and that the results for $g=2,3,4$ can be
easily obtained from \eqn{G=4gen} (or \eqn{G=4spec}),
\eqn{G=5} and \eqn{G=6} by means of
the `ascent prescription' described in section \ref{ss:descent}.
That prescription is given by equations \eqn{D3} resp. \eqn{substccd}
which `integrate' the descent equations
by converting ${\cal S}$--invariant
functions with ghost number $G$ to ${\cal S}$--invariant functionals with
ghost number $g=G-2$. As the analysis in section~\ref{ss:fullcoho} shows,
the following results are valid for any given action of
the form \eqn{genclact}.
\bsk

\noindent {\it $g=-1$: Rigid symmetries and conserved Noether currents.}

For ${\cal S}$--invariant functionals with ghost number $-1$ we have
to start from \eqn{G=1a}, \eqn{G=1b}. The only term for which the
ascent prescription \eqn{substccd} can lead to
$ dx^2\equiv dx^+ dx^- = -dx^- dx^+ $
is the antifield dependent one, as one needs a term quadratic in
ghosts. The only solutions in the cohomology  are then
linear combinations of
\beq W^{-1}_\AAA=\int d^2x \, X^*_\m \, \zeta_\AAA^\m(X)\label{a7}\eeq
where the $\zeta_\AAA$ are the Killing vectors of the
target space, satisfying \eqn{Hcomp0}.
The interpretation of these solutions is well--known:
according to \cite{BBH}
the nontrivial ${\cal S}$--invariant functionals with
ghost number $(-1)$ correspond one-to-one to the
nontrivial rigid symmetries of the classical action generated
by field transformations which are {\it local}, i.e. polynomial
in the derivatives of all fields.
We conclude that the linearly independent solutions
of \eqn{Killingzeta} provide {\em all} nontrivial rigid symmetries
of that type which leave
the corresponding action functional
\eqn{genclact} invariant.\footnote{A
rigid symmetry is called trivial in this context if
the field transformations reduce on--shell
to gauge transformations, possibly with field dependent parameters.}
In particular this implies that any rigid symmetry
generated by local field transformations is
independent of the two dimensional metric,
and of derivatives of the
matter fields and does not contain explicit dependence on the
coordinates $x^\a$ of the two dimensional base manifold.
For instance, Ka\v{c}--Moody symmetries do not occur here since
they are non--local in the space--time metric or zweibein field,
see remarks in appendix~\ref{ss:covconstKil}.
That the Killing vectors indeed generate
rigid symmetries can be easily verified, see e.g.
appendix~\ref{ss:Killingv}.
We also note that the corresponding
conserved Noether currents $j^\a$ whose divergence
vanishes on-shell can be
obtained from  the 1--forms $\om^0_1$
arising from \eqn{G=1b} by the ascent procedure \eqn{substccd} through the
identification \cite{BBH}
\begin{equation}
\om^0_1|_{X^*=0}\equiv dx^\a \ep_{\a\be}j^\be\ ;
\quad \ep_{-+}=-\ep_{+-}=1\ .
\end{equation}
We obtain
\bea & &j_a^\pm=\ft 1{1-y}\left[
(\zeta_{\mu a}\pm b_{\mu a})\nabla_\mp X^\mu
-h_{\mp\mp}(\zeta_{\mu a}\mp b_{\mu a})\nabla_\pm X^\mu
\right]\nonumber\\
& \LRA &
j_a^\a=\left(\sqrt g g^{\a\be}\zeta_{\mu a}+\ep^{\a\be} b_{\mu a}\right)
\6_\be X^\mu\ ; \quad \ep^{+-}=-\ep^{-+}=1\ .
\label{noethercurrents}\eea
One can check that this agrees with \eqn{delScl}.
\bsk

\noindent {\it $g=0$: Action and background charges.}

The antifield--independent solutions with
ghost number 0 arise from \eqn{G=2} and have the same form as the action
itself,
\beq
W^0_{(0)}=\int d^2x\, \left( \ft12\sqrt{g}\, g^{\alpha\beta}N_{\mu\nu}(X)
\partial_\alpha X^\mu
\cdot \partial_\beta X^\nu
+F_{\mu\nu}(X)\partial_+X^{\mu}\cdot \partial_- X^{\nu } \right)
\label{afgh0}\end{equation}
where $N_{\mu\nu}$ and $F_{\mu\nu}$ are arbitrary symmetric resp.
antisymmetric functions.
The equation \eqn{freedom2} implies now that \eqn{afgh0} is
cohomologically trivial iff
\begin{equation}
N_{\mu\nu}(X)+F_{\mu\nu}(X)=\partial_{[\mu}B'_{\nu]}(X)+
D_\mu^- H'_\nu(X) +D_\nu^+ H'_\mu(X)\ ,
\label{trivGB}
\end{equation}
for some $B'_\mu$ and $H'{}^\mu$. In particular,
two actions \eqn{genclact} differing only by a shift in
$G_{\mu\nu}+B_{\mu\nu}$ given by \eqn{trivGB}
are thus cohomologically equivalent. Indeed they should be regarded also
as physically equivalent since contributions $\partial_{[\mu}B'_{\nu]}$
to $B_{\mu\nu}$ give rise to a total derivative in the Lagrangian
while the other
contributions in \eqn{trivGB} are generated by infinitesimal target space
reparametrizations $X^\m\rightarrow X^\m+ H'{}^\mu(X)$.

The antifield--dependent solutions with ghost number 0 arise via the ascent
prescription \eqn{substccd} from \eqn{G=2b}. One gets
\begin{eqnarray}
W^0_{\AAA^\pm} = \int d^2x\left[
X^*_\mu\left(\partial_\pm \xi^\pm+h_{\mp\mp}\partial_\pm\xi^\mp\right)
-\ft2{1-y}\partial_\pm h_{\mp\mp} \cdot  \nabla_\pm X^\nu \cdot
G_{\n\m}(X)\right] \cdot\zeta^\m_{\AAA^\pm}(X)
\label{Mhha}
\end{eqnarray}
where the $\zeta_{\AAA^\pm}$ are special (`covariantly constant')
Killing vectors of the target space,
satisfying \eqn{Hcomp1}.
Hence, the solutions \eqn{Mhha} correspond
one--to--one to these Killing vectors whose existence and particular
form depends on the choice of $G_{\m\n}$ and $B_{\m\n}$.

The interpretation of \eqn{Mhha} is familiar in the
chiral gauge. Taking $h_{++} = 0$, dropping the corresponding $\xi
^-$ ghost, and specialising to $G_{\mu \nu} = \delta _{\mu \nu }$,
we get
\[
\int d^2x (X_\mu ^* \partial _+ \xi^+ -
  2\partial _+ X_\mu \cdot \partial _+ h_{--})\cdot\zeta_+^\mu\ ;
\quad  \zeta_+^\mu=\la^{\AAA^+}\zeta_{\AAA^+}^\mu\]
in which one recognises the so--called `background charge' terms (see
\cite{hiding,stefantoine} for their inclusion in the BV formalism).
To reproduce the well--known form of these background charge terms in the
conformal gauge, one has to include both chiralities, and add
an appropriate BRST--trivial term.
Therefore, $W^0_{\AAA^\pm}$ constitute the generalisation of this chiral
gauge treatment, and will be called background charge terms henceforth.

As we show in detail in \cite{paper2} these
background charge terms have in general two different
interesting applications: a) appropriate linear combinations of them
can be used to construct generalizations
of the corresponding action \eqn{genclact}
(consistent deformations
in the terminology of \cite{BH})
such that the generalized action is invariant
under corresponding extensions of the BRST (resp. gauge)
transformations \eqn{gaugetrafo}, and
b) other linear combinations represent indeed background charges in the
usual sense, i.e. they
can cancel (matter field independent) anomalies if regarded
formally of order $\hbar^{1/2}$.

In fact we show in \cite{paper2} that the actions obtained
from a) generalize the well-known Liouville actions.
\bsk

\noindent {\it $g=1$: Anomalies.}

\eqn{G=3a} provides two types of solutions:
matter field independent ones arising from the $\om_\pm^3$, and
matter field dependent ones arising from the
$\om_{X\pm}^3$. The former read, after performing a partial
integration,
\beq W_\pm^1=\mp 2\int d^2x\, c^\pm\partial_\pm^3 h_{\mp\mp}
=\mp 2 \int d^2x\, (\xi^\pm+h_{\mp\mp}\xi^\mp)\partial_\pm^3 h_{\mp\mp}\ ,
\label{chanom}\eeq
whereas the latter are given by
\begin{equation}
W_{X\pm}^1= \int d^2x\, \ft{1}{ 1-y}
\left(\partial_\pm \xi^\pm+h_{\mp\mp}\partial_\pm\xi^\mp\right) \cdot
 \nabla_+ X^\mu \cdot \nabla_- X^\nu\cdot
\left( N^\pm_{\mu\nu}(X)+F^\pm_{\mu\nu}(X)\right) \ .\label{chanomXdep}
\end{equation}
Some of these are cohomologically trivial. This is the case if
\begin{equation}
N_{\m\n}^+(X) +F_{\m\n}^+(X) = -2D^+_\n H'{}_\m^+(X)\ ;\qd
N_{\m\n}^-(X) +F_{\m\n}^-(X) = -2D^-_\m H'{}_\n^- (X)
\label{trivialanos}
\end{equation}
respectively,
with $D_\m^\pm$ as in \eqn{D+-def}
($H'{}^{\m \pm}(X)$ are arbitrary functions).

The physical interpretation of the solutions
\eqn{chanom} and \eqn{chanomXdep} is well-known: they are
the candidate anomalies. Those which are of the form \eqn{trivialanos}
can still be cancelled by local counterterms.
In section \ref{ss:Weylanom}
we will show that
particular linear combinations of these anomaly candidates indeed
reproduce the well-known Weyl anomalies.
\bsk

Finally we conclude that \eqn{a7},\eqn{afgh0},\eqn{Mhha},\eqn{chanom}
and \eqn{chanomXdep} provide, up to the
(locally) trivial solutions \eqn{trivGB} and \eqn{trivialanos},
a complete list of ${\cal S}$--invariant
functionals with ghost numbers $-1,0,1$. More precisely,
they represent all the inequivalent nontrivial cohomology classes of
these ghost numbers (neglecting ``topological'' solutions which
are locally but not globally trivial).
\section{Weyl anomalies and the dilaton}\label{ss:Weylanom}

The expressions \eqn{chanom} and \eqn{chanomXdep} provide the
candidate anomalies, up to the ${\cal S}$ variations of local counterterms.
All these solutions of the consistency condition can be grouped in two
chirality classes (`right' and `left' ones),
given by $\{W^1_+,W^1_{X+}\}$ and
$\{W^1_-,W^1_{X-}\}$ respectively.
Since the theories under consideration are
governed by left--right symmetric actions \eqn{genclact},
at most left--right symmetric combinations of the
solutions \eqn{chanom} and \eqn{chanomXdep}
are expected
to occur as true anomalies of the theories.
We will therefore now compute those linear
combinations of solutions \eqn{chanom} and \eqn{chanomXdep}
which are left--right symmetric. It will
turn out that, by
subtracting appropriate cohomologically trivial pieces,
all of them can be cast in the
form $\int d^2x (c\, \Om)$ where $c$ denotes the Weyl ghost and
$\Om$ is a density which does not
depend on antifields at all. This form suggests
to interpret them as candidate Weyl anomalies. The latter are of course
the only anomalies that can be present if
one uses a regularization scheme which preserves the diffeomorphism
invariance.

The left--right symmetric combination of the solutions
\eqn{chanom} reproduces precisely \eqn{A0}, up to a trivial solution:
\begin{equation}
W^1_+-W^1_- +{\cal S}M^0 ={\cal
A}_0 \end{equation}
where the counterterm $M^0$ is given by
\begin{equation}
M^0=\int dx^2\! \,  \ft 1{1-y}\left[-\nabla_+ L\cdot \nabla_-L +
\partial_-h_{++}\cdot \left( 2\nabla_-L-r_-
\right) +\partial_+h_{--}\cdot \left(
2\nabla_+L-r_+\right) \right]\ .
\end{equation}
(We have used \eqn{SL} and \eqn{eR}.)

To get the left--right symmetric combinations of the
chiral solutions \eqn{chanomXdep} we have to impose
\begin{equation}
 N^{+}_{\mu\nu} +F^{+}_{\mu\nu} =N^{-}_{\mu\nu}
+F^{-}_{\mu\nu}\equiv f_{\mu\nu}\ .
\label{+=-}
\end{equation}
Then the left--right symmetric matter field dependent
candidate anomalies
are given by the sum $W^1_X=W^1_{X+}+W^1_{X-}$ which indeed
can be transformed to a Weyl anomaly,
\begin{eqnarray}
\lefteqn{W^1_X-{\cal S}
\int d^2x\,  \ft1{1-y}\,L\ \nabla_+X^\mu\cdot \nabla_-X^\nu\cdot
f_{\mu\nu}(X)} \nonumber\\
&&= -\int d^2x\, c\left( \ft12 \sqrt{g}\,
g^{\alpha\beta}\partial_\alpha X^\mu
\cdot \partial_\beta X^\nu\cdot f_{(\mu\nu)}(X)
+\partial_+ X^{\mu}\cdot \partial_- X^{\nu }f_{[\mu\nu]}(X)\right) \ .
\label{AX}\end{eqnarray}
In fact, the anomalies of the general action \eqn{genclact}
have been investigated in \cite{strinbf},
for invertible $G_{\m\n}$, including a non--Weyl
invariant dilaton term, and all above types of anomalies
do appear there.
The dilaton term will be discussed below. Dropping it
for the moment,
they get in \cite{strinbf}, up to one loop,
anomalies of the form \eqn{AX} with
$f_{(\mu\nu)}$ and $f_{[\mu\nu]}$ given by
\begin{equation}
f_{(\mu\nu)}(X) = R_{\mu\nu}(X)-\ft14
H_\mu{}^{\lambda\sigma}(X)H_{\nu\lambda\sigma}(X)\ ;
\qquad  f_{[\mu\nu]}(X) =D_\lambda H^\lambda{}_{\mu\nu}(X)
\label{finstrinbf}
\end{equation}
where $D_\m$ denotes the target space covariant derivative
defined with torsionless connection $\Gamma_{\m\n}{}^\rho(X)$
and $R_{\mu\nu}(X)$ is the corresponding Ricci tensor
of the target space.
They also
get an anomaly of the form \eqn{A0} with coefficient (the second
term is now a two--loop contribution)
\begin{equation}
\frac{D-26}{48\pi^2}+\frac{\alpha'}{16\pi^2}\left\{
-R(X)+\ft1{12}H^2(X)\right\}  \label{A0strinbf}
\end{equation}
(where $(4\pi\alpha')^{-1}$ was put in front of the action, and the
expansion in $\alpha'$ is thus the loop expansion). It was noted
in \cite{strinbf}
that the vanishing of the functions in \eqn{finstrinbf} already
implies that \eqn{A0strinbf} is a constant. According to our
analysis it is anyway only this constant which is cohomologically
nontrivial, and thus is the relevant part of the result.

Let us now discuss how the dilaton terms of \cite{strinbf} arise
in our results.
As we pointed out in sections \ref{ss:fullcoho} and \ref{ss:interpr},
not all solutions \eqn{chanomXdep} are nontrivial but among them
there are trivial ones given by \eqn{trivialanos}.
Furthermore recall that these are the only
trivial solutions. Let us now investigate the trivial left--right
symmetric solutions \eqn{AX}. \eqn{+=-} imposes
\beq
 D^+_\n H'{}_\m^+ =D^-_\m H'{}_\n^-\qd\LRA\qd
D_\mu^- \zeta_\nu +D_\nu^+ \zeta_\mu +
\partial_\mu b_\nu -\partial_\nu b_\mu=0
\label{weyl1}\eeq
where
\beq \zeta^\m=H'{}^{\m -}-H'{}^{\m +}\ ;\qd
b_\m=H'{}_\m^-+H'{}_\m^+\ . \label{weyl2}\eeq
\eqn{weyl1} states that $\zeta^\m$ solves the Killing vector equations
\eqn{Hcomp0}.
Hence, it is a linear combination of the
Killing vectors $\zeta_\AAA^\m$,
and $b_\m$ is the corresponding linear combination
of $b_{\m\AAA}$, up to
a piece $2\6_\m\Ph$ containing an arbitrary function $\Ph(X)$
which drops out of \eqn{weyl1} (a factor of 2 has been introduced
to compare with the results of \cite{strinbf}),
\beq \zeta^\m=\la^\AAA \zeta_\AAA^\m\ ;\qd
b_\m=\la^\AAA b_{\m\AAA}+2\6_\m\Ph\ .\label{weyl3}\eeq
If we insert this result in the triviality
condition $f_{\m\n}=-2D^+_\n H'{}_\m^+$ and assume $G_{\m\n}$
to be invertible in order to make contact with
\cite{strinbf}, we find that
\eqn{AX} is trivial if
\bea \det (G_{\m\n})\neq 0:& & f_{(\m\n)}=-2D_\m\6_\n\Ph
-\la^\AAA D_{(\m}b_{\n)\AAA}\ ;\nonumber\\
& & f_{[\m\n]}=- H_{\m\n}{}^\rho\6_\rho\Ph
-\la^\AAA(\6_{[\m}\zeta_{\n]\AAA}
+\ft12\, H_{\m\n}{}^\rho b_{\rho\AAA})
\label{weyl4}\eea
where
\[ D_\m\6_\n\Ph=\6_\m\6_\n\Ph -\Gamma_{\m\n}{}^\rho\6_\rho\Ph\ .\]
In absence of Killing vectors \eqn{weyl4} reduces precisely to
the anomaly cancellation condition found in \cite{strinbf}%
\footnote{We have a difference in the factor in front of the
$H_{\m\n}{}^\rho\6_\rho\Ph$--term.}.
Notice however that in presence of Killing vectors we find
in fact that the anomaly cancellation condition is more general than
the one imposed in \cite{strinbf}.
It should also be noted that the covariantly constant Killing vectors
drop out of \eqn{weyl4} due to \eqn{b=zeta}, i.e. these Killing
vectors do {\it not} contribute to that anomaly cancellation condition
(rather, they provide the background charges!).

Finally we compute the counterterm whose ${\cal S}$--variation
leads to the anomaly cancellation \eqn{weyl4}.
To that end we recall that the latter arose from
\eqn{counterfunction} where we have to use \eqn{BHrel}. Hence, the
{\it function} whose ${\cal S}$--variation leads to \eqn{weyl4}
is given by
\beq
\eta=-T^*_\m(H'{}^{\mu +}\THE ++H'{}^{\mu -}\THE -)
+(T^\m +R^\m)H'{}_{\mu }^+\THE +-(T^\m -R^\m)H'{}_{\mu }^-\THE -\ .
\label{weyl5}\eeq
The integrand of the
{\it counterterm} we are looking for arises from \eqn{weyl5}
through the ascent prescription \eqn{D3} which converts
$\eta$ to a 2--form with ghost number 0. The resulting
counterterm is
\bea W^0
=-\int d^2\! x\, H'{}^+_\m(X)\left[X^{*\m}(\6_+\xi^++h_{--}\6_+\xi^-)
-\ft 2{1-y}\6_+h_{--}\cdot\nabla_+X^\m\right]
\nonumber\\
-\int d^2\! x\, H'{}^-_\m(X)\left[X^{*\m}(\6_-\xi^-+h_{++}\6_-\xi^+)
-\ft 2{1-y}\6_-h_{++}\cdot\nabla_-X^\m\right]
\label{gencounterterm}
\eea
where one has to insert the
expressions for $H'{}^{\m \pm}$ which result from \eqn{weyl2} and
\eqn{weyl3}, i.e.
\beq H'{}^{\pm}_\m=\6_\m\Ph
+\ft 12 \lambda^a(b_{\m a}\mp \zeta_{\m a})\ .\label{gencountertermb}\eeq
Then $W^0$ is the general form of the counterterm which
can cancel the left-right symmetric  anomalies $W_X^1$ in \eqn{AX}
at the one loop level when added to the action and
multiplied with $\hbar$. The reader can check that
the covariantly constant Killing vectors $\zeta_{a^+}$
contribute only to $H'{}^{+}$ (but not to $H'{}^{-}$) whereas
the $\zeta_{a^-}$
contribute only to $H'{}^{-}$.  Hence, these Killing vectors
occur in \eqn{gencounterterm} only through the functionals
\eqn{Mhha}. Since the latter are ${\cal S}$--invariant,
the covariantly constant Killing vectors do not
contribute to ${\cal S}W^0$ at all, in accordance with the above
observation that they drop out of the anomaly cancellation
condition \eqn{weyl4}.

Let us finally discuss those terms in
\eqn{gencounterterm} which contain the ``dilaton'' $\Ph(X)$.
After performing a partial integration they read
\bea W_\Ph^0&=&
\int d^2\! x\left[ -2\Ph (\nabla_+\ft1{1-y}\6_+h_{--}
+\nabla_-\ft1{1-y}\6_-h_{++})\right.\nonumber\\
&&\phantom{\int d^2\! x}\left.
-X^{*\mu} \6_\mu\Ph
(\partial_\a \xi^\a+h_{--}\partial_+\xi^-+h_{++}\partial_-\xi^+)
\right]\ .\label{weyl7}\eea
Using \eqn{eR}, \eqn{totderiv} and \eqn{SL} and
partial integrations we can cast \eqn{weyl7} in the form
\bea
W^0_\Ph
&=&\int d^2\! x\left[ -2\Ph (\nabla_-\ft1{1-y}\nabla_+L
-\ft 12\, eR)
-X^{*\mu} \6_\mu\Ph ({\cal S}L-\xi^\a \6_\a L-c)
\right]\nonumber\\
&=&\int d^2\! x\left[ \sqrt{g}(
\Ph R+g^{\a\be}\6_\a\Ph\cdot\6_\be L)
-X^{*\mu} \6_\mu\Ph ({\cal S}L-\xi^\a \6_\a L-c)
\right]\ .\label{a)}\eea
Finally we split off an ${\cal S}$-exact piece in the
last term in \eqn{a)} and end up with
\bea
W_\Ph^0&=&\int d^2\! x\left[ \sqrt g\, \Ph R
     -\ft{2}{1-y}\, L\nabla_+X^\nu \cdot \nabla_-X^\mu
     \cdot D^+_\mu \partial_\nu \Ph
\right.
\nonumber\\
&&\phantom{\int d^2\! x}
\left.+\partial_\mu\Ph\cdot X^{*\mu} c
+{\cal S}(L\, X^{*\nu}\6_\nu\Ph)
\right]
\label{weyl8old}
\eea
where the last term may be omitted since it does not contribute
to ${\cal S}W_\Ph^0$ at all.
Combining this with eq.(\ref{AX}), and using eq.(\ref{weyl4}),
we see that the dilaton dependence
of the counterterm that can cancel the matter
field dependent Weyl anomaly is just
\beq
W^0_{\Ph,Weyl}=
\int d^2x \left( \sqrt g\, \Ph R - c\, X^{*\mu} \partial_\mu \Ph \right)\ .
\label{weyl8}
\eeq

\section{Conclusions and final remarks}\label{ss:conclusion}

We investigated the BRST--antibracket cohomology for
two-dimensional theories with
given field content (two-dimensional metric and scalar matter fields)
and given gauge invariances (Weyl and diffeomorphism invariance).
We have solved that cohomology completely both on local functions and
on local functionals, where the latter arises from the
former via the descent equations. Neglecting global aspects,
we found that nontrivial cohomology exists only for ghost numbers
ranging from 0 to 6 in the case of local functions resp. from
$(-1)$ to 4 in the case of local functionals. In particular we
obtained the following results:
\ben
\item  The most general classical action
functional describing the models in question is given by \eqn{genclact}.

\item
The rigid symmetries of the models which are generated
by local field transformations (i.e. by field transformations which are
polynomial in the derivatives of the fields) correspond one-to-one to the
target space isometries, i.e. they are given by the independent
Killing vectors of the target space, solving \eqn{Killingzeta}.
In particular, Ka\v{c}--Moody symmetries are {\it not} present
among these symmetries since they are non-local in the two-dimensional
metric. They are only symmetries, strictly speaking,
after gauge--fixing the metric.

\item
The background charges are obtained from the
covariantly constant Killing vectors of  the target space.
There are in general two types of such Killing vectors,
distinguished by the connection ($\Gamma^+_{\m\n,\rho}$
resp. $\Gamma^-_{\m\n,\rho}$) which occurs in the respective
equation \eqn{constKilvec} defining these Killing vectors.
The general form of the corresponding
background charge terms in the BV--formalism
is given by \eqn{Mhha}.

\item\label{res:anoms}
There are two types of candidate anomalies. Both are
independent of antifields (up to cohomologically trivial contributions),
and both are subdivided in two chirality classes.
Those of the first type do not depend on the matter fields at all and are
represented by the two solutions \eqn{chanom} which are cohomologically
nontrivial and inequivalent. The
left-right symmetric combination of these two candidate anomalies
provides the Weyl anomaly \eqn{A0}.
The candidate anomalies of
the second type involve the matter fields
and are given by \eqn{chanomXdep}. They
depend on arbitrary functions
$N^\pm_{\mu\nu}$ and $F^\pm_{\mu\nu}$ of the matter fields, and are
cohomologically trivial if and only if these functions are
of the form \eqn{trivialanos}.

\item
The general conditions for the absence of
matter field dependent Weyl anomalies
are given by \eqn{weyl4}, expressing which of the corresponding
BRST--cocycles are cohomologically trivial. On the one hand these conditions
reproduce the dilaton terms well-known in the literature \cite{strinbf}.
On the other hand they involve further terms which, to our knowledge,
have not been discussed in the literature yet. These additional
terms occur in presence of isometries of the target space
and involve the corresponding Killing vectors.
The general form of the counterterm which can cancel the
matter field dependent Weyl anomalies
is given by \eqn{gencounterterm}, with $H'{}^\pm_\m$ as in
\eqn{gencountertermb}.
The part of this counterterm involving the dilaton can be
cast in the form \eqn{weyl8}.
Hence, the dilaton need not be introduced by hand but shows up
naturally within the cohomological analysis (and in the counterterm), and
there may exist novel anomaly free target space manifolds with
suitable isometries.
\een

Our presentation has been completely target space covariant. We
started by a covariant transformation rule on the $X$ coordinate,
i.e. it was independent of the choice of coordinates. Then we took
the most general solution for our action. This was then covariant
too. Therefore the cohomology problem was also treated covariantly.

As far as we know, our computation is the first {\em complete} computation
of the cohomology considered. Previous work \cite{BanLaz,OSS,buchb}
contains partial results, and is to some extent inaccurate.
In particular concerning the anomalies, we disagree with
\cite{BanLaz} where it is claimed that {\it all} matter field dependent
candidate anomalies become cohomologically trivial
when the antifields are taken into account. We have given explicitly,
eq. (\ref{chanomXdep}),
the form of the remaining {\em nontrivial} candidates, see also the
discussion under result
\ref{res:anoms} above.
In \cite{OSS} the splitting of all types of candidate
anomalies  in pairs of two cohomologically inequivalent
solutions with different chirality does not stand out.
Furthermore,  the form
of the matter field dependent candidate anomalies given in
\cite{OSS} is {\it not} the most general one, in that only candidate
anomalies are presented there which are Lorentz invariant
in target space. In \cite{buchb} the classical action is not the most
general one in that the torsion term is not present. Also, the chiral
splitting of the matter dependent anomalies is not found either.

After the preliminary report of part of our  work in \cite{cam},
some of our methods have been used also by \cite{Tataru}.
A first criticism on this work is that it  ignores the indices
of the matter fields and therefore overlooks the subtleties
stemming from (anti-) symmetrization
of these indices in $D>1$ target space dimensions.
But even in the case $D=1$ the results in table 2 of \cite{Tataru}
are not the same as ours in
table~\ref{tbl:numbersol}. To compare these tables, one must omit in our
table the zero modes indicated
by '$H$', as they arise from antifield dependent terms
which have not been taken into account in \cite{Tataru} (see discussion
below). Then  table~\ref{tbl:numbersol} would give us for $D=1$ as
number of solutions
involving arbitrary functions for $G=$2, 3, 4, 5, 6 respectively
1, 2, 2, 2, 1. This still differs from table 2 in \cite{Tataru} for $G=4$:
in fact their first two types of solutions can be shown to be identical
cohomologically in the case $D=1$, using the
``counterterm'' ${\cal S}\left[T^\mu \Theta^+ \Theta^-
{B'}_\mu^{+-}(X)\right]= {\cal S}\left[(c^+ X^\mu_{1,0}+ c^-
X^\mu_{0,1})c^+_0 c^-_0 {B'}_\mu^{+-}(X)\right]$.

A more serious criticism, which also applies to \cite{BanLaz,OSS},
is that the antifields are not taken into account fully.
This implies that they investigate strong BRST
cohomology and that their results are in fact gauge--dependent.

To clarify this difference we recall some points about gauge
fixing and BRST in the BV framework (for short reviews, see
\cite{bvber,Tonin}).
All field quantities occur in field-antifield pairs.
The terminology used throughout this paper is that we indicated
as `fields'
all those which have non--negative ghost numbers, while the
`antifields' are those with the negative ghost numbers. This is
referred to as the `classical basis'. Using this basis,
the `BRST'--operator is given by $\Omega=s$,
introduced in \eqn{defdkts}.
Another possible choice is to choose as
fields a set  that has no zero modes in the propagators.
This is referred to as the `gauge--fixed basis'.
For such a basis to exist, it is necessary that the extended action is
proper, although this does not guarantee that the change of basis
can be done in a local and covariant way. The
latter sometimes requires the introduction of extra trivial sectors,
although
this is not necessary in our case, where in the gauge fixed basis the
fields can be chosen to be
\begin{equation}
\{\Phi^A\}=\{ X^\mu, c^+,c^-, b^{++}=h^{++\,*},
b^{--}=h^{--\,*}\}\ .
\end{equation}

Of course, the antibracket cohomology does not depend on the
basis in which it is computed, i.e. our results remain
valid also in the gauge--fixed basis. What changes, however, is
the BRST--operator $\Omega$. On the fields it is defined through
\begin{equation}
\Omega F(\Phi)\equiv \left. {\cal S}F\right|_{\Phi^*=0}\ .
\end{equation}
In general $\Omega^2\approx 0$, where $\approx$ means equality up to
field equations, namely the field equations  of the extended action
with the appropriate
antifields set equal to zero. (In our case $\Omega^2= 0$.)
It can be proven in general \cite{Henncoho,stefantoine} that the
`weak cohomology' of $\Omega$ (in the definition of that cohomology all
equalities are replaced by $\approx$) for local functions is equal
to the cohomology of ${\cal S}$. For integrals of local functions,
this statement holds in the classical basis also for non--negative
ghost numbers, but there is no such statement for the gauge--fixed
basis. This problem could be investigated using the descent
equations if for the gauge--fixed basis the cohomology of local
integrals can be related to that of local functions, as is the case
for the cohomology  of ${\cal S}$.

The work cited above was concerned with the BRST cohomology.
The antibracket cohomology, which we have calculated, is related by
the considerations above to
the {\it weak} BRST cohomology (in the classical basis).
It is the relevant one for
anomalies, physical states,~...~. This remains true if, as in our case,
$\Omega^2=0$.  Thus our
results are more complete than those of
\cite{BanLaz,OSS,Tataru}
where antifields have not been taken into account seriously.
This confirms once more that the inclusion of the antifields in the
cohomological analysis gives more insight
into the properties of a theory than the
antifield independent (strong) BRST cohomology alone and is thus
superior to latter, even if the gauge algebra is closed.
It constitutes another good reason for computing the antibracket
cohomology directly, keeping all the antifields, as we have done.

The advantage of our treatment stands out if one considers
that the Killing vectors enter in the cohomological
analysis only if one includes the antifields (resp. investigates
the weak cohomology). The same holds for the
dilaton terms.
Our results show that the isometries (Killing vectors)
of the target space play an important role in the theory.
They provide all the rigid symmetries of the models,
all background charges and occur in the most general anomaly cancellation
condition. In fact we show in the companion paper \cite{paper2}
that they also give rise to interesting deformations of the models,
possibly providing new non-critical string theories.

\vspace{1cm}

\section*{Acknowledgments.}
This work is carried
out in the framework of the European Community Research Programme
"Gauge theories, applied supersymmetry and quantum gravity", with a
financial contribution under contract SC1-CT92-0789.
\newpage

\appendix
\section{Useful formulae}\label{app:useform}

Let us first remark that we use symmetrization and
antisymmetrization of indices with `total weight 1', i.e.
\begin{equation}
A_{[\mu} B_{\nu]}=\ft12 \left( A_\mu B_\nu - A_\nu B_\mu\right) \
;\qquad
A_{(\mu} B_{\nu)}=\ft12 \left( A_\mu B_\nu + A_\nu B_\mu\right) \ .
\end{equation}

We use covariant derivatives in the world--sheet
\begin{eqnarray}
\nabla _+&\equiv &\partial _+  -h_{++}\partial
_- +\lambda (\partial _-h_{++}\cdot )\ ; \nonumber\\
\nabla _-&\equiv &\partial _-  -h_{--}\partial
_+ -\lambda (\partial _+h_{--}\cdot )
\label{defnabla}\end{eqnarray}
and $\lambda $ is the number of lower $+$ indices of the expression
on which the operator acts ($-$ the number of lower $-$ indices +
the number of upper $-$ indices $-$ the number of upper + indices).
To express ordinary derivatives in terms of covariant ones, we have
\begin{equation}
\partial_\pm Z^{(\lambda)}=\frac{1}{1-y}\left(\nabla_\pm
+ h_{\pm\pm}\nabla_\mp
\mp \lambda  r_\pm\right)Z^{(\lambda)}\ , \label{invdefnabla}
\end{equation}
where  $Z^{(\lambda)} $ is an arbitrary tensor of weight $\lambda$,
and
\begin{equation}
r_\pm= \partial_\mp h_{\pm\pm} - h_{\pm\pm}\partial_\pm h_{\mp\mp}\ .
\end{equation}
Eq.(\ref{invdefnabla}) is often used for the ghosts:
\begin{equation}
\nabla_\pm c^\pm+ h_{\pm\pm} \nabla_\mp c^\pm =
(1-y)\partial_\pm c^\pm -c^\pm r_\pm \ .
\end{equation}
Other useful formulae concerning the covariant derivatives are the
commutators
\begin{equation}
[\nabla_\pm , \partial_\mp ] Z^{(\lambda)}=
\mp \lambda \partial_\mp^2 h_{\pm\pm}
\cdot Z^{(\lambda)}
\end{equation}
and the following identities:
\bea
& &\nabla_+\ft1{1-y}\nabla_- Z^{(0)}=
\nabla_-\ft1{1-y}\nabla_+ Z^{(0)}=
\ft12 \partial_\alpha\left( \sqrt{g}\,g^{\alpha\beta}\partial_\beta
Z^{(0)}\right)\ ;\label{totderiv}\\
&&\nabla_\a (c^\a\ft1{1-y}e Z^{(0)})
=\6_\a(\xi^\a \sqrt g Z^{(0)})\ .
\eea

For the inverse of the metric and some other conversions of
functions of the metric to the chiral basis we have
\begin{eqnarray}
\sqrt{g}\,g^{+-}&=& \frac{g_{+-}}{\sqrt{g}}=\frac{1+y}{1-y}\ ;\qquad
\sqrt{g}\,g^{\pm\pm}=
-\frac{g_{\mp\mp}}{\sqrt{g}}=\frac{-2h_{\mp\mp}}{1-y}\ ;
\nonumber\\
1+y&=&  \frac{2g_{+-}}{g_{+-}+\sqrt{g}}\ ;\qquad
1-y=  \frac{2\sqrt{g}}{g_{+-}+\sqrt{g}}\ .
\end{eqnarray}

For the ghosts, important translation formulae are
\bea
&&\xi^\pm=\ft1{1-y}\, (c^\pm-h_{\mp\mp}c^\mp)\ ;\nonumber\\
&&\nabla_\pm c^\mp=\xi^\a \6_\a h_{\pm\pm}+
\6_\pm \xi^\mp-(h_{\pm\pm})^2\6_\mp\xi^\pm+
h_{\pm\pm}(\6_\pm \xi^\pm-\6_\mp \xi^\mp)
\ ;\nonumber\\
&&\left( \xi^\alpha \partial_\alpha +\frac{\lambda}{1-y}(\xi^+r_+
-\xi^-r_-)\right)
Z^{(\lambda)} = \ft1{1-y}c^\alpha\nabla_\alpha
Z^{(\lambda)}\ .
\eea

The Riemann tensor is defined by
\beq
\ft12 \, R^\a{}_{\be\gamma\delta}=
\6_{[\gamma}\Gamma_{\delta]\be}{}^\a
+\Gamma_{\varepsilon[\gamma}{}^\a\Gamma_{\delta]\be}{}^\varepsilon\ .
\eeq
The curvature scalar can be written as
\beq
\ft12 e R=\ft12 e R^{\a\be}{}_{\be\a}=
\nabla_- \ft1{1-y} \nabla_+ L - \nabla_- \ft1{1-y}
\partial_-
h_{++}-\nabla_+\ft1{1-y} \partial_+ h_{--}\ .  \label{eR}
\eeq
where we introduced
\begin{equation}
L= \ln \frac{e}{1-y}=\ln \frac{g_{+-}+\sqrt{g}}{2}\ . \label{defL}
\end{equation}

For the fields,
the first few of the basis of chiral tensor fields, defined in \eqn{S8}, are
\begin{eqnarray}
X^\mu_{0,0}&=& X^\mu\ ;\nonumber\\
(1-y)X^\mu_{1,0}&=& \nabla_+ X^\mu  \ ;\qquad
(1-y)X^\mu_{0,1}= \nabla_- X^\mu \ ; \nonumber\\
(1-y)X^\mu_{1,1}&=& \nabla_+\ft1{1-y}\nabla_- X^\mu  =
\nabla_-\ft1{1-y}\nabla_+ X^\mu =
\ft12 \partial_\alpha\left( \sqrt{g}\,g^{\alpha\beta}\partial_\beta
X^\mu\right) \ ;
\nonumber\\
(1-y)X^\mu_{2,0}&=& \nabla_+\ft1{1-y}\nabla_+ X^\mu -
\ft1{1-y}r_+\cdot \nabla_+ X^\mu \ ; \nonumber\\
(1-y)X^\mu_{0,2}&=& \nabla_-\ft1{1-y}\nabla_- X^\mu -
\ft1{1-y}r_-\cdot \nabla_- X^\mu\ .  \label{valueT}
\end{eqnarray}
Another useful equation is
\begin{equation} (1-y) X^\mu_{1,0} X^\nu_{0,1}=
\ft{1}{ 1-y}
 \nabla_+ X^\mu\cdot  \nabla_- X^\nu =
\frac{1}{2}\sqrt{g}\, g^{\alpha\beta}\partial_\alpha X^\mu
\cdot \partial_\beta X^\nu
+\partial_+ X^{[\mu}\cdot \partial_- X^{\nu ]}\ ,\label{actsas}
\end{equation}
and the last term is covariantly written as
$\frac{1}{2}\ep^{\alpha\beta}
\partial_\alpha X^\mu\cdot \partial_\beta X^\nu$.

In target space we introduce the connections with torsion
\bea 2\Gamma_{\rho\si,\mu}&\equiv &
\6_\rho G_{\si\mu}+\6_\si G_{\rho\mu}
      -\6_\mu G_{\rho\si}\ ;\label{gammadef}\\
H_{\rho\si\mu}&\equiv &\6_\rho B_{\si\mu}+\6_\si B_{\mu\rho}+
                 \6_\mu B_{\rho\si}\ ; \nonumber\\
\Gamma_{\rho\si,\mu} ^\pm &\equiv &
\Gamma_{\rho\si,\mu}\pm \ft12 H_{\rho\si\mu}
= \Gamma_{\sigma\rho,\mu}^\mp
\ .\label{Hdef}
\label{Gamma+-def} \eea
There are two types of covariant derivatives:
\begin{equation}
D^\pm_\mu V_\nu\equiv \partial_\mu V_\nu  -\Gamma^\pm _{\mu\nu,\rho}V^\rho
\ , \label{covdertsp}\label{D+-def}
\end{equation}
and we have that
\begin{equation}
2\partial_{[\mu}V_{\nu]}=
D_\mu^- V_\nu -D_\nu^+ V_\mu\ . \label{antisymcovder}
\end{equation}
It is understood here
that the fundamental quantities are the $V^\m$ rather than the
$V_\m$ since we do not assume $G_{\m\n}$ to be
invertible and thus cannot use it to raise indices but only to
lower them. When written completely in terms of the $V^\m$,
\eqn{D+-def} reads
\begin{equation}
D^\pm_\mu V_\nu= G_{\n\rho}\6_\m V^\rho
   +\Gamma^\pm_{\m\rho,\n}V^\rho\ .
\label{c12cca}
\end{equation}
Note that  \eqn{c12cca} would be obvious
if  we could define covariant derivatives on vectors with
upper indices. However that requires an invertible metric.

\section{${\cal S}$--transformations in Beltrami basis}
\label{ss:brsttrafos}

On the variables introduced in section \ref{ss:simplectr}
the part $s$ of ${\cal S}$ acts according to
\begin{eqnarray}
sc^+&=&c^+\partial_+c^+\ ;\label{s00}\\
sh_{++}&=&\nabla_+ c^-\ ;\label{s1}\\
sX^\mu&=& c^+X^\mu_{1,0}+c^-X^\mu_{0,1}\ ;\label{s2}\\
s\7X^*_\mu&=&c^+\ft1{1-y}
(\nabla_+-r_+)
\7X^*_\mu+\partial_+c^+\cdot \7X^*_\mu\nonumber\\
& &+c^-\ft1{1-y}
(\nabla_--r_-)
\7X^*_\mu+\partial_-c^-\cdot \7X^*_\mu\ ;\label{s3}\\
sh^{*++}&=&c^-(\partial_-h^{*++}-
h_{--}\7X^*_\mu X^\mu_{0,1})
+c^+\7X^*_\mu X^\mu_{0,1}+2\partial_- c^-\cdot h^{*++}\, ;
\label{s4}\\
sc^*_-&=&c^-\partial_-c^*_-+2\partial_-c^-\cdot c^*_-
\label{s5}
\end{eqnarray}
with $\7X^*_\mu$ as in \eqn{defhat}. The transformations of $c^-,\,
h_{--},\, h^{*--}$ and $c^*_+$ are obtained
from those of $c^+,\, h_{++},\, h^{*++}$ and $c^*_-$
by interchanging all $+$ and $-$ indices.
The weights \eqn{weights} are read off \eqn{s2}--\eqn{s5} since
these equations take the form
\begin{equation}
s Z = w_+ \partial_+ c^+\cdot  Z + w_- \partial_- c^-\cdot Z +
c^+ L^+_{-1} Z + c^- L^-_{-1} Z\ . \label{sZ}
\end{equation}
This allows e.g. to determine
\begin{eqnarray}
{\cal S}X^\mu_{1,0}&=&\partial_+ c^+\cdot  X^\mu_{1,0} + c^+
X^\mu_{2,0}
+ c^- X^\mu_{1,1}\nonumber\\
&=&\ft1{1-y} \left( \nabla_+ c^+ +c^-\nabla_-
+h_{++}\nabla_- c^+\cdot \right) X^\mu_{1,0}\ ,
\end{eqnarray}
and to show that \eqn{actsas} is a density:
\begin{equation}
{\cal S}\left(\ft{1}{ 1-y}
 \nabla_+ X^\mu\cdot  \nabla_- X^\nu \right) =\nabla_\alpha\left(
 c^\alpha
 \ft{1}{ (1-y)^2} \nabla_+ X^\mu\cdot  \nabla_- X^\nu \right)\ .
\label{Sscact}\end{equation}

The part $\dkt$ of ${\cal S}$ is non-vanishing only on the antifields
and for the general classical action \eqn{genclact} given by
\begin{eqnarray}
\dkt \hat X^*_\mu&=&-2G_{\mu\nu}X^\nu_{1,1} -2\Gamma_{\rho\nu,\mu}^-
               X^\rho_{1,0}X^\nu_{0,1} =
 -2G_{\mu\nu}X^\nu_{1,1} -2\Gamma_{\rho\nu,\mu}^+
               X^\nu_{1,0}X^\rho_{0,1}\ ;\label{delta1}\\
\dkt h^{*++}&=&-G_{\mu\nu}X^\mu_{0,1}X^\nu_{0,1}\ ;\label{delta2}\\
\dkt c^*_-&=&-\nabla_+ h^{*++}+ \7X^*_\mu\nabla_- X^\mu
\label{delta3}
\end{eqnarray}
with $\Gamma_{\rho\nu,\mu}^\pm $ as in \eqn{gammadef}.
$\dkt h^{*--}$ and $\dkt c^*_+$ are obtained from $\dkt h^{*++}$
and $\dkt c^*_-$ by interchanging all $+$ and $-$ indices.

The useful quantity
\eqn{defL} transforms according to
\bea
{\cal S}L
&=& \xi^\alpha\partial_\alpha L + \partial_\alpha c^\alpha
-\ft 1{1-y}c^\alpha r_\alpha+c\nonumber\\
&=&\xi^\alpha\partial_\alpha L
+\partial_\alpha \xi^\alpha+h_{--}\6_+\xi^-+h_{++}\6_-\xi^++c \ .
\label{SL}
\eea
The transformation of $eR$ reads
\bea {\cal S}(eR)
&=&2\nabla_- \ft1{1-y}\nabla_+ c+\nabla_\a (c^\a\ft1{1-y}eR)
\nonumber\\
&=&\6_\a(\sqrt g g^{\a\be}\6_\be c+\xi^\a \sqrt g R)\ .
\label{SeR}
\eea
\section{${\cal S}$--exactness of target space
repara\-me\-tri\-za\-tions}\label{ss:repara}

Two actions \eqn{genclact} which differ only by a
(regular) target space
reparametrization should be regarded as physically equivalent.
This fits nicely with the fact that two actions
related by an infinitesimal target space reparametrization
\beq X^\m\rightarrow X^\m+f^\m(X)\label{000}\eeq
are cohomologically equivalent since their difference is
${\cal S}$--exact. This statement is implied by the following
more general one:
\begin{lemma}
The difference of two (local) classical actions $S_0[\phi]$,
$S_0[\phi+\de\phi]$ related by a (local)
infinitesimal field redefinition $\de\phi^i=f^i(\phi,\6\phi,\ldots)$
is ${\cal S}$--exact
in the space of (local) functionals of the fields
and antifields if it
is invariant under ${\cal S}$:\footnote%
{${\cal S}$ itself is defined with the extended
action of which $S_0[\phi]$ constitutes the part in the fields of
zero ghost number.}
\beq S_0[\phi+\de\phi]-S_0[\phi]={\cal S}\Ga[\PH,\PH^*]
\qd \LRA \qd {\cal S}\left( S_0[\phi+\de\phi]-S_0[\phi]\right) =0 \ .
\label{1}\eeq
Here  $\PH$
denotes collectively all fields (including the
ghosts or ghosts for ghosts, ..., corresponding to the gauge
symmetries of $S_0$).
\label{lem:target}
\end{lemma}

{\bf Proof:} The implication $\then$ follows from the nilpotency of
${\cal S}$.
In order to prove the implication $\Leftarrow$ we remark
$S_0[\phi+\de\phi]-S_0[\phi]=\int d^dx\,
( S_0[\phi]\dr/{\delta\phi^i})\de\phi^i$ which implies
\beq S_0[\phi+\de\phi]-S_0[\phi]=\de_{KT}\int d^dx\, \phi^*_i\de\phi^i
={\cal S}\int d^dx\, \phi^*_i\de\phi^i+W[\PH,\PH^*]\label{2}\eeq
where $\de_{KT}$ denotes the Koszul--Tate differential
($\de_{KT}\phi^*_i= S_0[\phi]\dr/{\delta\phi^i}$). It contains all
the terms of ${\cal S}\phi^*$ which have no antifields. Therefore
$W$, a (local) functional of  $\Phi$, $\Phi^*$ of ghost number zero,
contains an antifield and a ghost in all its terms. Since
${\cal S}\left( S_0[\phi+\de\phi]-S_0[\phi]\right) =0$ holds by assumption,
we conclude from \Gl{2}, using ${\cal S}^2=0$,
that $W$ is ${\cal S}$--invariant.
General theorems on the cohomology
of ${\cal S}$ \cite{Henncoho,Henncohob,stefantoine}
then imply that
\beq W={\cal S}\om[\PH,\PH^*] \label{3}\eeq
where $\om$ is a (local) functional. This completes the
proof of \Gl{1} since we have
\[ S_0[\phi+\de\phi]-S_0[\phi]={\cal S}\lb \om+\int d^dx\,
\phi^*_i\de\phi^i\rb.\]
\QED

One may now verify that lemma \ref{lem:target}
applies to the target space reparametrizations
\eqn{000} since $S_{cl}[X^\m+f^\m(X),g_{\a\be}]$
is ${\cal S}$--invariant. That
$S_{cl}[X^\m+f^\m(X),g_{\a\be}]-S_{cl}[X^\m,g_{\a\be}]$ is
indeed ${\cal S}$--exact can be seen from \eqn{freedom2}
and \eqn{metricchange}.

A few comments on the content of the above lemma seem to be
in order here. Notice that the requirement
${\cal S}\left( S_0[\phi+\de\phi]-S_0[\phi]\right) =0$ imposes
a highly nontrivial condition on the variations $\de\phi^i$.
For instance, if the gauge transformations form a closed algebra,
then it requires $S'_0[\phi]\equiv S_0[\phi+\de\phi]$ to be invariant
under exactly the {\it same} gauge transformations of the $\phi^i$
that leave
$S_0[\phi]$ itself invariant (recall that for closed algebras
${\cal S}S_0[\phi]=0$ holds due to the gauge invariance of
$S_0[\phi]$). Furthermore it should be noted that the
symmetries of $S_0[\phi]$ (both the rigid and the gauge symmetries) are
a subset of the transformations $\de\phi^i$ satisfying the
above condition. Namely, if $\de\phi^i$ is a symmetry, then
one even has $\de S_0[\phi]\equiv S_0[\phi+\de\phi]-S_0[\phi] =0$ which is
evidently a stronger condition than ${\cal S}(\de S_0[\phi])=0$.
This suggests to call
transformations $\de\phi^i$ which fulfill ${\cal S}(\de S_0[\phi])=0$
{\it generalized symmetries} or {\it pseudo--symmetries}.

Finally we remark that the lemma \ref{lem:target} applies also
to theories which do not possess a (nontrivial) gauge symmetry at all.
However, in that
particular case there are no conditions on the pseudo--symmetries
since $S_0[\phi+\de\phi]-S_0[\phi]$ is ${\cal S}$--invariant
for arbitrary field redefinitions $\de\phi^i$ because
then ${\cal S}$ reduces to $\de_{KT}$ and thus vanishes on all $\phi^i$.

\section{Lie derivatives and Killing vectors}\label{ss:killing}
In this appendix we collect properties of Lie derivatives, Killing
vectors, and finally special Killing vectors which are
covariantly constant. Most of these properties were found already in
\cite{sigmators} (where the target--space metric has been assumed to
be invertible), but we stress that we will not
assume that $G_{\m\n}$ is invertible. Instead, especially for the
properties of covariantly constant Killing vectors we will use
\eqn{localsymm}, i.e. assumption (iii) of section \ref{ss:action}.
\subsection{Lie derivatives}
The Lie derivative along a vector $H^\m$ is defined, for example
for a 2--tensor $Y_{\mu\nu}$, by
\beq \cL_H Y_{\m\n}\equiv H^\rho \6_\rho Y_{\m\n}+\6_\m H^\rho\cdot
Y_{\rho\n} + \6_\n H^\rho\cdot Y_{\m\rho}\ .\label{c12a}\eeq
The Lie--derivative commutes with the ordinary differential.
For the metric and the antisymmetric tensor $B_{\mu\nu}$ there are
the identities
\begin{eqnarray}
{\cal L}_H G_{\mu\nu}&=&2D_{(\mu}H_{\nu)}
=2\6_{(\m}H_{\n)}-2\Gamma_{\m\n,\rho}H^\rho\ ;\label{LieGB1}\\
{\cal L}_H B_{\mu\nu}&=&-2\partial_{[\mu}\left( B_{\nu]\rho}H^\rho
\right) + H_{\mu\nu\rho}H^\rho\ ;\label{LieGB2}\\
{\cal L}_H H_{\mu\nu\rho}&=&3\partial_{[\mu}\left( H_{\nu\rho]\sigma}
H^\sigma\right) \ ,
\label{LieGB3}\end{eqnarray}
where in \eqn{LieGB3}
we used the Bianchi identity
for the curl of $B_{\mu\nu}$.
The commutator of two Lie derivatives gives a new Lie derivative:
\begin{equation}
\left[{\cal L}_H, {\cal L}_K\right]={\cal L}_L\qquad\mbox{with}\qquad
L^\mu=H^\nu\partial_\nu K^\mu -K^\nu\partial_\nu H^\mu\ ,
\end{equation}
and $L^\mu$ is called the Lie bracket of $H$ and $K$.
\subsection{Killing vectors}\label{ss:Killingv}
Killing vectors $\zeta^\mu$ are defined by the condition that there
exists a vector $b_\mu$ such that
\begin{equation}
D_\mu^- \zeta_\nu +D_\nu^+\zeta_\mu +2\partial_{[\mu}b_{\nu]}=0
\label{Killingzeta}\end{equation}
with $D_\mu^\pm$ as in \eqn{D+-def}.
Splitting this condition in its symmetric and antisymmetric part
using \eqn{LieGB1} and \eqn{LieGB2}, we have
\begin{eqnarray}
0&=&\cL_\zeta G_{\m\n}  \label{isoma1}\\
0&=&H_{\mu\nu\rho}\zeta^\rho+2\partial_{[\mu}b_{\nu]}=
\cL_\zeta B_{\m\n}+ 2\6_{[\mu} \7b_{\nu]}
\qd\mbox{with}\qd \7b_\m=b_\m+B_{\m\n}\zeta^\n\ .\label{isoma2}
\end{eqnarray}
\eqn{isoma2} can also be written without making reference to
a function $b_\mu$ as ${\cal L}_\zeta H_{\mu\nu\rho}=0$.

Killing vectors of a given metric and torsion
generate rigid symmetries of the corresponding
action \eqn{genclact}. In general, replacing
$X^\mu$ by $X^\mu+f^\mu(X) \epsilon$ for arbitrary $f(X)$ and
infinitesimal
$\epsilon$ leads (up to a total derivative) to a classical action with
the replacement
\begin{equation}
G_{\mu\nu}+B_{\mu\nu} \quad\rightarrow \quad
G_{\mu\nu}+B_{\mu\nu} +\epsilon (D_\mu^-f_\nu +D_\nu^+f_\mu) \ ,
\label{metricchange}
\end{equation}
and an extra term
\begin{equation}
\int d^2x\, \sqrt{g}\, g^{\alpha\beta}G_{\mu\nu}(X)
f^\nu (X) \partial_\alpha X^\mu\cdot \partial_\beta \epsilon \ .
\end{equation}
Therefore if $f^\mu$ is replaced by $\zeta^\mu$, satisfying
\eqn{Killingzeta}, the action is invariant provided $\epsilon$ does not
depend on the world--sheet coordinates.
For future reference, if
$\epsilon$ depends on the world--sheet coordinates, we can use
\eqn{clactBdE} to obtain ($\ep^{+-}=-\ep^{-+}=1$)
\begin{equation}
\delta S_{cl}=
\int d^2x\, \left( \sqrt{g}\, g^{\alpha\beta}\zeta_\mu(X)-
\ep^{\alpha\beta}b_\mu(X)\right)
\partial_\alpha X^\mu
\cdot \partial_\beta \epsilon\ .
\label{delScl}
\end{equation}

The commutator
of two (infinitesimal) rigid symmetries is again a rigid
symmetry\footnote{A trivial symmetry cannot occur
in this case since the Killing vectors do not
involve partial derivatives of the $X$'s whereas both the equations
of motion as well as the gauge transformations would necessarily
introduce derivatives of the $X$'s.}.
Therefore the Lie bracket of two Killing vectors gives a new Killing
vector. Introducing a basis of the Killing vectors $\zeta^\mu_\AAA$,
we thus have that
\beq
\zeta_{[\AAA\BB]}^\m=\zeta_\AAA^\n\6_\n\zeta_\BB^\m-\zeta_\BB^\n
\6_\n\zeta_\AAA^\m
\label{c12bb}\eeq
defines again a Killing vector (or vanishes), i.e. it satisfies
again \eqn{isoma1} and \eqn{isoma2} with
\begin{eqnarray}
 \7b_{\m [\AAA\BB]}&=&\lie {\AAA}\7b_{\m \BB}
-\lie {\BB}\7b_{\m \AAA} \qquad\mbox{or}\nonumber\\
b_{\m [\AAA\BB]}&=&\lie {\AAA}b_{\m \BB}
-\lie {\BB}b_{\m \AAA}-H_{\mu\nu\rho}\zeta^\nu_a \zeta^\rho_b\ ,
\label{hatb}
\end{eqnarray}
the latter modulo an irrelevant total derivative
which drops out of \eqn{isoma2}.
In \eqn{hatb} we have used the abbreviation
\beq \lie {\AAA}=\lie {\zeta_{\AAA}}\ .\eeq

\subsection{Covariantly constant Killing vectors}\label{ss:covconstKil}
We shall now derive some useful properties of the
special Killing vectors
\eqn{basisspecialK}. They are defined by
\begin{equation}
D_\mu^\pm \zeta_{\nu a^\pm } =0\ ,\label{constKilvec}
\end{equation}
i.e. they are covariantly constant.
This definition is equivalent to the Killing equations \eqn{isoma1}
and \eqn{isoma2}
with the extra condition
\beq b_{\mu a^\pm}=\mp \zeta_{\mu a^\pm}\ ,
\label{b=zeta}\eeq
i.e. \eqn{constKilvec} is equivalent to
\beq \cL_{a^\pm} G_{\m\n}=0 \ ;\qd
H_{\mu\nu\rho}\zeta^\rho_{a^\pm}=\pm 2\partial_{[\mu}\zeta_{\nu]a^\pm}\ .
\label{constKilvec2}\eeq

These Killing vectors determine the Ka\v{c}--Moody symmetries. Indeed,
in these cases \eqn{delScl} shows that the action is invariant for
transformations with parameters $\epsilon^{a^\pm}$ satisfying
\begin{equation}
\left( \sqrt{g}\, g^{\alpha\beta}\pm
\ep^{\alpha\beta}\right) \partial_\beta \epsilon^{a^\pm} =0\ .
\label{epscond}\end{equation}
One can always find such $\epsilon^{a^\pm}(x)$ for any given metric
$g_{\a\be}(x)$
(In the zweibein
formalism the above equation reduces\footnote{using $\eta^{+-}=1$.}
to $e_\mp^\alpha
\partial_\alpha \epsilon^{a^\pm} =0$).
Hence, given an action \eqn{genclact} with a {\em fixed} metric
$g_{\a\be}(x)$ (keeping $X^\mu$ still arbitrary), $\delta_{\epsilon} X^\m=
\epsilon^{a^\pm}(x) \zeta^\m_{a^\pm}(X)$ generates chiral
(Ka\v{c}--Moody) symmetries of that action,
where $\epsilon^{a^\pm}$ are solutions of
\eqn{epscond}. However,
in our actions the metric $g_{\alpha\beta}$ is a field and thus
has to be regarded in \eqn{epscond} as a variable rather than as a
specific function of the world sheet coordinates. One can then still
solve \eqn{epscond} for $\epsilon^{a^\pm}$ but the solutions involve
infinitely many derivatives of the $g_{\a\be}$ and
are thus nonlocal. Hence, diffeomorphism invariant actions
\eqn{genclact} do not possess Ka\v{c}--Moody symmetries generated by
local field transformations, contrary to sigma models
with non--gauged world--sheet diffeomorphisms, or to the
gauge--fixed theory.
\bsk

Now we derive some useful properties for the
scalar products of the covariantly constant Killing vectors.
We define
\beq P_{a^\pm b^\pm}=\zeta^\m_{a^\pm}G_{\m\n}\zeta^\n_{b^\pm}\ ;
\qd P_{a^+ b^-}=\zeta^\m_{a^+}G_{\m\n}\zeta^\n_{b^-}\ .
\label{products}\eeq
Using \eqn{c12cca}, \eqn{covdertsp} and \eqn{constKilvec}
one easily verifies that $P_{\AAA^+\BB^+}$ and $P_{\AAA^-\BB^-}$
are constant:
\beq \6_\m P_{\AAA^\pm\BB^\pm}=
\6_\m \zeta_{\rho \AAA^\pm}\cdot \zeta^\rho_{\BB^\pm}
+\zeta^\n_{\AAA^\pm}G_{\n\rho}\6_\m \zeta^\rho_{\BB^\pm}=
\zeta^\n_{\AAA^\pm}\zeta^\rho_{\BB^\pm}
(\Gamma^\pm_{\m\rho,\n}-\Gamma^\pm_{\m\rho,\n})
=0\ .
\label{M++result}\eeq
Similarly one verifies that
\beq \6_\m P_{\AAA^+\BB^-}=
\zeta^\n_{\AAA^+}\zeta^\rho_{\BB^-}
(\Gamma^+_{\m\rho,\n}-\Gamma^-_{\m\rho,\n})=
-H_{\m\n\rho}\zeta^\n_{\AAA^+}\zeta^\rho_{\BB^-}\ .
\label{M+-comp}\eeq
\begin{lemma}
The
$\lie {\AAA^+},\lie {\AAA^-}$ span a Lie algebra
which is the direct sum of two subalgebras $\{\lie {\AAA^+}\}$
and $\{\lie {\AAA^-}\}$.
\end{lemma}
{\bf Proof:}
We have to prove that
\beq [\lie {\AAA^\pm},\lie {b^\pm}]=\la^{\CC^\pm}{}_{\AAA^\pm\BB^\pm}
\lie {\CC^\pm}\ ;\qd [\lie {\AAA^+},\lie {\AAA^-}]=0\ ,\eeq
for some constants
$\la^{\CC^+}{}_{\AAA^+\BB^+}$ and $\la^{\CC^-}{}_{\AAA^-\BB^-}$.
This is equivalent to showing that
the Lie bracket of any two
$\zeta_{\AAA^+}$'s is again a linear combination
of the $\zeta_{\AAA^+}$'s (analogously for the $\zeta_{\AAA^-}$'s) and
 the Lie bracket of
$\zeta_{\AAA^+}$ and $\zeta_{\BB^-}$ vanishes for all pairs $(\AAA^+,\BB^-)$, i.e.
\bea \zeta_{[\AAA^+\BB^+]}^\m &=&
\zeta_{\AAA^+}^\n\6_\n\zeta_{\BB^+}^\m-\zeta_{\BB^+}^\n\6_\n\zeta_{\AAA^+}^\m
=\la^{\CC^+}{}_{\AAA^+\BB^+}\, \zeta_{\CC^+}^\m\ ;\label{lie++}\\
\zeta_{[\AAA^-\BB^-]}^\m &=&
\zeta_{\AAA^-}^\n\6_\n\zeta_{\BB^-}^\m-\zeta_{\BB^-}^\n\6_\n\zeta_{\AAA^-}^\m
=\la^{\CC^-}{}_{\AAA^-\BB^-}\, \zeta_{\CC^-}^\m\ ;\label{lie--}\\
\zeta_{[\AAA^+\BB^-]}^\m &=&
\zeta_{\AAA^+}^\n\6_\n\zeta_{\BB^-}^\m-\zeta_{\BB^-}^\n\6_\n\zeta_{\AAA^+}^\m
=0\ .\label{lie+-}\eea

We note that \eqn{lie+-} is proved in one line
by means of \eqn{c12cca} if $G_{\m\n}$ is invertible:
\beann \det (G_{\m\n})\neq 0\ \then\
\6_\n\zeta_{\BB^\pm}^\m=-\Gamma^\pm_{\n\rho}{}^\m\zeta_{\BB^\pm}^\rho\
\then\
\zeta_{[\AAA^+\BB^-]}^\m=\zeta_{\AAA^+}^\n\zeta^\rho_{\BB^-}
(-\Gamma^-_{\n\rho}{}^\m+\Gamma^+_{\rho\n}{}^\m)=0\ .\eeann
For general $G_{\m\n}$ the proof of \eqn{lie++} and \eqn{lie+-}
is more involved.
We first compute, using \eqn{c12cca} and  \eqn{constKilvec},
\[
G_{\m\n}\zeta_{[\AAA^+\BB^\pm]}^\n = \zeta_{\AAA^+}^\n\zeta_{\BB^\pm}^\rho
(-\Gamma^\pm_{\n\rho,\m}+\Gamma^+_{\rho\n,\m})\ .\]
This gives
\bea G_{\m\n}\zeta_{[\AAA^+\BB^+]}^\n &=&
-H_{\m\n\rho}\zeta_{\AAA^+}^\n\zeta_{\BB^+}^\rho\ ;\label{lie1}\\
G_{\m\n}\zeta_{[\AAA^+\BB^-]}^\n &=& 0\ .\label{lie2}\eea
{}From \eqn{lie2} and $\lie {[\AAA^+\BB^-]}G_{\m\n}=0$ we conclude that
\beq \Gamma_{\m\n,\rho} \zeta_{[\AAA^+\BB^-]}^\rho=0\ . \label{lie2a}\eeq
We finally compute $H_{\m\n\rho}\zeta_{[\AAA^+\BB^\pm]}^\rho$
in order to
verify the second equation \eqn{constKilvec2} for
$\zeta_{[\AAA^+\BB^\pm]}^\rho$. Since the latter is a
Killing vector, it satisfies \eqn{isoma2}, i.e.
\beq H_{\m\n\rho}\zeta_{[\AAA^+\BB^\pm]}^\rho
=-2\6_{[\m} b_{\n][\AAA^+\BB^\pm]}\label{liez1}\eeq
with
\bea b_{\m [\AAA^+\BB^\pm]}&=&\mp\lie {\AAA^+}\zeta_{\m \BB^\pm}
+\lie {\BB^\pm}\zeta_{\m \AAA^+}
-H_{\mu\nu\rho}\zeta^\nu_{\AAA^+} \zeta^\rho_{\BB^\pm}\nonumber\\
&=&-(1\pm 1)\zeta_{\m [\AAA^+\BB^\pm]}
-H_{\mu\nu\rho}\zeta^\nu_{\AAA^+} \zeta^\rho_{\BB^\pm}
\label{liez2}\eea
where we used \eqn{hatb}, \eqn{b=zeta} and \eqn{c12bb}.
By means of \eqn{lie1} respectively \eqn{M+-comp} we conclude from
\eqn{liez2}
\beq b_{\m [\AAA^+\BB^+]}=-\zeta_{\m [\AAA^+\BB^+]}\ ;
\qd b_{\m [\AAA^+\BB^-]}=\6_\m P_{\AAA^+\BB^-}\ .\label{liez3}\eeq
Inserting this result in \eqn{liez1} we get
\bea
H_{\m\n\rho}\zeta_{[\AAA^+\BB^+]}^\rho
&=& \6_\m\zeta_{\n [\AAA^+\BB^+]}-\6_\n\zeta_{\m [\AAA^+\BB^+]}\ ;
\label{lie7}\\
H_{\m\n\rho}\zeta_{[\AAA^+\BB^-]}^\rho &=&0\ .\label{lie8}\eea
\eqn{lie7} and $\lie {[\AAA^+\BB^+]}G_{\m\n}=0$ show that
$\zeta_{[\AAA^+\BB^+]}$ solves \eqn{constKilvec2} and hence must be
a linear combination of the $\zeta_{\AAA^+}$'s. This proves
\eqn{lie++} (of course \eqn{lie--} can be proved analogously).
\eqn{lie2}, \eqn{lie2a} and \eqn{lie8} imply $\zeta_{[\AAA^+\BB^-]}^\m=0$
due to \eqn{localsymm}.
This proves \eqn{lie+-}.
\QED
\subsection{Non--chiral covariantly constant Killing vectors}
\label{ss:nonchconstKill}
Consider now a Killing vector $k^\mu$ which is covariantly constant
for both covariant derivatives \eqn{covdertsp}.
For such constant vector, we find that there is no torsion
in this direction
\begin{equation}
H_{\mu\nu\rho}k^\rho=0 \ ,
\label{nonchconstKill}
\end{equation}
and furthermore  because of \eqn{antisymcovder}
\begin{equation}
k_\mu\equiv G_{\mu\nu}k^\nu=\partial_\mu\Lambda\ ;\qquad
\partial_\mu\partial_\nu\Lambda-\Gamma_{\mu\nu,\rho}k^\rho=0\ .
\label{d2Lambda0}
\end{equation}
Note that according to \eqn{nonchconstKill}
one has $b_\m(X)=\6_\m b(X)$ in \eqn{isoma2}. Therefore it
is clear from \eqn{delScl} that if the metric would be degenerate
such that $k_\mu=0$ for non--zero $k^\mu$, then $S_{cl}$
would have an additional local symmetry in contradiction to assumption
(iii) of section \ref{ss:action}.


\begin{thebibliography}{99}
\bibitem{WZcc} J. Wess and B. Zumino, {\em Phys. Lett.} {\bf B37} (1971)
95;\\
W.A. Bardeen and B. Zumino, {\em Nucl. Phys.} {\bf B244} (1984) 421.
\bibitem{BBH} G. Barnich, F. Brandt and M. Henneaux,
{\em Commun. Math. Phys.} {\bf 174} (1995) 57, hep-th/9405109.
\bibitem{FrBrStructure} F. Brandt, `Structure of BRS--invariant
local functionals', preprint NIKHEF-H 93-21,
hep-th/9310123.
\bibitem{BanLaz} G. Bandelloni and S. Lazzarini, {\em Journ.
Math. Phys.} {\bf 34} (1993) 5413; {\em Journ. Math. Phys.}
{\bf 36} (1995) 1, hep-th/9410190.
\bibitem{OSS}
M. Werneck de Oliveira, M. Schweda and S.P. Sorella, {\em Phys. Lett.}
{\bf B315} (1993) 93, hep-th/9305148.
\bibitem{Tataru}
L. T\u{a}taru and I.V. Vancea, `BRST cohomology in Beltrami
parametrization', hep-th/9504036; P.A. Blaga, L. T\u{a}taru and
I.V. Vancea, `BRST cohomology for 2D gravity', hep-th/9504037.
\bibitem{buchb} I.L. Buchbinder, B.R. Mistchuk, V.D. Pershin, {\em Phys.
Lett.} {\bf B353} (1995) 457, hep-th/9502087.
\bibitem{cam} F. Brandt, W. Troost and A. Van Proeyen,
in {\em Geometry of Constrained
Dynamical Systems}, proc. of a conference held at the
Isaac Newton Institute,  Cambridge, June 1994,
ed. J.M. Charap, (Cambridge Univ. Press, 1995), p. 264, hep-th/9407061.
\bibitem{turkproc} W. Troost and A. Van Proeyen,
in {\em Strings and Symmetries}, Lecture Notes in Physics,
Vol. 447, Springer-Verlag,
eds. G. Aktas, C. Saclioglu, M. Serdaroglu, p. 183,
hep-th/9410162.
\bibitem{strinbf} E. Fradkin and A. Tseytlin, {\em Nucl. Phys.} {\bf B261}
(1985) 1;\\
C.G. Callan, D. Friedan, E.J. Martinec and M.J. Perry,
{\em Nucl. Phys.} {\bf B262} (1985) 593.
\bibitem{paper2} F. Brandt, W. Troost and A. Van Proeyen,
`Background charges and consistent continuous deformations of
$2d$ gravity theories', preprint KUL--TF--95/33, hep-th/9510195.
\bibitem{BH}
G. Barnich and M. Henneaux, {\em Phys. Lett.} {\bf B311} (1993) 123.
\bibitem{Becchi} C. Becchi, {\em Nucl. Phys.} {\bf B304} (1988) 513.
\bibitem{BBBeltr} L. Baulieu and M. Bellon, {\em Phys. Lett.} {\bf B196}
(1987) 142.
\bibitem{BV} I.A. Batalin and G.A. Vilkovisky, {\em Phys. Rev.} {\bf
D28} (1983) 2567 (E:{\bf D30} (1984) 508).
\bibitem{anombv} W. Troost, P. van Nieuwenhuizen and A. Van Proeyen,
{\em Nucl. Phys.} {\bf B333} (1990) 727.
\bibitem{GomisParis}  J. Gomis, J. Par\'{\i}s and S. Samuel,
{\em Phys. Rep.} {\bf 259} (1995) 1; hep-th/9412228.
\bibitem{BVboek} W. Troost and A. Van Proeyen, {\em An introduction
to Batalin--Vilkovisky Lagrangian quantization},
Leuven Notes in Math. Theor. Phys., in preparation.
\bibitem{grav} F. Brandt, N. Dragon and M. Kreuzer,
{\em Nucl. Phys.} {\bf B340} (1990) 187.
\bibitem{APL} A.M. Vinogradov, {\em  Sov. Math. Dokl.} {\bf
18} (1977) 1200,
{\bf 19} (1978) 144, {\bf 19} (1978) 1220~; F. Takens, {\em J. Diff.
Geom} {\bf 14} (1979) 543~;
M. De Wilde, {\em Lett. Math. Phys.} {\bf 5} (1981) 351~; W.M.
Tulczyjew,
{\em Lecture Notes in Math.} {\bf 836} (1980) 22~; P. Dedecker and
W.M. Tulczyjew, {\em
Lecture Notes in Math.} {\bf 836} (1980) 498~; T. Tsujishita, {\em
Osaka J.of Math.} {\bf 19}
(1982) 311;
L. Bonora and P. Cotta-Ramusino, {\em Comm. Math. Phys.}
{\bf 87} (1983) 589;
P.J. Olver,
{\em Applications of Lie Groups to Differential Equations},
Graduate Texts in Mathematics, volume 107, Springer Verlag (New York:
1986);
F. Brandt, N. Dragon and M. Kreuzer,
{\em Nucl. Phys.} {\bf B332} (1990) 224;
R.M. Wald, {\em J. Math. Phys.} {\bf 31} (1990)
2378;
M. Dubois-Violette, M. Henneaux, M. Talon and
C.M. Viallet {\em Phys. Lett.} {\bf B267} (1991) 81;
L.A. Dickey, {\em Contemp. Math.} {\bf 132} (1992) 307;
I.M. Anderson, {\em Contemp. Math.} {\bf 132} (1992) 51,
{\em The variational bicomplex},
(Academic Press, Boston, 1994).
\bibitem{Takens} F. Takens, {\em J. Differential
Geometry} {\bf 14} (1979) 543; I.M. Anderson and
T. Duchamp, {\em Amer. J. Math.} {\bf 102} (1980) 781.
\bibitem{Torre} C.G. Torre, {\em Class. Quantum Grav.} {\bf 12} (1995)
L43, hep-gr-qc/9411014.
\bibitem{bbhgrav} G. Barnich, F. Brandt and M. Henneaux,
{\em Nucl.\ Phys.} {\bf B455} (1995) 357, hep-th/9505173.
\bibitem{sugra} F. Brandt, `Lagrangian densities and anomalies in
four-dimensional supersymmetric
theories' (in German), RX--1356, Ph.D. thesis (Hannover, 1991), unpublished;
{\em Class.~Quant.~Grav.} {\bf 11} (1994) 849.
\bibitem{let}
F. Brandt, N. Dragon and M. Kreuzer, {\em Phys. Lett.} {\bf B231} (1989) 263.
\bibitem{infSCd2} A. Van Proeyen, in {\em Super Field Theories}, eds.
H.C. Lee et al., Plenum Press, 1987;\\
K. Schoutens, {\em Nucl. Phys.} {\bf B292} (1987) 150;\\
A. Sevrin and J.W. van Holten, {\em Nucl. Phys.} {\bf B292} (1987) 474.
\bibitem{Henncoho} J. Fisch, M. Henneaux, J. Stasheff and C.
Teitelboim, {\em Commun. Math. Phys.} {\bf 120} (1989) 379;\\
J. Fisch and M. Henneaux, {\em Commun. Math. Phys.} {\bf 128} (1990) 627.
\bibitem{pssym}
G. Moore and P. Nelson, {\em Phys. Rev. Lett.} {\bf  53} (1984)
117;\\
C.M. Hull and A. Van Proeyen,
{\em Phys. Lett.} {\bf B351} (1995) 188,  hep-th/9503022.
\bibitem{hiding}F.~De~Jonghe, R.~Siebelink and W.~Troost,
{\em Phys.\ Lett.}\ {\bf B306} (1993) 295.
\bibitem{stefantoine} S. Vandoren and A. Van Proeyen, {\em Nucl. Phys.} {\bf
B411} (1994) 257.
\bibitem{bvber} A.  Van Proeyen, in Proc.  of the Conference {\it
Strings \& Symmetries 1991}, Stony Brook, May 20--25, 1991, eds.  N.
Berkovits et al., (World Sc.  Publ.  Co., Singapore, 1992), p. 388;\\
W. Troost and A. Van Proeyen,
in {\em Strings 93}, proceedings of the
Conference in Berkeley, CA,  24-29 May 1993,
eds. M.B. Halpern, G. Rivlis and A. Sevrin, (World Sc. Publ. Co.,
Singapore), p. 158; hep-th/9307126.
\bibitem{Tonin} M. Tonin, {\em Nucl. Phys.} {\bf B} (Proc. Suppl.) {\bf
29B,C} (1992) 137.
\bibitem{Henncohob}
M. Henneaux, {\em Commun. Math. Phys.} {\bf 140} (1991) 1.
\bibitem{sigmators} B. de Wit and P. van Nieuwenhuizen, {\em Nucl. Phys.}
{\bf B312} (1989) 58;\\
C.M. Hull and B. Spence, {\em Phys. Lett.} {\bf B232} (1989) 204; {\em Nucl.
Phys.} {\bf B353} (1991) 379.
\end{thebibliography}
\end{document}